Aristotle University of Thessaloniki
School of Sciences
Department of Physics
Section of Astrophysics Astronomy and Mechanics
Mechanics

# Marios Tsatsos

Supervisor assoc. Professor Simos Ichtiaroglou

Dissertation

# Theoretical and Numerical Study of the Van der Pol equation

# Abstract

*Summary:* In this work, we present the basic theoretical efforts that are known in order to deal with non-trivial solutions of the Van der Pol oscillator, such as theory of average, successive approximations and symbolic dynamics. We also construct a set of diagrams (bifurcation, 2D and 3D Fourier power spectra) and maps, based on numerical investigations, corresponding to the expected theoretical results. Furthermore we examine closely the existence of chaotic attractors, both theoretically (with symbolic dynamics) and numerically (period doubling cascades). We construct sounds based on the Fourier spectra, each one corresponding to a periodic, an almost periodic and a chaotic solution and moreover a video showing the variation of the Poincarè map together with the corresponding sounds, as one of the parameters of the system alters. Last we outline modern applications and modeling with Van der Pol oscillator.

All diagrams and numerics have been made with Mathematica (version 5.2) and C++ programming language.

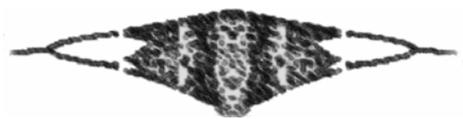

# PREFACE

The present was worked out as a dissertation (diploma thesis) under the supervision of the associate Professor Simos Ichtiaroglou (Physics Department, Aristotle University of Thessaloniki), during the academic year 2005-2006. Our goal is to study both theoretically and numerically the non-trivial solutions and comprehend, in general, the behavior of a specific non linear differential equation of second order;

the van der Pol equation $\quad \ddot{x} = -x - a(x^2-1)\dot{x} + b\cos t \quad$ (0.0)

This equation is one of the most intensely studied systems in non-linear dynamics. Many efforts have been made to approximate the solutions of this equation or to construct simple maps that qualitatively describe important features of the dynamics.

The solutions of this equation are oscillations, which may be periodic or non periodic, as well. We can mark off two cases: the unforced, which is autonomous (the independent variable t does not explicitly occurs in the differential equation, or in (0.0) *b* equals to zero)) and the forced ($b \neq 0$) oscillator, which is non autonomous.

In *Chapter 1* we focus on the van der Pol equations from a historical perspective and we emphasize on the significance of the van der Pol equations with a short reference to some modern applications of the van der Pol system.

In *Chapter 2* and *Chapter 3* we present some basic theorems which are quite useful in order to deal with non linear differential equations. We implement these theorems in the van der Pol equations; in the autonomous equation (*Chapter 2*) the existence of a limit cycle is the main characteristic of the system while in the non autonomous (*Chapter 3*) more complicated attractors occur.

In *Chapter 4* we construct a set of diagrams and numerics, which describe the behavior of our equation, for different values of the set of parameters, according to the predictable theoretical results. The non autonomous equation displays entrainment, quasi-periodicity, and chaos. The characteristics of these different modes are discussed as well as the transitions between the modes. Moreover period doubling cascade is shown, which is considered as a possible route to chaos.

Last in *Chapter 5* some useful adding notes can be found, regarding the procedures that were used in the previous chapters. Finally in *Appendix* we present the codes, which we constructed in Mathematica, in order to plot all the diagrams.

Finally I would like to thank my teachers, Simos Ichtiaroglou and E. Meletlidou, for their help and patience showing to me, student Odysseas-Jamal Maayta for his remarks and last, but not least, student Christos Gentsos for helping me with programmes and plots. Many of them would not have been made without his help.

# Contents

# *Chapter 1*
# INTRODUCTION

## WHY VAN DER POL D.E.'S SYSTEM IS IMPORTANT TO STUDY?

Balthazar van der Pol (1889-1959) was a Dutch electrical engineer who initiated modern experimental dynamics in the laboratory during the 1920's and 1930's. He, first, introduced his (now famous) equation in order to describe triode oscillations in electrical circuits, in 1927. The mathematical model for the system is a well known second order ordinary differential equation with cubic nonlinearity – the Van der Pol equation. Since then thousands of papers have been published achieving better approximations to the solutions occurring in such non linear systems. The Van der Pol oscillator is a classical example of self-oscillatory system and is now considered as very useful mathematical model that can be used in much more complicated and modified systems.

But, why this equation is so important to mathematicians, physicists and engineers and is still being extensively studied?

*Historical Perspective*

During the first half of the twentieth century, Balthazar van der Pol pioneered the fields of radio and telecommunications [1,2, 3, 4, 5, 6]. In an era when these areas were much less advanced than they are today, vacuum tubes were used to control the flow of electricity in the circuitry of transmitters and receivers. Contemporary with Lorenz, Thompson, and Appleton, Van der

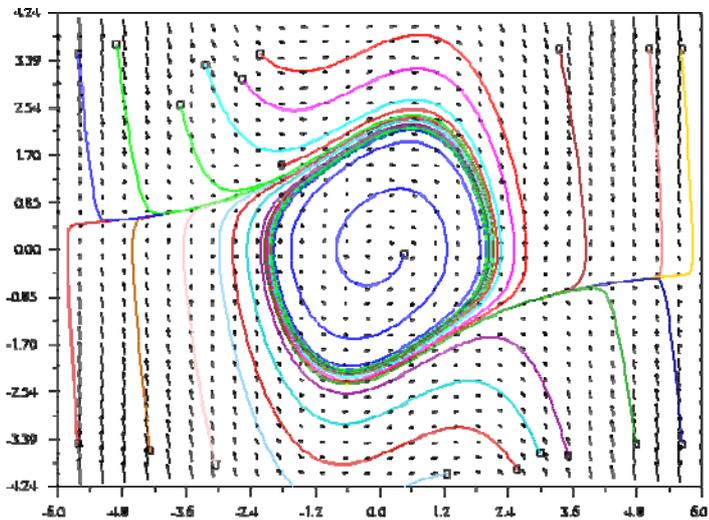

*Figure showing the behavior (in phase space) of a self-oscillatory system*

Pol, in 1927, experimented with oscillations in a vacuum tube triode circuit and concluded that all initial conditions converged to the same periodic orbit of finite amplitude. Since this behavior is different from the behavior of solutions of linear equations, van der Pol proposed a nonlinear differential equation

$$\ddot{x} + \mu(x^2 - 1)\dot{x} + x = 0, \tag{1.1}$$

commonly referred to as the (unforced) van der Pol equation [3], as a model for the behavior observed in the experiment. In studying the case $\mu \gg 1$, van der Pol discovered the importance of what has become known as *relaxation oscillations* [4].

The relaxation oscillations have become the cornerstone of geometric singular perturbation theory and play a significant role in the analysis presented here. Van der Pol went on to propose a version of (1.1) that includes a periodic forcing term:

$$\ddot{x} + \mu(x^2 - 1)\dot{x} + x = a\sin(\omega t) \qquad (1.2)$$

In a similar equation, he and van der Mark first noted the existence of two stable periodic solutions with different periods for a particular value of the parameters and observed noisy behavior in an electrical circuit modelled with (1.2) [6]. Van der Pol further speculated that (1.2) also had this property. Van der Pol and Van der Mark [6] in their investigations of the oscillator behaviour in the relaxation oscillation regime found that the subharmonical oscillations are appeared during changes of natural frequency of the system. **Moreover, the authors** (in the September 1927 issue of the British journal Nature), **noted appearing of "irregular noise" before transition from one subharmonical regime to another. It seems to be it was one of the first observations of chaotic oscillations in the electronic tube circuit.** Their paper is probably one of the first experimental reports of chaos - something that they failed to pursue in more detail.

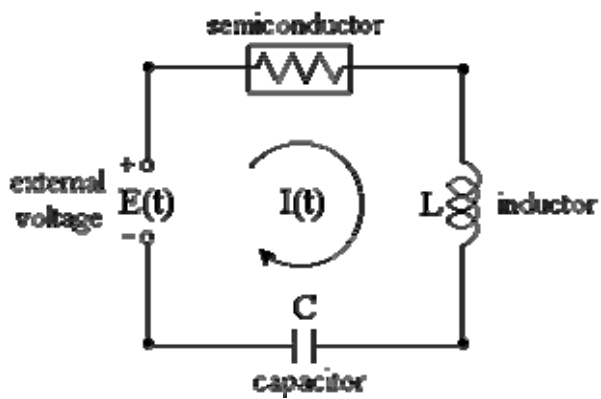

Van der Pol's work on nonlinear oscillations and circuit theory provided motivation for the seminal work of Cartwright and Littlewood. In 1938, just prior to World War II, the British Radio Research Board issued a request for mathematicians to consider the differential equations that arise in radio engineering. Responding to this request, Cartwright and Littlewood began studying the forced van der Pol equation and showed that it does indeed have bistable parameter regimes. In addition, they showed that there does not exist a smooth boundary between the basins of attraction of the stable periodic orbits. They discovered what is now called chaotic dynamics by detailed investigation of this system [2, 7, 8, 9, 10].

Later, in 1945, Cartwright and Littlewood [11] during their analyzing the Van der Pol equation with large value of nonlinearity parameter have shown that the singular solutions exist. In 1949 Levinson [12], analytically analyzing the Van der Pol equation, substitutes the cubic nonlinearity for piecewise linear version and shows that the equation has singular solutions in the case also. In this version, in piecewise linear model of sinusoidally forced Van der Pol oscillator, has been shown, on the basis of numerical simulation, that the transition from periodic oscillations to chaotic ones is possible [13]. The transition takes place through cascades of period doubling. It was revealed that chaotic behaviour is appeared in the Van der Pol equation with smooth nonlinearity by period doubling bifurcations [14] and crossroad area is observed in synchronization tongues [13].

*Modern Applications*

Since its introduction in the 1920's, the Van der Pol equation has been a prototype for systems with self-excited limit cycle oscillations. The classical experimental setup of the system is the oscillator with vacuum triode. The investigations of the forced Van der Pol oscillator behaviour have carried out by many researchers. The equation has been studied over wide parameter regimes, from perturbations of harmonic motion to relaxation oscillations. It was much attention dedicated to investigations of the peculiarities of the Van der Pol oscillator behaviour under external periodic (sinusoidal) force and, in particular, the synchronization phenomena and the dynamical chaos appearing (see e.g., [14, 15, 16]). The Van der Pol equation is now concerned as a basic model for oscillatory processes in physics, electronics, biology, neurology, sociology and economics [17]. Van der Pol himself built a number of electronic circuit models of the human heart to study the range of stability of heart dynamics. His investigations with adding an external driving signal were analogous to the situation in which a real heart is driven by a pacemaker. He was interested in finding out, using his entrainment work, how to stabilize a heart's irregular beating or "arrhythmias". Since then it has been used by scientists to model a variety of physical and biological phenomena. For instance, in biology, the van der Pol equation has been used as the basis of a model of coupled neurons in the gastric mill circuit of the stomatogastric ganglion [18, 19] (see also Appendix). The Fitzhugh–Nagumo equation is a planar vector field that extends the van der Pol equation as a model for action potentials of neurons. In seismology, the van der Pol equation has been used in the development a model of the interaction of two plates in a geological fault [20].

After this short historical review, about the significance of the Van der Pol equation, we are ready to begin investigating its behavior in the aspect of …Mathematics!

# PART A – THEORETICAL STUDY

## Chapter 2

### THE AUTONOMOUS VAN DER POL EQUATION

*A simple form of the theory of Averaging*

We are dealing with a perturbation problem (like the Van der Pol equation) in the standard form, which we write as

$$\frac{dx}{dt} = \varepsilon f(t,x) + \varepsilon^2 g(t,x,\varepsilon) \quad x(t_0) = x_0$$

Suppose that $f$ is T-periodic, then it seems natural to average[i] $f$ over t, (while holding x constant). So we consider the averaged equation

$$\frac{dy}{dt} = \varepsilon f^0(y) \, , \quad y(t_0) = x_0$$

with

$$f^0(y) = \frac{1}{T}\int_0^T f(t,x)dt.$$

Making some assumptions, we can establish the following asymptotic result, as long as we introduce a definition:

*Definition*

Consider the vectorfield $f(t,x)$ with $f$: $R \times R^n \to R^n$, Lipschitz-continuous in x on $D \subset R^n$, $t \geq 0$; $f$ continuous in t and $x$ on $R^+ \times D$. If the average

$f^0(x) = \lim_{T\to\infty} \frac{1}{T} \int_0^T f(t,x)dt$  exists $f$ is called a *KBM – vectorfield* (KBM stands for Krylov, Bogoliubov and Mitropolsky).

*Theorem (general averaging)*

Consider the initial value problems

$\dot{x} = \varepsilon f(t,x) \, , \quad x(0) = x_0$

with $f$: $R \times R^n \to R^n$ and

$\dot{y} = \varepsilon f^0(y) \, , \quad y(0) = x_0$

$x, y, y_0 \in D \subset R^n$, $t \in [0, \infty)$, $\varepsilon \in (0, \varepsilon_0]$. Suppose

a)     $f$ is a KBM-vectorfield with average $f^o$;

---

[i] If $\varepsilon << 1$, $dx/dt$ will vary very slowly as t changes. That's why we refer in t as "fast time" and in $\varepsilon t$ as "slow time". With the averaging method we can eliminate the fast time and obtain an outline of the slow time evolution

b)   $y(t)$ belongs to an interior subset of $D$ on the time-scale $\frac{1}{\varepsilon}$;

then $x(t) - y(t) = O(\delta^{1/2}(\varepsilon))$ as $\varepsilon \to 0$ on the time scale $\frac{1}{\varepsilon}$,

where

$$f^0(x) = \lim_{T \to \infty} \frac{1}{T} \int_0^T f(t,x)dt,$$

$$\delta(\varepsilon) = \sup_{x \in D} \sup_{t \in [0, \frac{L}{\varepsilon})} \varepsilon \left| \int_0^t [f(\tau,x) - f^0(x)]d\tau \right|.$$

## Implementation of the theory of Averaging to the autonomous Van der Pol Equation

Van der Pol equation is written (as mentioned before) as
$$\ddot{x} = -x - a(x^2 - 1)\dot{x} + b\cos\omega t \qquad (2.1)$$
and in the case that we have no excitation it becomes
$$\ddot{x} = -x + a(1-x^2)\dot{x} \quad \text{or} \quad \ddot{x} = -x + af(x,\dot{x})$$
for an arbitrary damping function $f$ which is sufficiently smooth in $D \subset R^2$. Equation (2.1) can equivalently be written in the form:

$$\begin{cases} \dot{x} = y - a\left(\frac{x^3}{3} - x\right) \\ \dot{y} = -x + b\cos(\omega t) \end{cases} \xrightarrow{\text{for } b=0} \begin{cases} \dot{x} = y - a\left(\frac{x^3}{3} - x\right) \\ \dot{y} = -x \end{cases} \qquad (2.2)$$

The initial values are $x(0)$ and $y(0)$. This is a quasilinear system[ii]. We use the phase-amplitude transformation to put the system in the standard form. [Sanders, 21]

$$\begin{aligned} x &= r\cos(t+\psi) \\ y &= -r\sin(t+\psi) \end{aligned} \qquad (2.3)$$

Note that the second equation of (2.3) is not the derivative of the first one. It's nothing but a transformation, still not accidentally chosen.

The perturbation equations become
$$\begin{aligned} \frac{dr}{dt} &= -a\sin(t+\psi)f(r\cos(t+\psi), -r\sin(t+\psi)), \quad r(0) = r_0 \\ \frac{d\psi}{dt} &= -\frac{a}{r}\cos(t+\psi)f(r\cos(t+\psi), -r\sin(t+\psi)), \quad \psi(0) = \psi_0 \end{aligned} \qquad (2.4)$$

---

[ii] *quasilinear* is a system in the form: $\dfrac{dx}{dt} = A(t)x + \alpha g(t,x;\alpha)$

We note that the vectorfield is $2\pi$-periodic in t and that if $f \in C^1(D)$ we may average the right hand side as long as we exclude a neighborhood of the origin. Since the original equation is autonomous, the averaged equation depends only on $r$ and we define the two components of the averaged vectorfield as follows

$$f_1(r) = \frac{1}{2\pi}\int_0^{2\pi} \sin(t+\psi)f(r\cos(t+\psi), -r\sin(t+\psi))dt$$
$$= \frac{1}{2\pi}\int_0^{2\pi} \sin(\tau)f(r\cos(\tau), -r\sin(\tau))d\tau \tag{2.5}$$

and $\quad f_2(r) = \frac{1}{2\pi}\int_0^{2\pi} \cos(\tau)f(r\cos(\tau), -r\sin(\tau))d\tau.$ \hfill (2.6)

An asymptotic approximation can be obtained by solving

$$\frac{d\tilde{r}}{dt} = -\alpha f_1(\tilde{r}), \quad \frac{d\tilde{\psi}}{dt} = -\alpha \frac{f_2(\tilde{r})}{\tilde{r}} \tag{2.7}$$

with appropriate initial values, where the symbol ~ holds for the averaged values. This is a reduction to the problem of solving a first order autonomous system.

We specify this now for the damping equation, $f = (1-x^2)\dot{x}$, (2.8) which is the most common of Van der Pol equations.

The system (2.4) can equivalently be written in a matrix form:

$$\begin{pmatrix} \dot{r} \\ \dot{\psi} \end{pmatrix} = \begin{pmatrix} -\alpha \sin(t+\psi)f \\ \dfrac{a}{r}\cos(t+\psi)f \end{pmatrix} \tag{2.9}$$

Since the damping function f is odd[iii] and the cosinusoidal term is even their product is odd and its mean value, within a period, is zero, so the right hand of (2.6) turns to be zero.

Regarding the equation of (2.5) we easily find:
$f = (1-x^2)\dot{x} = -(1-r^2\cos^2\tau)r\sin\tau \quad$ and (2.5) becomes:

$$f_1(r) = -\frac{1}{2\pi}\int_0^{2\pi} r\sin^2(\tau)(1-r^2\cos^2\tau)d\tau$$

---

[iii] To prove this we only have to do the transformation (2.3).
Thus $f = (1-x^2)\dot{x} = -(1-r^2\cos^2(t+\psi))r\sin(t+\psi)$. It is obvious that the latter is odd, since it is a product of an even and an odd function.

$$= -\frac{1}{2\pi}\int_0^{2\pi}(r\sin^2\tau - r^3\sin^2\tau\cos^2\tau)d\tau = -\frac{r}{2\pi}\int_0^{2\pi}\sin^2\tau d\tau + \frac{r^3}{2\pi}\int_0^{2\pi}\sin^2\tau\cos^2\tau d\tau$$

$$= -\frac{r}{2\pi}\int_0^{2\pi}\frac{1-\cos 2\tau}{2}d\tau + \frac{r^3}{2\pi}\int_0^{2\pi}(1-\cos^2\tau)\cos^2\tau d\tau =$$

$$= -\frac{r}{4\pi}\int_0^{2\pi}d\tau + \overbrace{\frac{r}{2\pi}\int_0^{2\pi}\cos 2\tau d\tau}^{=0} + \frac{r^3}{2\pi}\int_0^{2\pi}\cos^2\tau d\tau - \frac{r^3}{2\pi}\int_0^{2\pi}\cos^4\tau d\tau$$

$$= -\frac{r}{4\pi}2\pi + \frac{r^3}{4\pi}\int_0^{2\pi}(\cos 2\tau + 1)d\tau - \frac{r^3}{2\pi}\int_0^{2\pi}\frac{1}{4}(\frac{\cos 4\tau}{2} + 2\cos 2\tau + 1)d\tau$$

$$= -\frac{r}{2} + \frac{r^3}{2} - \frac{3r^3}{8} = \frac{1}{8}r^3 - \frac{1}{2}r$$

So (2.7) becomes

$$\frac{d\tilde{r}}{dt} = -\alpha f_1(\tilde{r}) = a(\frac{1}{2}\tilde{r} - \frac{1}{8}\tilde{r}^3) = \frac{\alpha \tilde{r}}{2}(1 - \frac{1}{4}\tilde{r}^2), \quad \frac{d\tilde{\psi}}{dt} = 0 \quad (2.10)$$

If the initial value $r_0$ equals to 0 or 2 the amplitude $\tilde{r}$ remains constant for all time. To examine the stability type of the solution r=0 we have to see what happens to the system when we slightly perturb it, by ξ, from the equilibrium point, which corresponds to $r_0$=0:

r=0+ξ  and from (2.10) we obtain

$$\dot{\xi} = \frac{\alpha}{2}\xi + O(\xi^3),$$

and by neglecting second and greater order terms we have

$$\xi = \xi_0 e^{\alpha t/2}$$

this term goes to infinity, as time increases and the equilibrium point turns to be unstable.

In a similar way for the second equilibrium point we get:

r=2+ξ  ⇒  $\dot{\xi} = \alpha(-\xi + O(\xi^2) + O'(\xi^3))$

$$\overset{first\ order\ perturbation}{\Rightarrow} \quad \dot{\xi} = -\alpha\xi \Rightarrow \xi = \xi_0 e^{-\alpha t}$$

and $r_0 = 2$ corresponds to a stable solution.

By integrating (2.10) we find that

$$\tilde{r}(t) = \frac{2e^{\alpha t/2}}{\sqrt{e^{\alpha t} + 1/4r_0 - 1}} \quad \text{(see appendix)}$$

and $\tilde{r}$ tends to achieve a constant value ($\tilde{r} = 2$), while t increases, as can be seen in figure 1, for **all** values of initial conditions $r_0$. In other words, this critical value, $\tilde{r} = 2$ (for which we proved that it is a stable equilibrium point of (2.10)), seems to attract every point in the space of initial conditions.

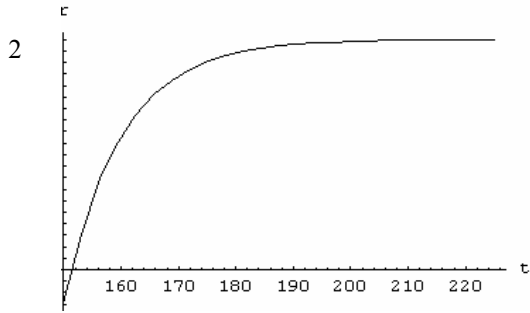

*Figure 2.1.*
the averaged amplitude r versus time, for a=0.1 and $r_0$=1

Up to now we found, using the theory of averaging, an expression for $\tilde{r}$. Consequently the "unperturbed" variable r equals to $\tilde{r} + O(\alpha)$.

Let's, return now to the x variable in order to depict its behavior.
We have, from (2.3),
$$x(t) = r_0 \cos(t + \psi) + O(a) \quad \text{on a time-scale } 1/\alpha$$
and for $r_0$=2
$$x(t) = 2\cos(t + \psi) + O(a) \tag{2.11}$$

In general we find
$$x(t) = \frac{r_0 e^{\frac{1}{2}at}}{(1 + \frac{1}{4}r_0^2(e^{at} - 1))^{1/2}} \cos(t + \psi_0) + O(\alpha) \tag{2.12}$$

For values $r_0$=0 and $r_0$=2 we have a stable and a time-periodic solution (2.11) respectively. For all other values of r the solutions tend toward the periodic solution (2.11) and we call its phase orbit a limit cycle (stable). In figure 2 we present this limit cycle and in figure 3 the time evolution of two orbits with different initial values (the first one start inside of the limit cycle while the second one starts out of it).

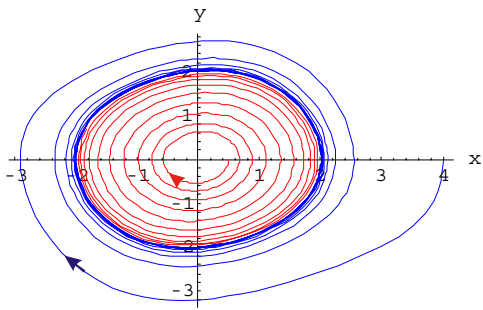

*Fig. 2.2*
limit cycle for the averaged
equation (2.12), with $\alpha$=0.1

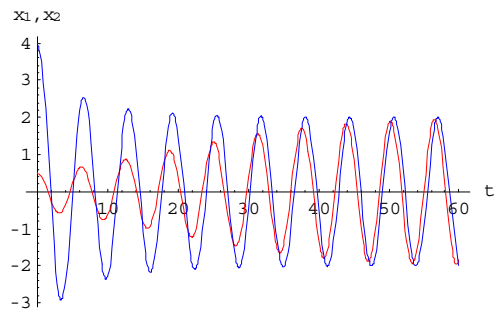

*Fig. 2.3*
time evolution for two trajectories,
of (2.12), with $\alpha$=0.1

*Second order Averaged equations*

We shortly present now the behavior of the equation system (2.2) or (2.4) using a second order approximation. It is now convenient to introduce the coordinate transformation

$$x = r\sin\phi$$
$$y = r\cos\phi \qquad (2.13)$$

so the inverse transformation will be

$$r = x^2 + y^2$$
$$\phi = \tan^{-1}\left(\frac{x}{y}\right) \qquad (2.14)$$

from which we obtain (after a few calculations)

$$\dot\phi = 1 + a\left[-\frac{1}{2}\sin 2\phi + \frac{1}{8}r^2(2\sin\phi - \sin 4\phi)\right]$$
$$\dot r = a\frac{r}{2}\left[1 + \cos 2\phi - \frac{r^2}{4}(1 - \cos 4\phi)\right] \qquad (2.15)$$

the 'second order averaged equations' [Sanders, 21] of this vectorfield are

$$\dot\phi = 1 - \frac{a^2}{8}(\frac{11}{32}r^4 - \frac{3}{2}r^2 + 1) + O(a^3)$$
$$\dot r = a\frac{r}{2}(1 - \frac{r^4}{4}) + O(a^3)$$

Neglecting the $O(\alpha^3)$ term, the equation for r represents a subsystem with attractor r=2. The fact that the $O(\alpha^3)$ depends on another variable as well (i.e. on φ) is not going to bother us in our estimate since φ(t) is bounded (the circle is compact). This means that, by solving the equation

$$\dot r = a\frac{r}{2}(1 - \frac{r^4}{4}), \qquad \tilde r(0) = r_0 = (x_0^2 + y_0^2)^{1/2}$$

we obtain an O(α)-approximation to the r-component of the original solution, valid on $[0,\infty)$. (The fact that the $O(\alpha^2)$ terms vanish in the r-equation does in no way influence the results). Using this approximation we can obtain an O(α)-approximation for the φ-component by solving

$$\dot{\tilde\phi} = 1 - \frac{a^2}{8}\left(\frac{11}{32}\tilde r^4 - \frac{3}{2}\tilde r^2 + 1\right), \qquad \tilde\phi(0) = \phi_0 = \arctan\left(\frac{x_0}{y_0}\right) \qquad (2.16)$$

Although this equation is easy to solve the treatment can be more simplified by noting that the attraction in the r-direction takes place on a time-scale 1/α, while the slow fluctuation of φ occurs in a time scale $1/\alpha^2$. This has a consequence that, in order to obtain an O(α) approximation $\tilde\phi$ for $\phi$ on the time scale $1/\alpha^2$ we may take r=2 in computing $\tilde\phi$. Thus we are left with the following simple system

$$\dot{\hat{\phi}} = 1 - \frac{a^2}{16}, \qquad \dot{\hat{\phi}}(0) = \phi_0,$$

$$\dot{\tilde{r}} = a\frac{\tilde{r}}{2}\left(1 - \frac{\tilde{r}^2}{4}\right), \qquad \tilde{r}(0) = r_0 \qquad (2.17)$$

For the general solution of the Van der Pol equation with $r_0 > 0$ we find

$$x(t) = \frac{r_0 e^{\frac{1}{2}at}}{(1 + \frac{1}{4}r_0^2(e^{at} - 1))^{1/2}} \cos\left(t - \frac{a^2}{16}t + \phi_0\right) + O(\alpha) \qquad (2.18)$$

There is no obstruction against carrying out the averaging process to any higher order to obtain approximations valid on longer time-scales, to be expressed in inverse power of α.

Comparing equations (2.12) and (2.18) we see that they are slightly different equations. In the periodic case, for example, (r=2 and the periodic solution is now described by (2.11)) the two equations have different angular frequencies and this difference is proportional to the square of α.

What we did above was to average the initial Van der Pol equations in order to estimate their behavior. Tacitly we accepted that perturbation (damping) parameter α was kept sufficiently small. Besides that the theorem of averaging, in the case of great α, cannot give satisfactory results; the greater the parameter α becomes the smaller the time-scale 1/α, in which the theorem holds.

In order to prove the existence of a stable limit cycle in the aforementioned case we will follow a simple and more intuitive approach (Guckengeimer-Holmes [22]). The **limit - large α** can once more be treated by perturbation methods, although this time the perturbation is singular.

Letting $\hat{y} = y/a$ and dropping the hats (2.2) becomes

$$\begin{cases} \dot{x} = a\left(y - \left(\frac{x^3}{3} - x\right)\right) \\ \dot{y} = -\frac{x}{a} \end{cases} \qquad (2.19)$$

Since $\alpha \gg 1 \gg 1/\alpha$, we have $|\dot{x}| \geq |\dot{y}|$ except in a neighborhood of the curve $c$ given by $y = x^3/3 - x$. Thus the family $H$ of horizontal lines y=constant approximates the flow of (2.19) away from $c$ increasingly well as $\alpha \to \infty$. Near $c$, and in particular when $|y - (x^3/3 - x)| = (1/a^2)$ both solution components are comparable and hence, after entering this *boundary layer*, solutions turn sharply and follow the flow $c$ until they reach a critical point of $c$(x=±1, y=∓2/3) where they must leave $c$ and follow $H$ to another point of $c$, see *figure 2.4.* Making these ideas precise, it is possible to find an annular region R into which the vector field is directed at each point, and which must therefore contain a closed orbit, by the Poincare-Bendixson theorem, since it contains no equilibria for $\alpha \neq 0$. To prove that this orbit is unique, we show that, since solutions spend most of their time near the stable branches of c, where trace *Df=-a(x²-1)<0,* any orbit within the annulus R must be asymptotically stable; hence only one such orbit can exist.

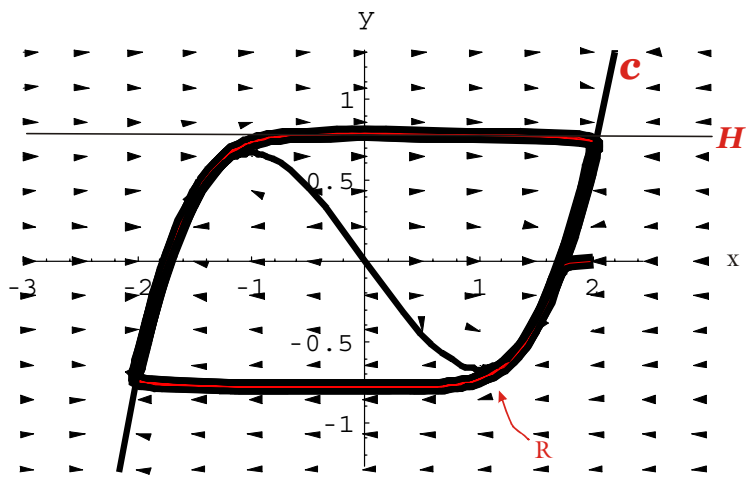

Fig. 2.4.
Relaxation Oscillations for *a=8*.

*Liénard's theorem*

We can obtain the same results using a theorem, which is set in a very simple form, and is called **Liénard's theorem**. According to this:

If      $f(x)$ is an even function for all $x$
and    $g(x)$ is an odd function for all $x$
and    $g(x) > 0$ for all $x > 0$
and    $F(x) = \int_0^x f(t)\,dt$ is such that $F(x) = 0$ has exactly one positive root, $\gamma$,

and
$F(x) < 0$ for $0 < x < \gamma$ and $F(x) > 0$ and non-decreasing for $x > \gamma$,

then

the system
$$\dot{x} = y, \quad \dot{y} = -f(x)y - g(x)$$
or, equivalently,
$$\frac{d^2x}{dt^2} + f(x)\frac{dx}{dt} + g(x) = 0$$
has a **unique limit cycle** enclosing the origin and that limit cycle is asymptotically stable.

When all of the conditions of Liénard's theorem are satisfied, the system has exactly one periodic solution, towards which all other trajectories spiral as $t \to \infty$.

Now, if we put $f(x) = -a(1-x^2)$ and $g(x) = x$, (with a>0), then Liénard's theorem becomes
$$\frac{d^2x}{dt^2} + a(x^2-1)\frac{dx}{dt} + x = 0, \quad \text{or}$$
$$\ddot{x} + a(x^2-1)\dot{x} + x = 0$$

which is same as equations (2.1).
Checking the conditions of Liénard's theorem:
$f(x) = -a(1-x^2) = a(x^2-1)$ is an even function.
$g(x) = x$ is an odd function, positive for all $x > 0$.
$$F(x) = \int_0^x -a(1-t^2)dt = -a\left[\left(t - \frac{t^3}{3}\right)\right]_0^x = +a\left(\frac{x^3}{3} - x\right)$$
$F(x) = 0$ has only one positive root, $\gamma = \sqrt{3}$.
$F(x) < 0$ for $0 < x < \sqrt{3}$ and $F(x) > 0$ and increasing for $x > \sqrt{3}$.

Therefore Van der Pol's equation possesses a unique and asymptotically stable limit cycle.

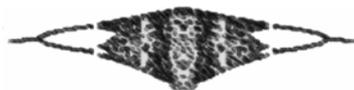

# Chapter 3
## THE NON-AUTONOMOUS VAN DER POL EQUATION

We are now concerned with the excited - by a periodic external force - Van der Pol oscillator. This is the more general case, so we rewrite equations (2.1) without setting b=0[iv]. In order to work out with this non-trivial differential equation we implement some remarkable theorems (skipping their proofs) which deal with more general systems. First let's see how we can establish the existence of periodic and non periodic solutions of (2.1) using the methods, such as the Implicit Function Theorem and others proposed by M. L. Cartwright and J. E. Littlewood [7,8] and Hale[23] (we examine more general differential equations than the Van der Pol one and all the results given below stand for our equations as well)

*Implicit Function Theorem (I.F.T.)*

The **Implicit Function Theorem** (extended to a function with several independent variables) can be expressed as follows:

*Let $F(x,y,\ldots,z,u)$ be a continuous function of the independent variables $x,y,\ldots z,u$ with continuous partial derivatives $F_x$, $F_y$, $\ldots$, $F_z$, $F_u$. Let $(x_0,y_0,\ldots,z_0,u_0)$ be an interior point of the domain of definition of F, for which*

$$F(x_0,y_0,\ldots,z_0,u_0) = 0 \text{ and } F_u(x_0,y_0,\ldots,z_0,u_0) \neq 0$$

*Then we can mark off an interval $u_0 - \beta \leq u \leq u_0 + \beta$ about $u_0$ and a rectangular region R containing $(x_0,y_0,\ldots,z_0,u_0)$ in its interior such that for every $(x,y,\ldots,z)$ in R, the equation $F(x,y,\ldots,z,u) = 0$ is satisfied by exactly one value of u in the interval $u_0 - \beta \leq u \leq u_0 + \beta$, For this value of u, which we denote by $u = f(x,y,\ldots,z)$ the equation $F(x,y,\ldots,z, f(x,y,\ldots,z)) = 0$ holds identically in R. (Courant and John [24])*

Now let's see what the I.F.T. yields for the vector field

$$\dot{x} = f(x,\mu) \quad x \in R^n \tag{3.1}$$

where μ is a parameter, μ∈R.
For the unperturbed system (μ=0) there is a unique periodic solution $x^*$, which depends on the initial conditions $x_0$

$$\left.\begin{array}{l} x^* = x(x_0,t,0) \\ x^*(t+T) = x^*(t) \end{array}\right\} \mu = 0 \tag{3.2}$$

$$t, T \in R, \quad x_0, x^* \in R^n$$

and $f(x_0,0)=0$

the periodicity condition is

---

[iv] but however kept small

$$x^*(x_0, T, 0) - x_0 = 0 \tag{3.3}$$

For the perturbed system ($\mu \neq 0$) the periodicity condition (3.3) becomes

$$F = X(x_0, T, \mu) - x_0 = 0 \tag{3.4}$$

According to the implicit function theorem

**if** $\quad \left| \dfrac{\partial F}{\partial x_0} \right|_{\mu=0} \neq 0$

or from (3.4) $\quad \left| \dfrac{\partial X(x_0, 0)}{\partial x_0} - I \right|_{\mu=0} \neq 0 \quad$ (3.5) where I is the n x n identity matrix,

**then** there is a sufficiently small interval $-\mu_0 < \mu < \mu_0$ (containing $\mu=0$) such that for every x, the equation $f(x,\mu)=0$ is satisfied by exactly one value of $\mu$ in this interval.

If one can prove relation (3.5) then knows that there exists one periodic solution of the vector field (3.1).

### *Trajectories of non-autonomous Van der Pol are bounded*

*If $k \geq 1$, every solution of the equation* $\ddot{y} + kf(y)\dot{y} + g(y,k) = p(t)$

*ultimately satisfies* $|y| < B, \quad |\dot{y}| < Bk$, *where B is a constant, independent of k.*
[8]

Cartwright and Littlewood, studying the equation
$\ddot{y} + kf(y, \dot{y}) + g(y, k) = p(t) = p_1(t) + kp_2(t); \quad k > 0, f(y) \geq 1$

mentioned that *"in the special case that g(y)=y it is known that (for f' and p continuous, kf ≥ a>0) there is a single periodic trajectory to which every trajectory converges as $t \to \infty$."* . The proof of this is stated in N. Levinson [25].

### *Existence of recurrent trajectories*

Levinson provides a fluent method for showing the **existence of recurrent** trajectories in second order non-linear D.E.'s [12]

Let's follow his method.

We consider

$$\ddot{y} + f(y)\dot{y} + y = b\cos t \tag{3.6}$$

where *f(y)* is a certain polynomial and *b* is a constant restricted to belong to a certain set of intervals.

Among the solutions of (3.6) there is a family *F* of remarkably singular structure. Solutions, *y(t)*, of *F* have a maximum value of approximately 3. If the maximum occurs at $t = t_1$ then for $t > t_1$ and as long as $y > 1$, *y* is approximately of the form

$$(3-d)e^{-\rho(t-t_1)} - d\cos t \quad (3.7)$$

where *d* is a constant, $0 < d < 1$, and the constant $\rho > 0$ is small. Thus for $t > t_1$, *y* aside from the cosine term, decreases slowly. When *y* reaches the value 1 within an interval of *t* at most $2\pi$ in length, then it falls, to its minimum value of approximately -3. It then repeats its behavior, with opposite sign, slowly rising to *y* = -1 and from there rapidly reaching a maximum close to 3 again. This general pattern is repeated over and over again.

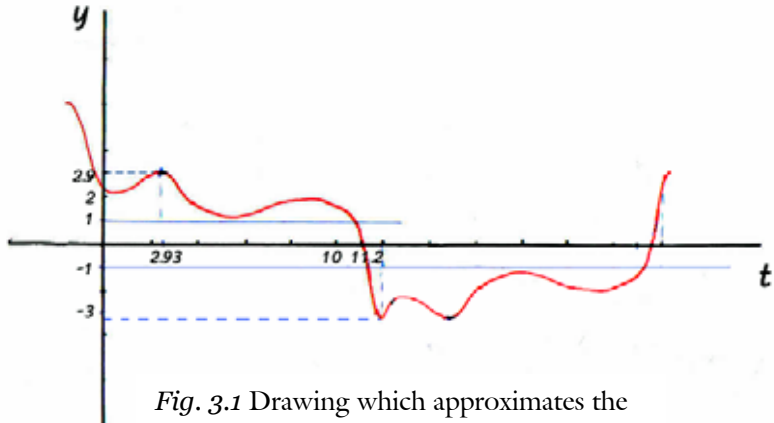

*Fig. 3.1* Drawing which approximates the behavior of (3.7), in Levinson's discussion

We denote the value of *t* where a solution of *F, y(t),* descending from its maximum of approximately 3 first crosses $y = 1$ as an *even* crossing point. We denote *t* where *y(t)* ascending from its minimum of approximately -3 first crosses $y = -1$ as an *odd* crossing point. For any solution of *F,* even and odd crossing points alternate as *t* increases. (It is the case that $y = 1$, $\dot{y} < 0$, only at even crossing points and $\dot{y} = -1$, $y > 0$, only at odd crossing points for the solutions of *F.*)

An even crossing point always lies in a short interval $t = \tau$ (mod $2\pi$), $0 < \tau < \tau_1 < 1/10$, which we shall call an *even base interval.* An odd crossing point always lies in a short interval $t = \pi + \tau$ (mod $2\pi$), $0 < \tau < \tau_1 < 1/10$, which we shall call an *odd base interval.* By crossing point we shall mean either an even or an odd crossing point. **Associated with (3.6) there is a large integer n. The spacing between the successive base intervals in which the crossing points of a solution of F lie is either $(2n-1)\pi$ or $(2n+1)\pi$. Moreover given any arbitrary sequence $\{d_k\}$, $-\infty < k < \infty$, where $d_k$ is either $(2n-1)\pi$ or $(2n+1)\pi$ there is a solution of F with crossing points that lie in base intervals with successive spacing $d_\kappa$, $-\infty < k < \infty$. Moreover the solution has no other crossing points.**

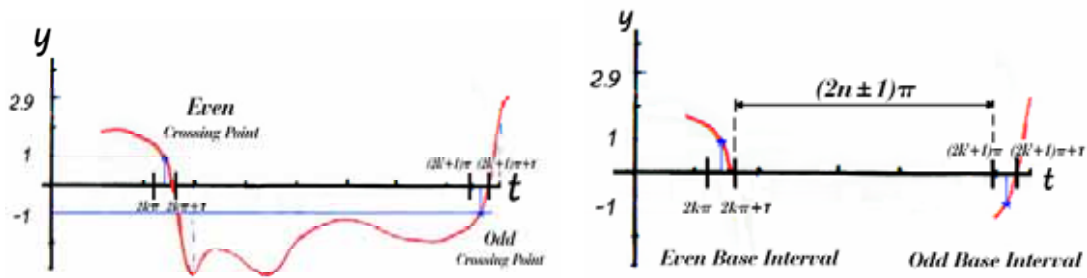

*Fig. 3.2* Diagrams showing Crossing Points (a) and Base Intervals (b)

Before going further it is useful to summarize this method:
First we sketched out a general pattern of the behavior of solutions of F, which are recurrent. This means that after an arbitrarily long time interval Δt the trajectory returns arbitrarily close to its starting point. The general pattern of F was described above, extensively. By choosing two crossing points in the aforementioned solution and two base intervals, in which those points lie, we constructed a sequence of successive spacing between those intervals. It is obvious that the existence of a periodic solution demands the existence of a periodic sequence $d_k$.

Reversely, it can be proved that

**THEOREM 3-1 to each such periodic sequence there is at least one periodic solution of F, the crossing points of which lie in base intervals having the designated periodic spacing. These periodic solutions we shall denote by Σ.**
(Levinson [12])

It is obvious that to a non-periodic sequence there are only non-periodic solutions.

What we do next is to show the existence of periodic, non periodic trajectories or even chaotic invariant sets by examining the properties of the sequence $\{d_k\}$, with the use of **Symbolic Dynamics**[v].

To make these ideas more specific we obtain a generic sequence as follows:
$$d_1 = (2n-1)\pi$$
$$d_2 = (2n+1)\pi$$
$$d_3 = (2n+1)\pi$$
$$d_4 = (2n-1)\pi$$

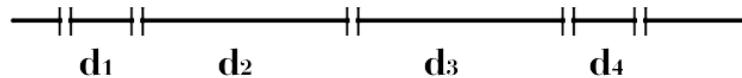

For each such sequence we choose one trajectory, whose existence is provided from the above theorem, and especially for each periodic sequence we choose the periodic trajectory. So, we form a set of all trajectories with a 1-1 correspondence to all double infinite sequences.

The number of all possible sequences is of the power of the continuum, since sequence $\{d_k\}$ involves an infinite number of choices of ±1 in defining $d_k$. In order to point out the existence of periodic sequences and, consequently, trajectories, we interpret our results with the use of Symbolic Dynamics. To do that we set a correspondence between the spacing (between successive base intervals) and 0 and 1, which are elements of the binary space $S=\{0,1\}$
$$(2n-1)\pi \longrightarrow 0$$
$$(2n+1)\pi \longrightarrow 1$$

Now the set D of all sequences $\{d_k\}$ is the set of all **double infinite** sequences of elements of S.

That is if $d \in D$ then $d = \{...d_{-i}...d_{-2}.d_{-1}.d_0.d_1.d_2...d_i...\}$ $d_i \in S$ or
for example $d = \{...0...100.101....1...\}$ $d_i \in \{0,1\}$

---

[v] This term characterizes the technique of describing the structure of trajectories of a dynamical system with the use of infinite sequences, which consist of (infinite or finite) symbols.

We can easily define in $D$ a metric $r(d,\tilde{d}) = \sum_{i=-\infty}^{\infty} \frac{|d_i - \tilde{d}_i|}{2^{|k|}}$, for which it can be proved that it has the properties of distance. We next define a mapping $D \to D$, which we call *Bernoulli Shift $\sigma$*, as follows

if $d = \{...0...100.101....1...\} \in D$

Bernoulli Shift is defined by

$\sigma(d) = \{...0...1001.01....1...\} \in D$

or $[\sigma(d)]_i = d_{i+1}$

In words, this mapping shifts the decimal point in a sequence to the right by one step.

The sequences which are invariant under $m$-iterations of Bernoulli shift are $m$-periodic. For example the sequences:

$\{...000.000...\}$, $\{...111.111...\}$ are periodic with period 1 (or fixed points) while the sequence

$\{...1010.1010...\} \xrightarrow{\sigma(d)} \{...10101.010...\} \xrightarrow{\sigma(d)} \{...1010.1010...\}$ is periodic with period 2, etc.

These sequences are periodic trajectories of $\sigma$. Since for every $m$ the number of these periodic trajectories is finite, $\sigma$ **should have a countable set of periodic trajectories, with all possible periods**. Furthermore $\sigma$ has a non-countable set of non-periodic trajectories (since there is a 1-1 correspondence of these non-periodic sequences to the set of irrational numbers). Thereby, of the continuum of sequences $\{d_k\}$ there is a countable infinity of periodic sequences. Most of the solutions of $F$ are certainly not periodic since most of the sequences $\{d_k\}$ are not periodic.

Moreover, by investigating this mapping, one can examine if **chaotic trajectories** occur, or, more precisely, chaotic invariant sets. There are three known conditions which have to be fulfilled in order to call an invariant set chaotic. These are:
- The periodic trajectories of $\sigma$ are dense in D
- $\sigma$ has sensitive dependence on initial conditions
- $\sigma$ is topologically transitive in D

It can be proved that there is an element $d \in D$ whose trajectory is *dense* in $S$, that is for every $d' \in D$ and $\varepsilon > 0$ there exists an integer $n$, so that

$$r(\sigma^n(d), d') < \varepsilon$$

so the first condition is fulfilled.

Furthermore, all three conditions can be proved [26]

Conclusively, the dynamics of $d$ in $D$ satisfies the following theorem:

**THEOREM 3-2** The Shift mapping $\sigma$ in $D$ has
  (a)  Countable infinity periodic trajectories, of arbitrarily large period
  (b)  Non-countable infinity non-periodic trajectories
  (c)  A dense trajectory

**Among the limiting solutions of (3.6), are two completely stable[vi] periodic solutions $\Gamma_1$ of least period $(2n - 1)2\pi$ and $\Gamma_2$ of least period $(2n + 1)2\pi$ and one completely unstable periodic solution Pu of period $2\pi$. The solutions $\Gamma_1$ and $\Gamma_2$ are of physical interest since they are stable solutions.**

Aside from the remarkable family *F* of solutions, (2.1) and (3.6) afford an example of another singular situation which before the paper of N. Levinson was not known to arise in differential equations. Let *($y_0$, $v_0$)* be a point $P_0$ in the *(y, v)* plane and let *y(t)* be a solution of (3.6) with initial values $y(t_0) = y_0$ and $\dot{y}(t_0) = v_0$. Let $y_1 = y(t_0 + 2\pi)$ and $v_1 = \dot{y}(t_0 + 2\pi)$ and $P_1$ be *($y_1$, $v_1$)*. Then the transformation of the *(y, v)* plane into itself defined by $TP_O = P_1$ (T stands for trajectory) is analytic as is its inverse. In other words, we have defined a mapping similar to Poincarè map. For this map it is known that an m-periodic trajectory corresponds to an m-set of points and an almost periodic trajectory to a closed connected curve, or to be more precise to a set of points that its *closure* is a closed connected curve .

Under iterations of *T* all points in the finite part of the *(y, v)* plane, except the point associated with *Pu*, the completely unstable periodic solution, tend to a closed connected point set *$K_0$*. *$K_0$* divides the plane into two parts, an interior which is an open simply connected point set containing *Pu* and an exterior which is open and becomes simply connected if the point at infinity is adjoined to it (Riemann's Sphere). Every point of *$K_0$* is a limit point of interior or exterior points. *$K_0$* is of course invariant under *T* as are the interior and exterior of *$K_0$*. Among the points of *Ko* are a continuum of points corresponding to the solutions *F* including 2. *Ko* also contains *(2n - 1)* points corresponding to $\Gamma_1$ and *(2n + 1)* points corresponding to $\Gamma_2$. Now *$K_0$* cannot be a simple Jordan curve for if it were, it would have a definite rotation number under *T* which would be the rotation number of all the points on *$K_0$*. That the points of *$K_0$* do not have a unique rotation number is clear both from the arbitrary structure of the solutions of *F* and also from the different periodicity of $\Gamma_1$ and $\Gamma_2$. Thus *$K_0$* is a singular "curve" with points that are neither exterior accessible nor interior accessible and moreover *$K_0$* is an attracting and compact connected set on which chaotic dynamics occur (as we have seen above in Bernoulli shift).

Whether such "curves" could arise from the transformation associated with a differential equation was not known. In fact Levinson [27] raised the question whether such "curves" could occur in connection with second order equations of the type under consideration. The "curve" *$K_0$* which occurs for (2.1) and (3.6) is the first known example arising from an actual differential equation.

*Existence and Construction of periodic solution*
*The method of Successive Approximations*

The problem of the existence of periodic solutions of weakly non-linear differential systems and the determination of the characteristic exponents of almost constant linear periodic differential systems can be reduced to the determination of periodic solutions of systems of the form

---

[vi] Levinson used the term "completely stable" instead of "stable". This term nowadays (such as the term "recurrent" and others) is not in use anymore. Even we looked for a definition of this term ("completely stable") and why this differs from the known term "stable" we couldn't find any, so we use the latter term.

$$\dot{x} = Ax + bX(t,x,b) \qquad (3.8)$$

where b is a parameter, t is a real variable, x, X are n vectors, A is a $n \times n$ constant matrix, X is periodic in t of period T, is continuous in t,z,b, has a continuous first derivative with respect to b and continuous first and second partial derivatives with respect to x for $-\infty < t < \infty$, $0 \leq \|x\| \leq R$, $0 \leq |b| \leq b_0$, $R > 0$, $b_0 > 0$. We shall briefly refer to system (3.8) satisfying all the above smoothness conditions as system (3.8). Such restrictive smoothness conditions on X are not necessary and are assumed only for simplicity.

System (3.8) will be said to be in *standard form* if the matrix A has the form

$$A = \text{diag } (0_p, B) \qquad (3.9)$$

where $0_p$ is the *pxp* zero matrix and B is *qxq* matrix, *q=n-p*, with the property that no solution of the equation

$$\dot{y} = By$$

is periodic of period T except y=0. The system x'=Ax is critical since there are nontrivial constant solutions.

We always assume that system (3.8) is in standard form and we want to find necessary and sufficient conditions for the existence of periodic solutions of (3.8), (3.9) of period T which, for b = 0, are periodic solutions of x = Ax of period T. The most general periodic solution of period T of this system is ***x*** = col(*a*,0) where *a* is an arbitrary constant *p* vector. The natural thing to do is to iterate on (3.8), (3.9) using as an initial approximation x = col(*a*,0). Unfortunately, the successive iterations will not in general be periodic, even though the final limit function may be periodic. The presence of these nonperiodic terms (the secular terms) in the successive approximations makes it very difficult to discuss the qualitative behavior of the solution. On the other hand, one does not expect to obtain a periodic solution starting with an arbitrary initial vector col (*a*,0). Consequently, one attempts to determine the vector a in such a way that no secular terms appear in the successive iterations. This is the essence of most of the methods for obtaining periodic solutions by successive approximations. As a final remark, one would also hope to postpone as long as possible the determination of the vector *a;* for otherwise, some of the qualitative properties of the solutions may be obscured in the shuffle. More precisely, we shall reduce the problem of determining periodic solutions of (3.8), (3.9) to the determination of a *p* vector *a*, which is a solution of a set of transcendental equations (determining equations) (bifurcation equations).

Let S be the space of all continuous n-vector functions of the real variable t, periodic in t of period T, with the uniform topology; that is if $f \in S$ the norm $v(f)$ is given by $v(f) = \sup\|f(t)\|$. If $f \in S, f = $col (g,h) where g is a p vector and *h* is a n-p vector, we define the operator $P_0: S \to S$ so that

$$P_0 f = col\left[\frac{1}{T}\int_0^T g(t)dt, 0\right] \qquad (3.10)$$

that is, $P_0$ projects a function $f \in S$ into a constant n vector all of whose components are zero except the first *p* which are the mean values of the first *p* components of the vector *f*. For given positive constants *c, d, c < d < R*, and a given constant *p* vector *a*, $\|a\| \leq c$, we define the set $S_o \subset S$ by the relation

$$S_0 = \{f \in S \mid P_0 f = col(a,0), v(f) \leq d\} \qquad (3.11)$$

Basic to our understanding of the general system (3.8) is the behavior of the solutions of the nonhomogeneous linear system for forcing functions in S. These results are embodied in Lemma 3-1.

**Lemma 3-1.** If $f \in S$, then the nonhomogeneous equation $\dot{z} = Az + f(t)$ with A as in (3.8) has a periodic solution of period T if and only if $P_0 f = 0$. Furthermore, there is a unique periodic solution $z^*(t)$ of period T with $P_0 z^* = 0$. If this unique solution is designated by $z^*(t) = \int e^{A(t-u)} f(u) du$ then there exists a constant K which is independent of f such that

$$\|z^*(t)\| = \left\| \int e^{A(t-u)} f(u) du \right\| \leq \frac{K}{T} \int_0^T \|f(u)\| du \leq K_v(f)$$

We now quote two quite helpful theorems[vii]

**THEOREM 3-3.** For given constants $d > 0$, $0 < c < R$, there is a $b_1 > 0$ with the following property: corresponding to each constant p vector $a$, $\|a\| \leq c$, and to each b, $|b| \leq b_1$, there is unique function

$$x^*(t) = x(t,a,b) \in S_0 \tag{3.12}$$

which has a continuous first derivative with respect to t, and whose derivative satisfies the relation

$$\dot{x}^* = Ax^* + bX(t, x^*, b) - bP_0 X(t, x^*, b) \tag{3.13}$$

where A, X(t,x,b) are the functions given in (3.8), (3.9). Furthermore, $x(t,a,0) = a^*$, $a^* = \text{col } (a,0)$, and x(t,a,b) has continuous first derivatives with respect to $a,b$. Also, if X(t,a,b) has continuous first derivatives with respect to some parameter, then the function x(t,a,b) also has continuous first derivatives with respect to this parameter. Finally $x^*(t)$ may be obtained by a method of **successive approximations**, given by formula (3.14) below.

The function $x^*(t) = x(t,a,b)$ is the uniform limit of the sequence of functions $\{x^{(k)}(t)\}$ defined by

$$\begin{aligned} x^{(0)} &= \text{col } (a,0) = a^* \\ x^{(k+1)}(t) &= \Im x^{(k)}(t) = a^* + b \int e^{A(t-u)} (I - P_0) X(u, x^{(k)}(u), b) du \\ k &= 0, 1, 2 ... \end{aligned} \tag{3.14}$$

**THEOREM 3-4.** Let x(t,a,b) be the function given by theorem 3-3 for all $\|a\| \leq c < R$, $0 \leq |b| \leq b_1$. If there exist a $b_2 \leq b_1$ and a continuous function $a(b)$ such that

---

[vii] For a more analytical treatment of these and proofs of the theorems, the reader is referred to J. Hale (p. 38-39) [23]

$$P_0 X(t, x(t,a(b),b), b) = 0 \quad \|a(b)\| \leq c \leq R \quad for \quad 0 \leq |b| \leq b_2 \quad (3.15)$$

then x(t,a(b),b) is a periodic solution of (3.8), (3.9) for $0 < |b| \leq b_2$. Conversely if (3.8), (3.9) has a periodic solution $\tilde{x}(t,b)$, of period T, continuous in b, $\|\tilde{x}(t,b)\| < R$, $0 \leq \|b\| \leq b_2 \leq b_1$ then $\tilde{x}(t,b) = x(t,a(b),b)$, where x(t,a,b) is given in theorem 3-3 and $a(b)$ satisfies (3.15). **Therefore, for $b_2$ sufficiently small, the existence of a continuous $a$(b) satisfying (3.15) is a necessary and sufficient condition for the existence of a periodic solution of (3.8), (3.9) of period T.**

    Let us now reinterpret the above theorems in terms of Fourier series and successive approximations. We wish to find a periodic solution of (3.8), (3.9) which for $b = 0$ is equal to col $(a,0)$, where $a$ is a constant $p$ vector. Let $x = (x_1, x_2)$, where $x_1$ is a $p$ vector,

$$x_1 = a + \sum_{k \neq 0} a_k e^{ik\omega t}$$

$$x_2 = \sum_k b_k e^{ik\omega t}$$

be a periodic function of period $T = 2\pi/\omega$. Then $P_0 x = $ col $(a,0)$. To obtain a periodic solution of system (3.8), (3.9) of the above form, the natural procedure is to substitute these functions in (3.8) and demand that the coefficient of the term in $e^{ik\omega t}$ vanish for all $k$. Theorem 3-3 asserts that for any $p$ vector $a$, and b sufficiently small, the coefficients $a_k$, $b_k$ can be determined as functions of $a$, b [by the method of successive approximations (3.14) in such a way that system (3.8) is satisfied except for a constant $p$ vector. Theorem 3-2 asserts that, if the p vector $a$ can be chosen so that equations (3.15) are satisfied, then one has a periodic solution of (3.8).

    Equations (3.15) will be referred to as the *determining equations* for the periodic solutions. Another term, which has been used in the literature to these equations, is bifurcation equations. If we let X in (3.8) be partitioned as X = col $(X_1, Y_1)$, where $X_1$ is a $p$ vector, $X_2$ is $q=n-p$ vector, then the determining equations (3.15) may be written more explicitly as

$$\frac{1}{T} \int_0^T X_1(t, x(t,a,b), b) dt = 0 \quad (3.16)$$

As an immediate consequence of the above procedure, we have the following result

    **Theorem 3-5.** If the equation $\dot{x} = Ax$ has no periodic solution of period T except the trivial solution x=0, then for *b* sufficiently small, there always exists a periodic solutions of (3.8) which approaches zero as $b \to 0$. Furthermore, in a neighborhood of x=0, this solution is unique.

## Implementation of theorems 3-3, 3-4 to the non-autonomous Van der Pol oscillator

From Theorems 3-3 and 3-4 we know that we can obtain a periodic solution of (3.8) if we find the function x*(t,a,b)* by the method of successive approximations in (3.14) and then solve the equations (3.15). However, from the practical point of view, one can obtain an approximation to x*(t,a,b)* to a given order in b and then see if it is possible to find a simple root of the determining equations (3.15) for this approximate function. If this can be done, then from the continuity properties of the function x*(t,a,b)*, it follows that, for b sufficiently small, the determining equations will have an exact solution close to the approximate solution so obtained.

In case the function $X$ in (3.8) is analytic in $x, b$, the procedure is extremely simple. In fact, one constructs the method of successive approximations as follows:
$x^{(0)} = \text{col}(a,0) = a^*$

$$x^{(k+1)}(t) \equiv a^* + b \int e^{A(t-u)}(I - P_0) X(u, x^{(k)}(u), b) du \quad (trunc \ b^{k+2}) \quad (3.17)$$
$$k = 0,1,2...$$

where (trunc $b^{k+2}$) signifies that all terms of order higher than $b^{k+1}$ are ignored. Then for any given value of $k, k = 0, 1, 2…$, one can check to see if it is possible to solve the approximate determining equations

$$P_0 X(t, x^k(t), b) \equiv 0 \quad (trunc \ b^{k+1}) \quad (3.18)$$

for some p vector $a_o$ (remember that $x^k(t)$ depends on the vector $a$). If equations (3.18) have a solution whose corresponding Jacobian matrix has a nonzero determinant, then the determining equations (3.15) will have a solution for $b$ sufficiently small and we then know from Theorem 3-2 that there is a periodic solution of (3.17) close to col $(a_0, 0)$ for $b$ sufficiently small.

The method of successive approximations (3.17), (3.18) is not the usual kind of power-series expansion in powers of the small parameter $b$. Many of the classical procedures assume that the solution $x$ and the vector $a$ are power series in $b$ and then successively determine the coefficients of the expansion of $a$ in such a way as to obtain a periodic solution. The solution $x$ is thus obtained as a power series in $b$ whose coefficients do not depend upon $b$.

In the above procedure, the function $x(t,a,b)$ may be determined to any degree of accuracy as a power series in $t$ with the coefficients depending upon the parameter $a$. In other words, the dependence of the function $x$ upon $a$ may be clearly exhibited. Afterward, the parameter $a$ is determined as a function of $b$ so that the determining equations are satisfied. The function $x$ may then be expanded as a new power series in $b$, but it may be more desirable to have the more compact form where the dependence on $a$ is preserved.

We will now apply this method in Van der Pol equations only to the first approximation in $b$.

Consider equations (2.1), (2.2), which is now more convenient to rewrite as

$$\begin{cases} \dot{x} = y \\ \dot{y} = -x + a(1-x^2)y + b\cos(\omega t + \vartheta) \end{cases} \quad (3.19)$$

where x,y are scalars, $b>0$, $a \neq 0$, $\omega$ are real numbers and $\omega^2 = 1 + b\sigma$, $\sigma \neq 0$. We wish to investigate whether or not this equation possesses a periodic solution of period $2\pi/\omega$ for $b$ sufficiently small.

To apply the previous results, (3.19) must be transformed to "standard" form. The transformation which achieves that (introduced by Van der Pol himself) is:

$$\begin{aligned} x &= z_1 \sin \omega t + z_2 \cos \omega t \\ y &= \omega(z_1 \cos \omega t - z_2 \sin \omega t) \end{aligned} \quad (3.20)$$

(which is actually the solution matrix of the harmonic oscillator) to obtain the new differential equations in $z_1$, $z_2$:

$$\begin{aligned} \dot{z}_1 &= \frac{a}{\omega}\left[\sigma x + (1-x^2)y + \frac{b}{a}\cos(\omega t + \vartheta)\right]\cos \omega t \\ \dot{z}_2 &= -\frac{a}{\omega}\left[\sigma x + (1-x^2)y + \frac{b}{a}\cos(\omega t + \vartheta)\right]\sin \omega t \end{aligned} \quad (3.21)$$

where x,y are given in (3.20) and $\sigma = (\omega^2-1)/b$.

By integrating numerically the above equations (see appendix) we obtain the figures shown below. In the first one the time evolution for $z_1$, $z_2$ is shown and the following diagrams are two enlargements of the first one. It can be easily seen that this system, since it is not averaged, possesses two time flows; the *slow-time* and the *fast-time* flow. In the previous chapter, using the averaging theorems, we "eliminated" - averaged over the fast-time.

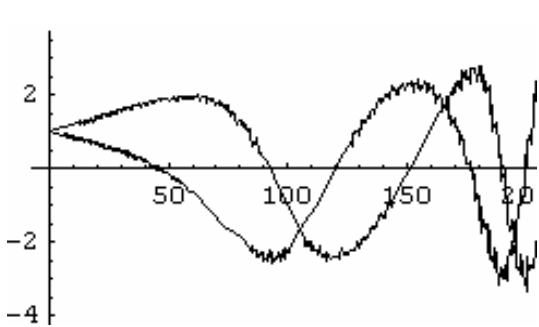
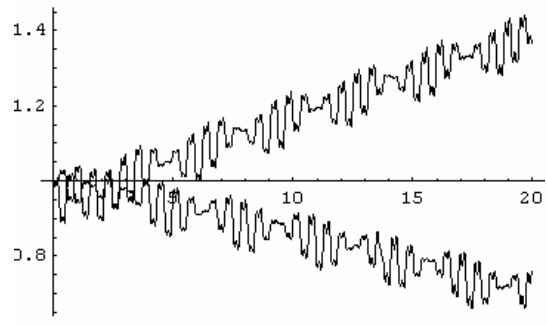

3.3. a                                   3.3. b

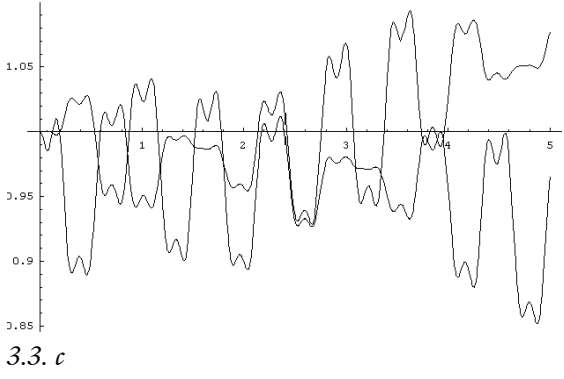
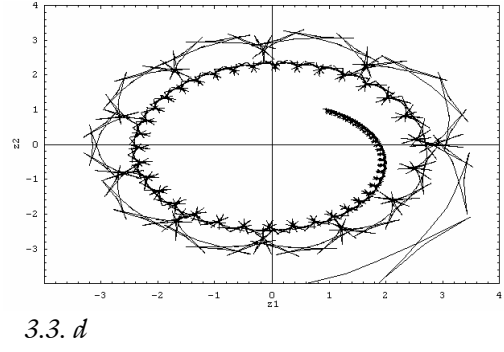

3.3. c                           3.3. d

*Fig. 3.3* a-c: Time evolution of $z_1$, $z_2$ of the non-Averaged, non-Autonomous system (3.21), in different time scales.
*Fig. 3.3* d : Phase portrait for $z_1$, $z_2$. Compare with Figures 2.2, 2.3.

Equation (3.21) is now a special case of system (3.8) with x = z = col ($z_1$, $z_2$) and the matrix A equal to the zero matrix. Then we may apply the previous corollaries. In particular, since our equations (3.21) are analytic in $b$, x and y for all x, y and $b$ sufficiently small, we may apply (3.17), (3.18). For k = 0, $x^{(0)} = a$ = col ($a_1, a_2$) and equations (3.18) for k = 0 are

$$\frac{\sigma}{\omega^2}a_2 + \left(1 - \frac{a_1^2 + a_2^2}{4}\right)a_1 + \frac{b}{a\omega^2}\cos\vartheta = 0$$
$$-\frac{\sigma}{\omega^2}a_1 + \left(1 - \frac{a_1^2 + a_2^2}{4}\right)a_2 + \frac{b}{a\omega^2}\sin\vartheta = 0 \quad (3.22)$$

If these equations have a solution $a_{10}$, $a_{20}$ and the corresponding Jacobian of the left-hand sides of these equations has a non-zero determinant, then, for $b$ sufficiently small, there is an exact solution of the determining equations (3.15) and thus a periodic solution of (3.19).

To analyze these equations, let $a_1 = r\cos\phi$, $a_2 = r\sin\phi$, to obtain the equivalent equations

$$r = \frac{b}{a\sigma}\sin(\vartheta - \phi)$$
$$\frac{\sigma}{\omega^2}\cot(\vartheta - \phi) + 1 = \frac{b^2}{4a^2\sigma^2}\sin^2(\vartheta - \phi) \quad (3.23)$$

Suppose that $0 < \sigma \leq b/2a$ (equivalently $b^2/4(\omega^2 - 1)^2 \geq 1$) and let $\xi = \theta - \phi$, $0 \leq \xi \leq \pi/2$. Then there is a unique solution $\xi(\sigma)$ of the last equation in (3.23). If $y(\sigma) = (1/\sigma)\sin\xi(\sigma)$, then the last equation in (3.23) is equivalent to

$$\frac{(1 - \sigma^2 y(\sigma))^{1/2}}{\omega^2 y(\sigma)} + 1 = \frac{b^2}{a^2 4}y^2(\sigma), \quad 0 < y(\sigma) \leq \frac{1}{\sigma} \quad (3.24)$$

Using this relation, it is possible to show that $y(\sigma)$ is monotonically decreasing and $y(\sigma) \to 2a/b$ as $\sigma \to b/2a$ from below. Also, $y(\sigma)$ approaches a limit $y^* = y^*(a,b,\omega)$ as $\sigma \to 0$ from above. The amplitude $r = b/a\; y(\sigma)$ is the unique positive solution of the cubic equation $r^3 - 4r - 4\dfrac{b}{a\omega^2} = 0$.

The symmetry in $\sigma$ implies that the amplitude r as a function of $\sigma$ is the one shown below (Fig 3.4).

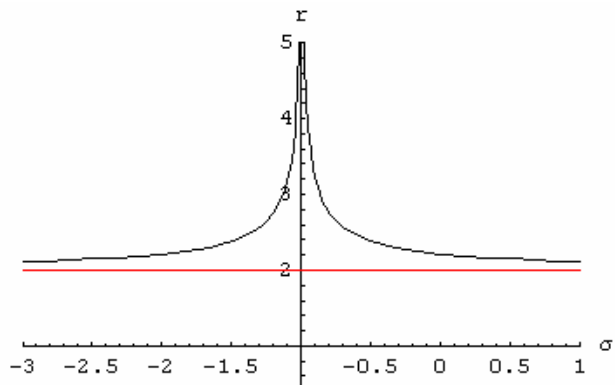

*Fig. 3.4*
variation of amplitude r while the frequency function $(\omega^2-1)/b$ (or $\sigma=\sigma(\omega)$) increases. The existence of an attracting set in r=2 is evident.

For $b = 0$, system (3.19) has a free frequency which is equal to 1; that is, **all** the solutions are periodic of period $2\pi$. On the other hand, if $b^2/4(\omega^2-1)^2 \geq 1$ for all $b$, $b \geq 0$, sufficiently small, then we have seen that there is a periodic solution of (3.19) with frequency $\omega$ and amplitude that is not small in $b$ even though the forcing function is small. **In other words, if $\omega^2 - 1$ is of order $b$, then the free frequency has been "locked" with the forcing frequency**. This phenomenon is sometimes referred to as the *locking-in phenomenon* or *entrainment of frequency*. [a more extended treatment of these phenomena can be found in appendix]

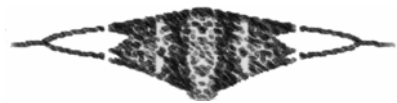

# PART B - NUMERICAL STUDY

## Chapter 4.

### A MORE THOROUGH VIEW OF THE VAN DER POL EQUATION

Up to now we have set the general (or some times more extensive) theoretical outline of the Van der Pol equation. So we know that in the autonomous equations a limit cycle occurs (a periodic solution with r=2). In the non-autonomous we expect to find periodic solutions as well, and moreover non-periodic (still bounded), almost-periodic and chaotic, which are indicated by the symbolic dynamic's procedure, that we presented in the previous chapter.

In this chapter we present a set of diagrams, constructed numerically, which provide precious help in obtaining a general overview of Van der Pol's equation behavior. Some of these diagrams correspond to the expected theoretical behavior, while others reveal valuable information which would be quite difficult or even impossible to gain by theoretical study.

All diagrams are made in Mathematica (versions 5.1 and 5.2) and C++ programming languages. The structure of these programs and codes are presented in the last part (appendix) of this work.

First we plot phase portraits and Poincarè sections for both cases (autonomous and non-autonomous). We have already shown a phase portrait in the previous chapter, when we studied the existence of a limit cycle (fig. 2.2).
We rewrite Van der Pol equation:

$$\begin{cases} \dot{x} = y \\ \dot{y} = -x + a(1-x^2)y + b\cos(\omega t + \vartheta) \end{cases} \quad (3.12)$$

We are going to investigate several cases of the autonomous Van der Pol equation, initially, by changing the damping parameter *a*, and see how phase portraits alter.

### *Numerics for the Autonomous Van der Pol equation*

We set in the above equations $b = 0$. Now the system of D.E.'s and the phase portraits, as well, are two dimensional so there is no need to use Poincarè sections. We obtain phase portraits for three different values of parameter *a*, as follows. In each diagram we present two trajectories, with different initial values ($\{x_0=0.5, y_0=0\}$ for the first and $\{x_0=4, y_0=0\}$ for the second trajectory). It can be clearly seen that the first trajectory starts from the inner domain of the limit cycle, while the second one from the outer domain. Both trajectories are approaching asymptotically the limit cycle, so (in a manner that) we can say that it is an attracting set (simple Jordan curve), which divides the phase plane into two parts.

Examining those three diagrams we see that the unstable fixed point (0,0) and limit cycle, as well, do not vanish, neither new fixed points occur, as the parameter *a* varies. In other words, the topology of the phase space does not change, as the parameter *a* increases and we expect no bifurcations. In fact, as the parameter *a* increases all the solutions tend to approach the limit cycle in a shorter and shorter time interval.

*Fig. 4* Limit cycles of (3.12) for b=0 and different values of a

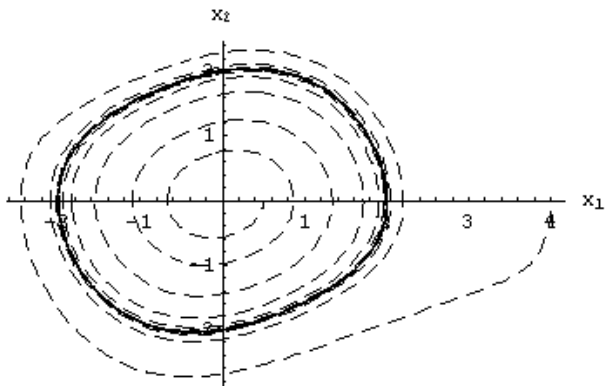

*Fig. 4.1a  a = 0.2*

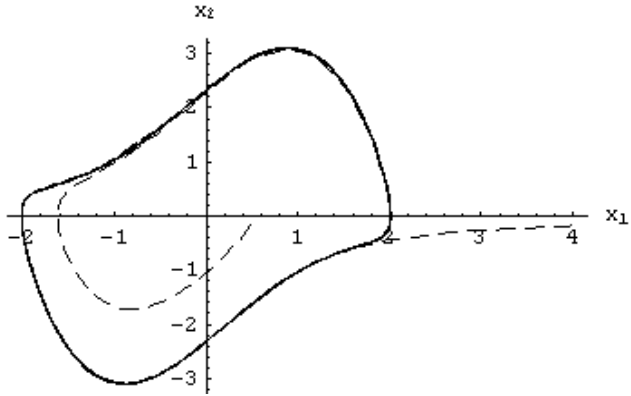

*Fig. 4.1b  a = 1.4*

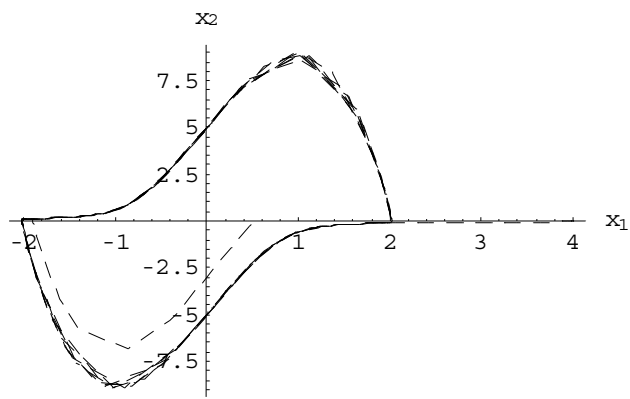

*Fig. 4.1c  a = 6*

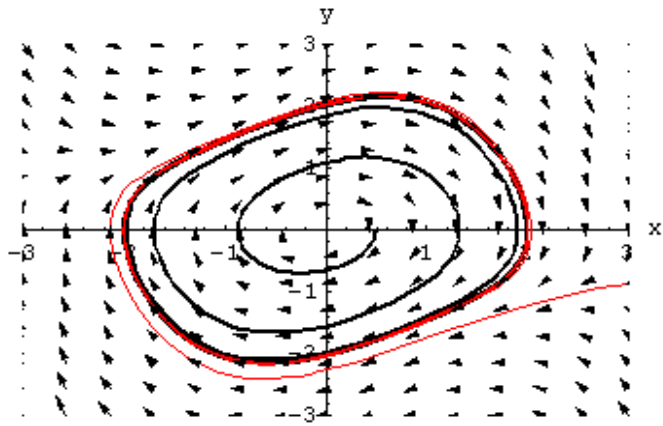

*Fig. 4.1d a = 0.4*

Phase Portrait with its vector field

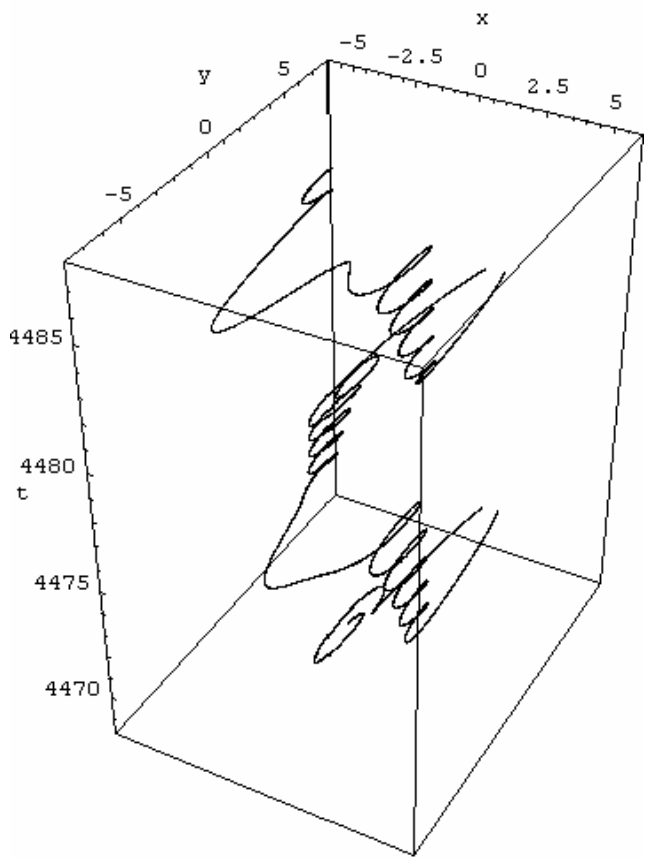

Three dimensional phase portrait for system (3.12), for *a*=5, *b*=15, *ω*=7 and Δt=22T

## *Numerics for the non - Autonomous Van der Pol equation*

Let's now investigate the case, in which an external excitation *p(t)* exists (*b ≠ 0* in (3.12)). This excitation has to be time periodic and the most common function for the *p(t)* is the (co)sinusoidal one. The phase portraits are now extended, which means that are three dimensional, {x,y,t} and we plot the 2-dimension projections of these. In order to study numerically this case we now need Poincarè sections, which we present below. As mentioned before, we expect to see both periodic and non-periodic trajectories and the existence of the former or the latter solutions, obviously, has to do with the set of parameters, which now are three[viii]; *a* for the damping function, *b* for the amplitude of the excitation and *ω* for the frequency of the excitation. We present quite many diagrams, for different values of these parameters, enough to outline the behavior of our equations. We moreover plot and present the time evolution of the solution x(t).

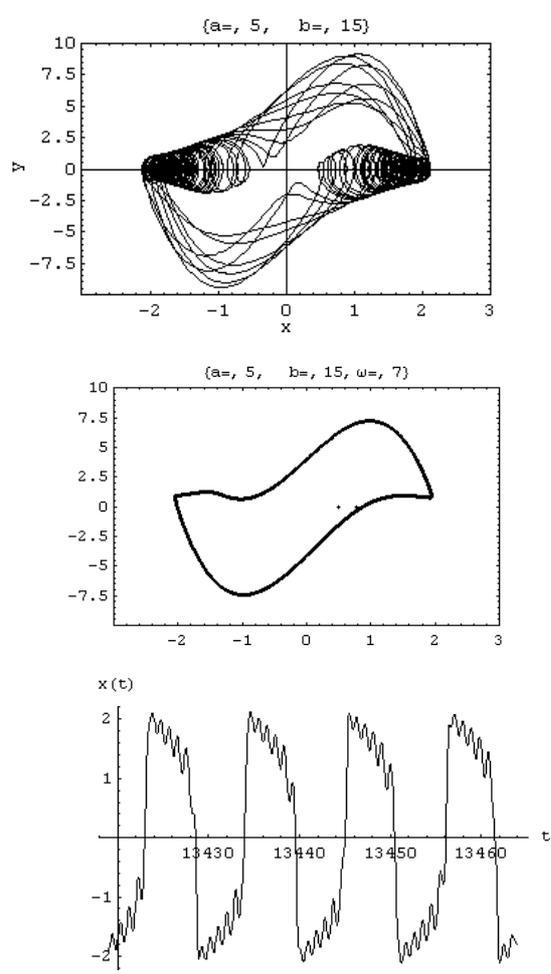

*Fig. 4.2* Phase portrait (projection of the extended 3D portrait), Poincarè map and the x-time series of (3.12) for values *a=*5, *b=*15 and ω=7, corresponding to **almost Periodic** orbit.

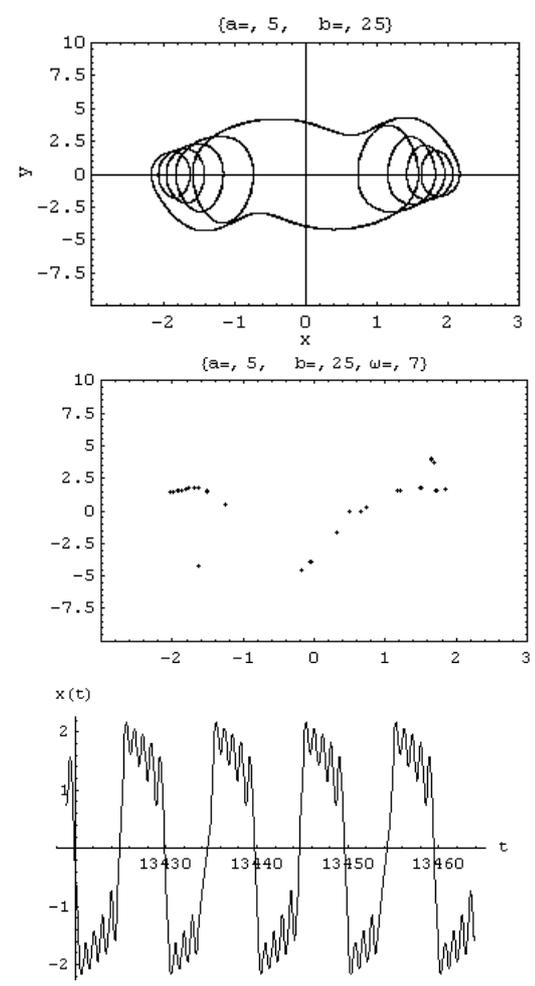

*Fig. 4.3* Phase portrait (projection of the extended 3D portrait), Poincarè map and the x-time series of (3.12) for values *a=*5, *b=*25 and ω=7, corresponding to **Periodic** orbit.

---

[viii] The forth parameter *θ* is set, for simplicity, equal to zero, without loss of generality

All these diagrams are plotted for large values of time ($t_{min}$ = 9900T, $t_{max}$ = 10000T, where T is the period of the exciting function) in order for the system to reach a stable state.

The right group of diagrams corresponds to a periodic solution. Poincarè section (second diagram) consists of distinct points; that is a periodic trajectory with period equal to the number of the points in Poincarè section. In the phase portrait (shown in the first diagram) the *phase trajectory* wouldn't cover densely all the interior area, even if it was evaluated for very large time intervals. In the third diagram (x-time series) it can clearly be seen that there is a superposition of two different frequencies are rationally dependent and the total motion is periodic.

Contrary to this, the left group of diagrams corresponds to an almost (or quasi-) periodic solution. The orbit in the Poincarè section is now a closed connected curve, so the trajectory never returns, after an n-period time, exactly to the same point where it was, but close to that. As an effect, the phase trajectory (first diagram) would cover densely the whole area of the (interior) phase space, if it was left evaluated for a quite large time interval t. In the third diagram (x-time series) there is, as well, a superposition of two different frequencies, whose ratio in this case, is an irrational number and the total motion is quasi-periodic.

In a similar way, we present two set of diagrams for greater values of *b*; a set of periodic and a set of almost periodic trajectories.

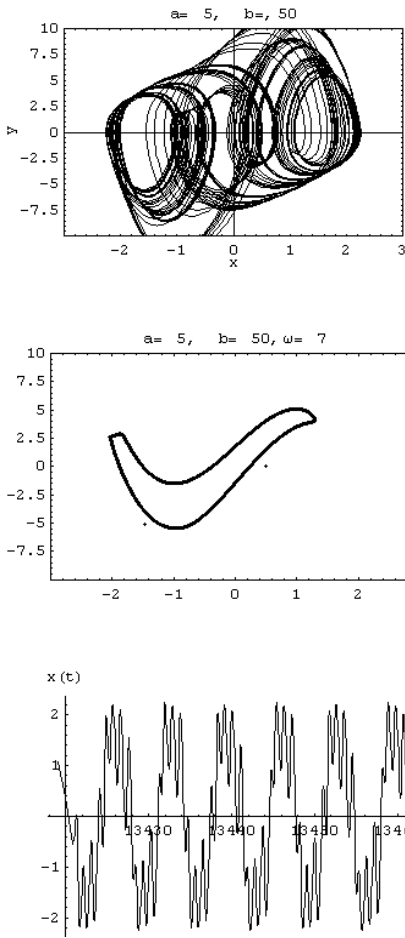

*Fig. 4.4* Phase portrait (projection of the extended 3D portrait), Poincarè map and the x-time series of (3.12) for values *a*=5, *b*=50 and ω=7, corresponding to **almost Periodic** orbit.

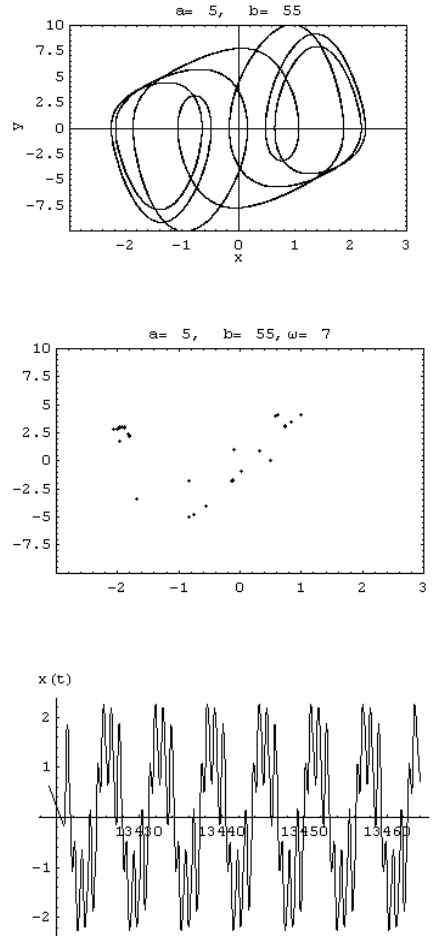

*Fig. 4.5* Phase portrait (projection of the extended 3D portrait), Poincarè map and the x-time series of (3.12) for values *a*=5, *b*=55 and ω=7, corresponding to **Periodic** orbit.

## *Bifurcation diagrams*

In the procedure above we kept parameters *a* and *ω* fixed and we increased *b* (from 15 to 25) to plot Poincarè sections. Periodic and almost periodic solutions occur alternately (for those specific values for *b*). In general, in the driven Van der Pol equations (*b≠0* in (3.12)) we might expect two possibilities. Firstly the two oscillations (the external force and the free oscillation) might continue independently, much like they do in the driven linear oscillator): solutions in this case are known as **drifting**. Alternatively the internal oscillator might be captured by the drive, so that there are oscillations at a single frequency (and harmonics): this case is known as **locked** or **entrained** [22, see also Appendix]. A sharp way to discriminate between these two possibilities is to ask what happens as a parameter (e.g. the drive amplitude or the dissipation) is smoothly varied.

So, to depict a more global behavior of the solutions of (3.12) and their dependence on parameters *a, b* and *ω* we plot the **bifurcation diagrams**[ix]. In fact a bifurcation diagram consists of points of the Poincarè section (only of the x(t) – solution, for great values of *t*) with respect to one parameter of the system (3.12). Thus we can trail the evolution of the behavior of solutions, as this parameter increases. For the diagrams, shown below, we used 50 points of the Poincarè section, for every value of *b*, with $t_{max}$=5000 (or 10000 T) and $t_{min}$ = $t_{max}$ – 50 T. The parameter *b* increases from *b*=0.01 to *b*=53, with step size=0.1 (530 different values) in the first one and from *b*=0.01 to *b*=80, with the same step size in the second diagram (800 different values for *b*).

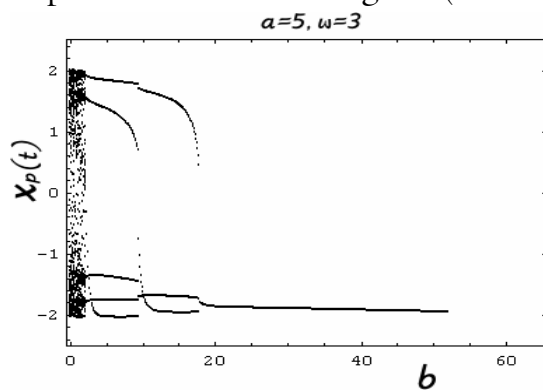 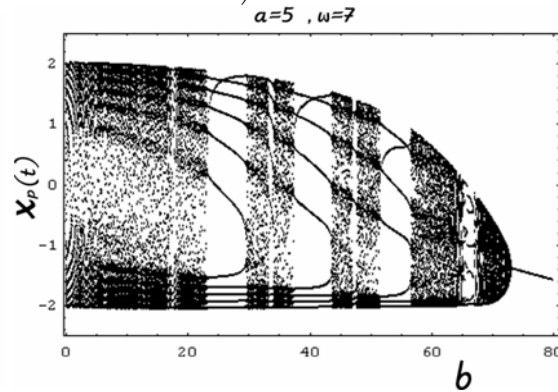

*Fig. 4.6a*  Bifurcation Diagram with respect to b, for *a=5, ω=3*

*Fig. 4.6b*  Bifurcation Diagram with respect to b, for *a=5, ω=7*

From this diagrams we can mark out that our system passes through periodic and non-periodic regions alternately, while the amplitude *b* increases. In other words, we have for some intervals of *b*, locking-in phenomena (providing periodic trajectories), with the period, of the solutions in that region, to be **odd**

---

[ix] We do not present a bifurcation diagram with respect to the parameter *a* due to the fact that it is out of interest, since no bifurcations occur. However, it can be easily plotted in Mathematica with a programme, same to the programmes constructed here in order to plot the above bifurcation diagrams, and can be found in appendix.

**multiple** of the driving period (sub-harmonics). When the driving amplitude turns to be sufficiently large and reaches a critical value ($b \approx 73.4$) the period of the solution equals to the driving period.

J.E. Littlewood, studying the equation $\ddot{y} - k(1-y^2)\dot{y} + y = b\mu k \cos(\mu t + a)$ mentioned that

"The equation has a considerable literature and experiments have suggested very interesting behaviour, especially in the case of large *a*. Stable motions occur which are subharmonics of large odd order (comparable with *a*), decreasing as *b* increases. Further, as *b* increases we have alternately one set of periodic stable motions, of order 2n-1, and two sets, of order 2n+1 and 2n-1, the shorter growing at the expense of the longer." [9]

A subharmonic is defined by the relation $\omega = \omega_E / m$, where $\omega_E$ is the external frequency and m is the order of the subharmonic. From this we obtain $T = 2\pi/\omega = m \cdot 2\pi/\omega_E = mT_E$.

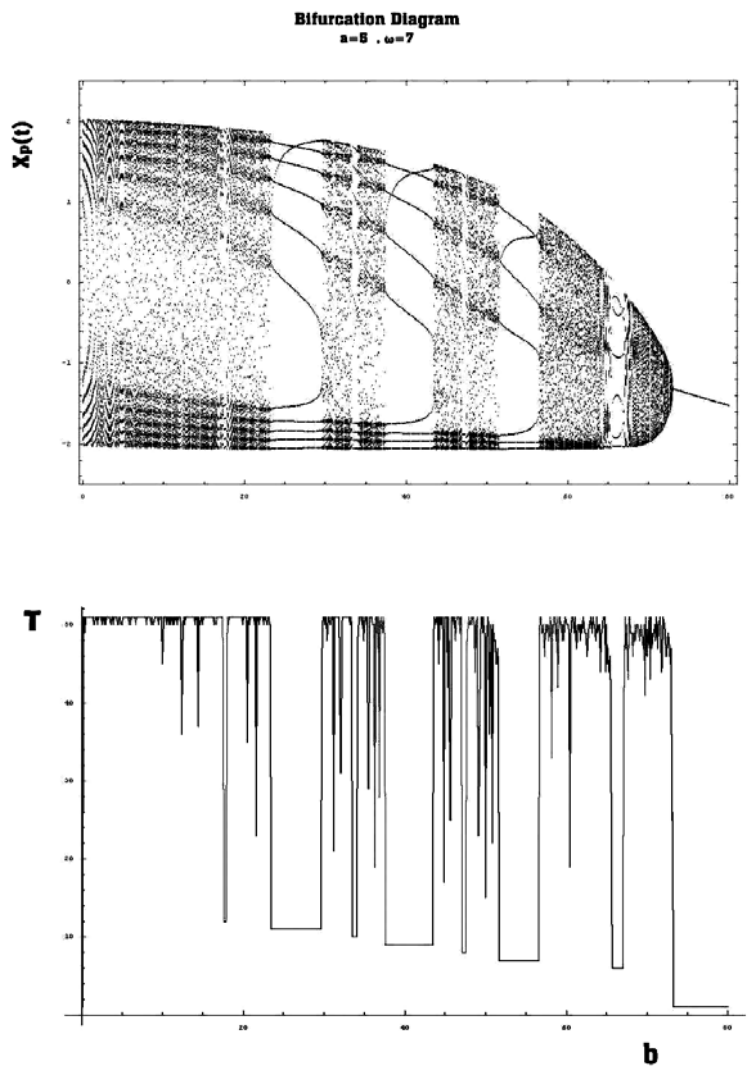

Consequently as b increases, m decreases and so does the period T of the system.

Our next diagram, showing exactly that behavior; bifurcations and the corresponding period, confirms the result coming from the theoretical study of the equation, by Littlewood, Levinson and others.

Moreover, comparing diagrams 4.6a and 4.7b we note that as the third parameter of the system ($\omega$) increases the entrained regions in the system become larger and more bifurcations occur (see also diagrams 4.8a and 4.11a)

*Fig. 4.7*  Bifurcation Diagram with respect to, *b* for *a=5*, $\omega = 7$ and the corresponding period

Next we present the bifurcation diagram with respect to the parameter $\omega$, which is the frequency of the exciting force. The parameter $\omega$ increases from $\omega = $ 0.1 to $\omega$=1 with step size=0.1 and from $\omega$=1 to $\omega$=7 with step size=0.001 (4100 different values in total).

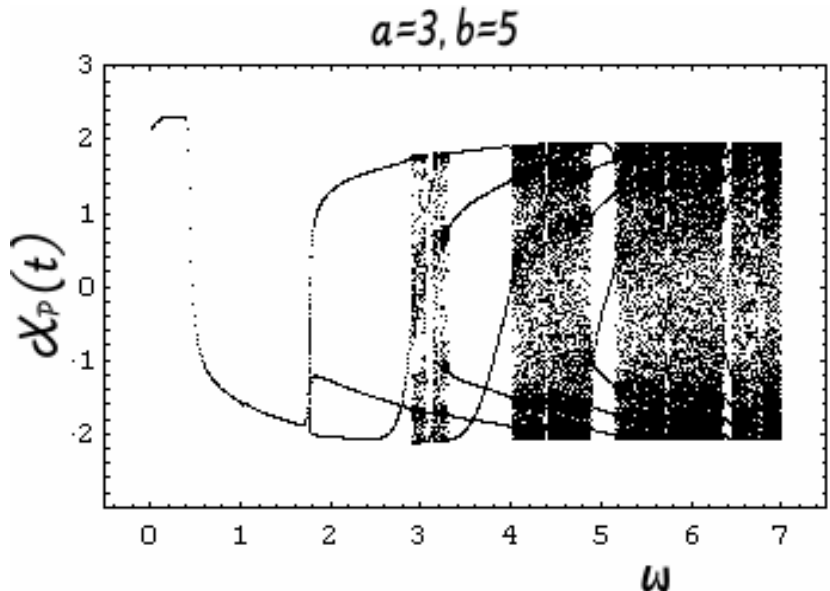

*Fig. 4.8a* Bifurcation Diagram with respect to $\omega$, for *a=3,* b =5

We now notice a similar behavior to the previous diagram but this time the bifurcations are inversed. The solutions seem to have a period equal to the exciting one for quite small values of $\omega$. As $\omega$ increases bifurcations occur, locked and drifting intervals alternates and again the period of the trajectory is an odd multiple of exciting period T (1, 3, 5, 7…). The periodic intervals occurring between non-periodic ones seem to become always narrower as the frequency $\omega$ increases. See also diagram 4.9.

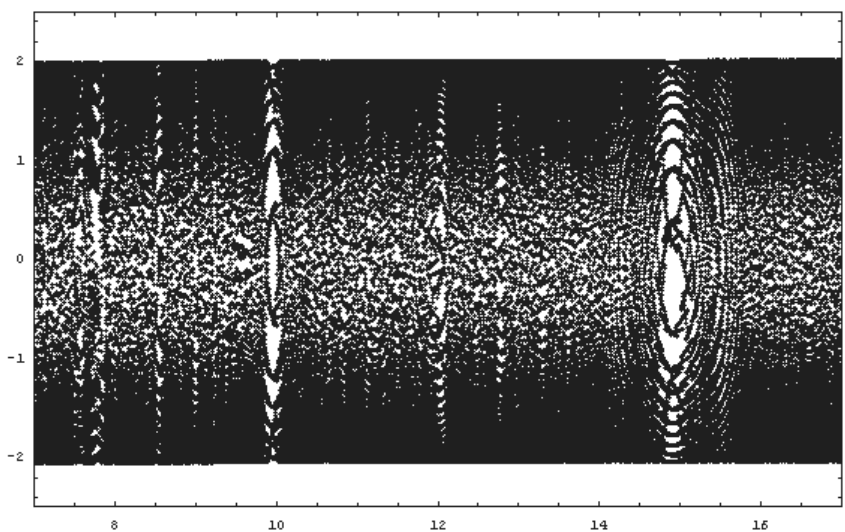

*Fig. 4.8b* Bifurcation Diagram with respect to $\omega$, for *a=3,* b =5, $\omega$ goes from 7 to 17

In diagram 4.8b the plot is the same as above, but this time for ω increasing from 7 to 17. In the greatest part of this diagram's interval (it seems that) no periodic trajectories occur. However we can not be sure about that until examining closely what happens in very short ω-intervals.

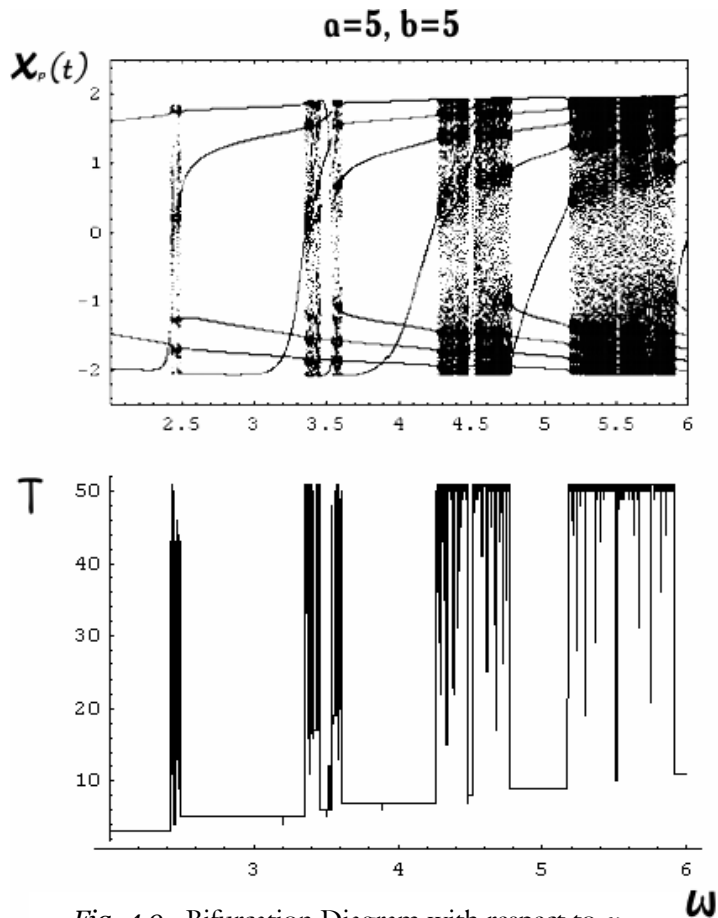

In diagram 4.9 we show how the period of the system varies as ω changes. The periodic and non periodic regions (entrainment and drifting regimes respectively) can be clearly seen. The values shown in this diagram indicate that the n period of the system is equal to n – times the external period, i.e. the first straight line in $T_d=3$ indicates that the system has a period equal to $T=3*2\pi/\omega$.

*Fig. 4.9* Bifurcation Diagram with respect to ω and the corresponding periods, for $a=5$, b =5

Last we present a bifurcation diagram (4.10), with respect to ω, but the parameter $a$ slightly increased ($a=5$). Parameter $b$ is chosen equal to 25, so from figure 4.7 we expect to obtain period trajectories (for values of ω close to 7). In this diagram the periodic intervals are now wider and they have compressed the non periodic ones. The bifurcations are similar to the previous diagram's ($a=3$, $b=5$) and the general pattern is also the same.

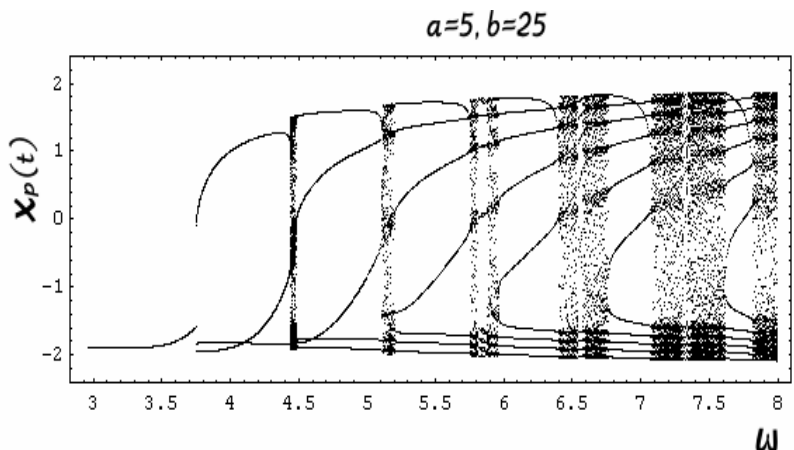

*Fig. 4.10.* Bifurcation Diagram with respect to ω, for $a=5$, b =25

## *Chaos in Van der Pol*

We examine next whether or not there is a region in the space of initial conditions in which chaotic dynamics occur. In the previous chapter we are known, from a theoretical point of view that one can prove – under a proper mapping for the successive intervals of the solution depicted – that there is a chaotic invariant set.

Before going further it is important to summarize the main efforts that have been done in order to prove the existence of chaotic trajectories.

In fact the Van der Pol equation, for many years, was not considered as a chaotic system by most of scientists. The study with Poincarè mapping and bifurcation diagrams of Van der Pol equations had not shown chaotic trajectories. Indeed, although Levinson [12] and Littlewood [9] gave the mathematical clue for chaotic behavior of driven Van der Pol oscillator, the experiments showed no chaotic behavior.

Parker and Chua [28] studied closely the behavior of electric circuits, which are described by Van der Pol equations. They reported a great number of simulations in computers, having replaced the characteristic of the circuit ($u=i^3/3 - i$) by a piecewise linear characteristic, which Levinson adopted in his analysis. By changing the parameters *a* and *b* they studied the asymptotic behavior of the solutions for each pair of parameters. They found the existence of periodic orbits and the results that are presented above. Parker and Chua, however, did not find chaotic solutions. Levi's [29] analysis showed that in initial conditions' space chaotic solutions form a set of measure zero, that is an infinitely thin set. Consequently it is impossible to trace them experimentally (in laboratory). It's not difficult, however, to explain qualitatively their existence. Chaotic solutions occur every time we have two fixed sub-harmonics with different periods, and lie exactly between the **domains of attraction** of the two sub-harmonics. Indeed, for a specific time interval they adopt one of these two periods, then they switch to the other, next they return to the first one etc., so that the whole behavior is completely random [30].

Nevertheless there are several (numerical) clues for the chaotic behavior of the Van der Pol oscillations. The most representative of them are:

- **The period doubling cascades**
- **Positive Lyapunov exponents**
- **Strange attractors and x-time series (showing sensitive dependence)**

## Period Doubling Cascades

The period doubling cascades are considered as a possible route to chaos. We have already seen these in the previous bifurcation diagrams. For example, comparing *Fig. 4.8a* and *Fig. 4.10* we notice that when the excitation amplitude *b* increases, the locking intervals in the diagram become wider and intervals with smaller periods compress or remove those with larger periods. Then the invariant torus is destroyed and (first-finite) period-doubling cascades occur. In order to see these cascades more evidently we choose some specific values for parameters *a* and b, and we plot again the bifurcation diagrams with respect to the exciting frequency $\omega$, but this time in a very small $\omega$-interval, using greater step size [14].

In Figures 4.10 we present the bifurcation diagram for values $a=b=5$ and $\omega$ increasing from 2 to 6. On the next two plots, which are enlargements of the first, one can easily see that between quasi-periodic intervals there exist small periodic ones with large periods. In the last plot period doubling phenomena are seen clearly. The superstructure of the bifurcation set of the driven non-linear oscillator arranges a specific fine structure of bifurcation curves in the parameter space in a repetitive order that is closely connected with the non-linear resonance of the system. The whole plot seems to be self-similar under scaling, which is a fractal property. The dependence of the periodicity of the system (number of fixed points) to the frequency $\omega$ seems to be quite sensitive, as we notice several bifurcations by changing the third or even the fifth decimal digit in $\omega$.

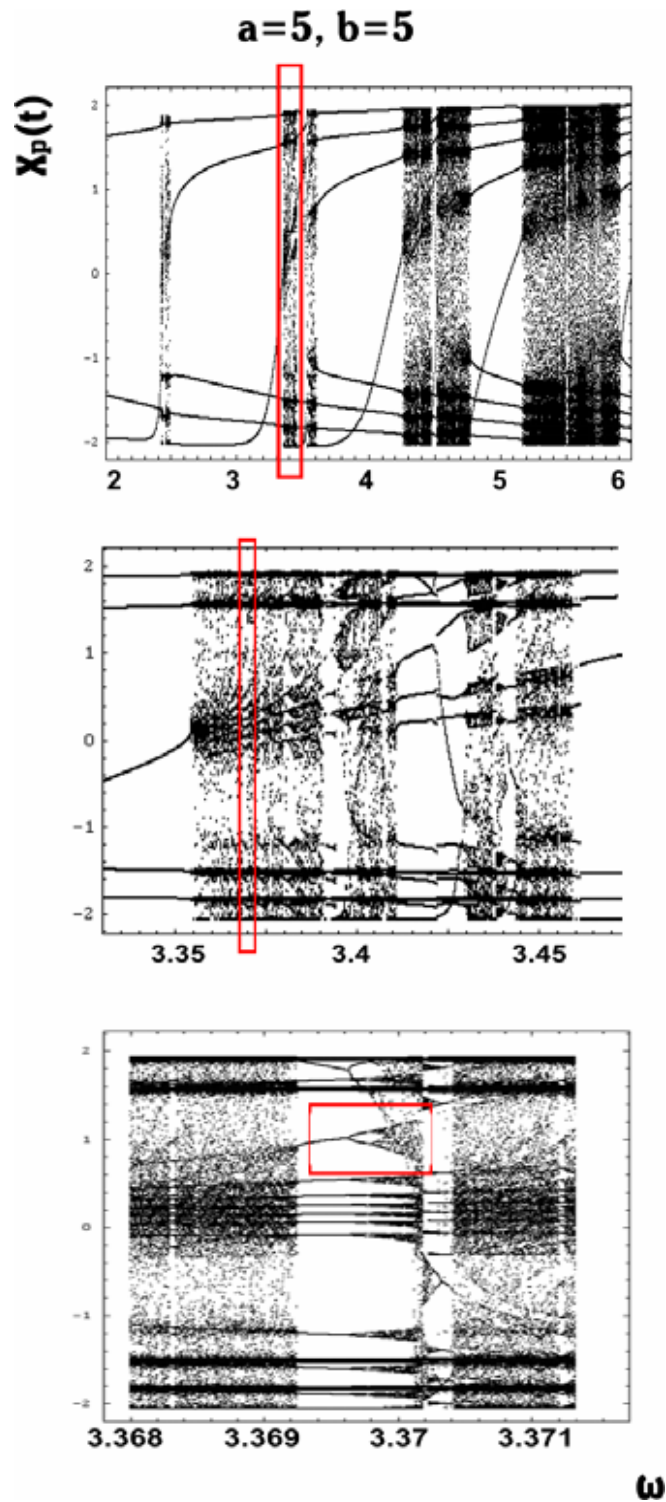

*Fig. 4.11.* Bifurcation Diagram and enlargements, of the marked regions, showing complete period doubling cascades into chaos.

We can also demonstrate the period doubling cascades by plotting phase portraits. To do this we keep fixed parameters *a* and *b* (both equal to 5) and we slightly change the frequency $\omega$. In the phase portraits below it is obvious that system passes through periodic and non-periodic regions, as $\omega$ increases. In the first group of diagrams $\omega$ varies from 2.457 to 2.469 with a step size that is not constant (0.001 to 0.003).

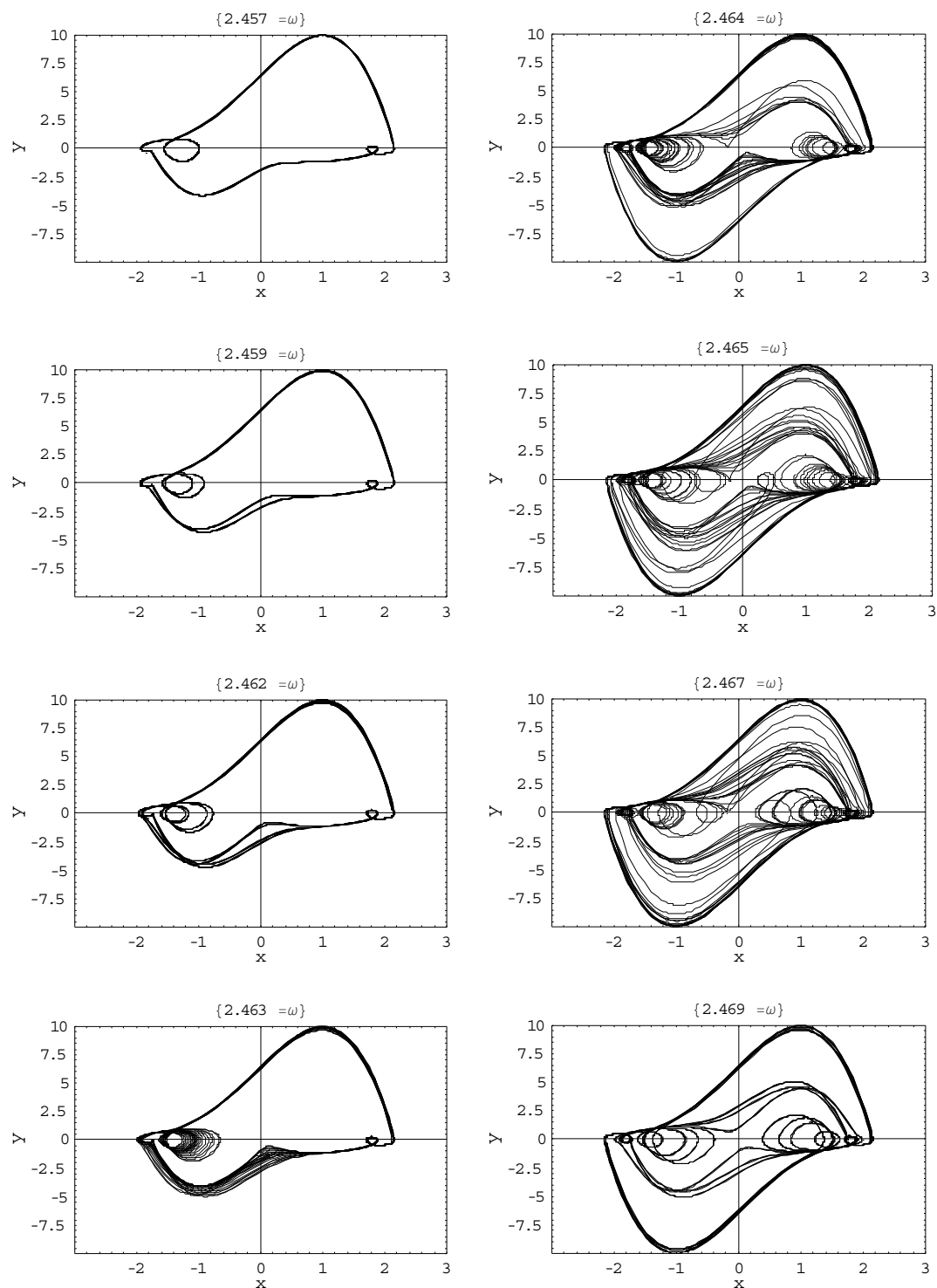

*Fig. 4.12* Phase Portraits for different values of $\omega$, showing period doubling cascades

In the second group of diagrams, values for ω are set from 3.365 to 3.375, corresponding thereby to *fig*.4.10. Parameters *a* and *b* are equal to 5.

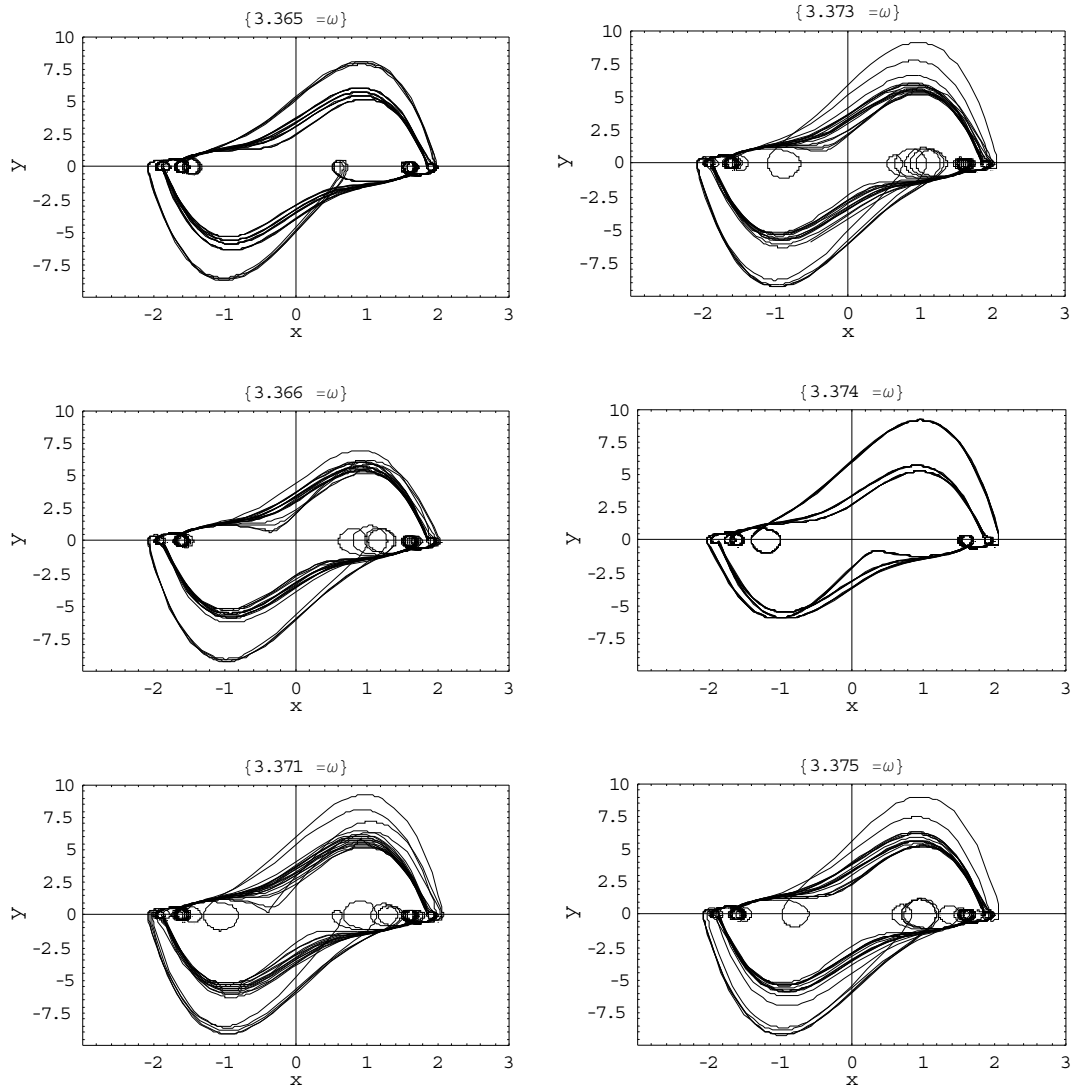

*Fig. 4.13* Phase Portraits for different values of ω, showing multiple period doubling cascades

It is obvious that in the second group of diagrams period doubling cascades occur in a faster and more complicated manner than in the first group. This is especially shown in the last two figures, where under a slight variation of ω (3.374 to 3.375) a great number of period doubling cascades occurs.

### *Lyapunov exponents*

Apart from period doubling cascades, a certain evidence for chaos in a non linear system is the existence of (at least) one positive *Lyapunov exponent*. A Lyapunov exponent $\lambda$ is, in fact, a measure of the divergence of two trajectories that start with (arbitrarily) close initial conditions. For one dimensional maps the exponent is simply the average $<\log|df/dx|>$ over the dynamics. There are now a number of exponents equal to the dimension of the phase space $\lambda_1, \lambda_2, \lambda_3$ where we choose to order them in decreasing value. The exponents can be intuitively understood geometrically: line lengths separating trajectories grow as $e^{\lambda_1 t}$ (where $t$ is the continuous time in flows and the iteration index for maps); areas grow as $e^{(\lambda_1+\lambda_2)t}$; volumes as $e^{(\lambda_1+\lambda_2+\lambda_3)t}$ etc. However, areas and volumes will become strongly distorted over long times, since the dimension corresponding to $\lambda_1$ grows more rapidly than that corresponding to $\lambda_2$ etc., and so this is not immediately a practical way to calculate the exponents.

It is obvious that if the largest of the Lyapunov exponents $\lambda$ comes positive then the trajectories will diverge exponentially after a short time interval, and the total behavior will be chaotic. Parlitz and Lauterborn [14], computed the largest Lyapunov exponent for the Van der Pol system, and, for $a=b=5$ and $\omega \geq 2.463$, they found it positive, which indicates chaotic behavior.

*Strange attractors*

Furthermore, in order to demonstrate the chaotic behavior of the Van der Pol system we can use Poincarè sections. We use the following values:
$a=3$, $b=5$ and $\omega=1.788$
$a=5$, $b=5$ and $\omega=3.37015$ and
$a=5$, $b=25$ and $\omega=4.455$
in all the cases above the largest Lyapunov exponents are also positive (see J. C. Sprott [31]).

In the following diagrams a part of the attractor is enlarged to emphasize its very thin structure, Fig. *4.14 -4.16*.

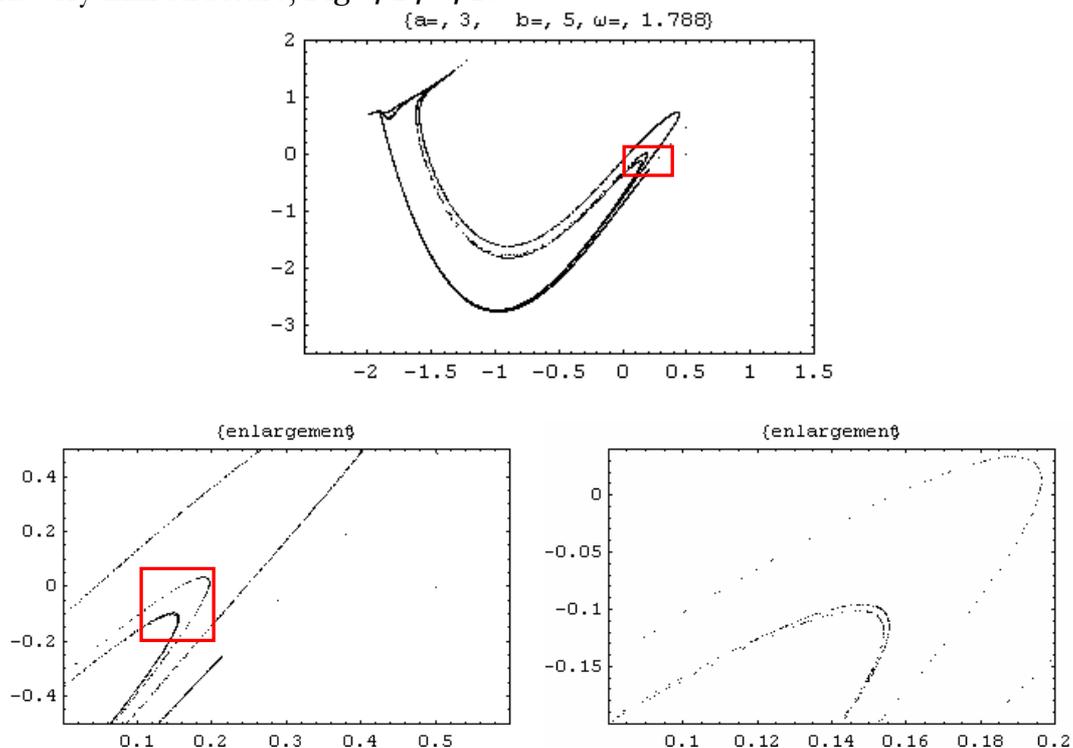

*Fig. 4.14* Poincarè section for $a=3$, $b=5$, $\omega=1.788$ and enlargements showing its very thin structure

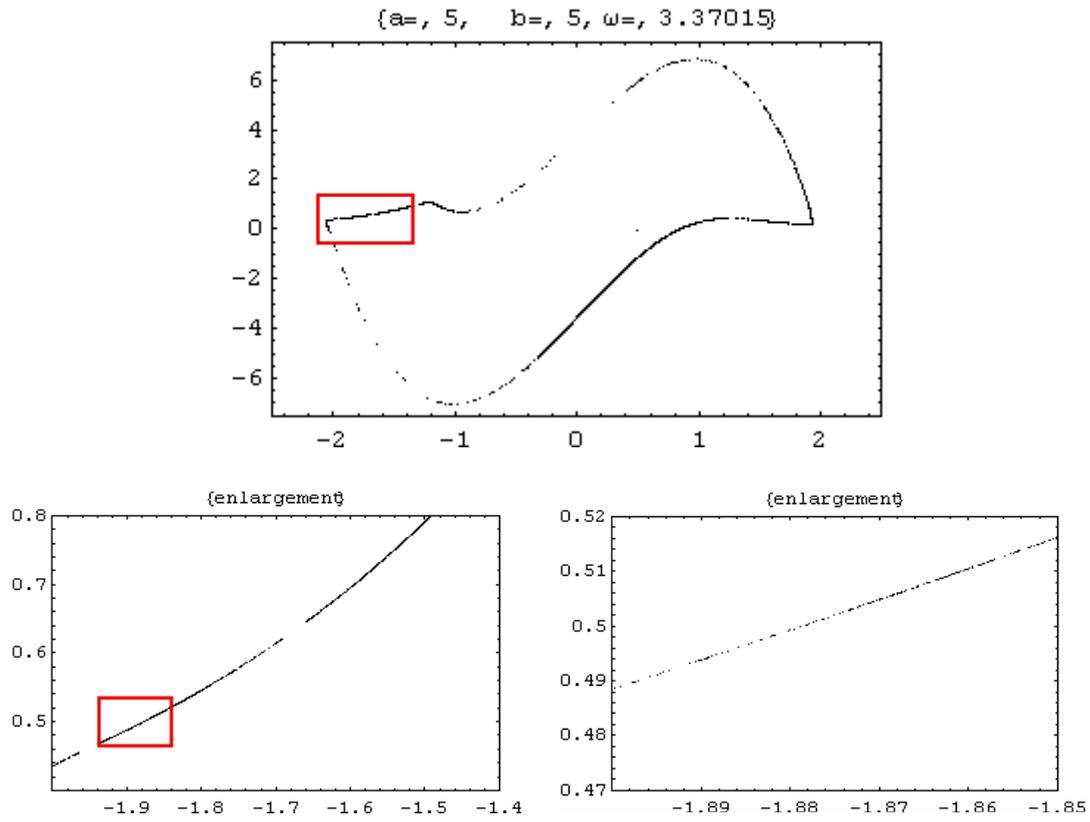

*Fig. 4.15* Poincarè section for *a*=5, *b*=5, *ω*=3.37015 and enlargements

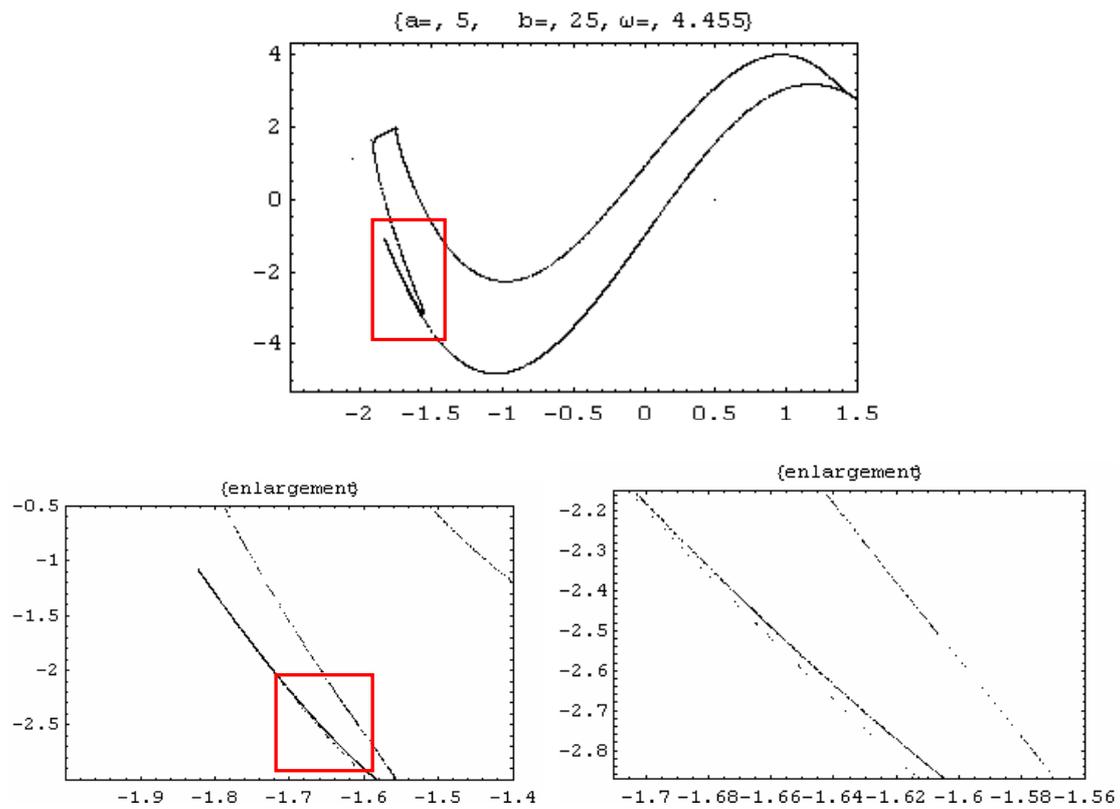

*Fig. 4.16* Poincarè section for *a*=5, *b*=25, *ω*=4.455 and enlargements

Some parts of these attractors (as it is specially shown in *fig. 4.14* and *4.16*) which seem to be simple curves, disclose their thin structure when they are enlarged. What looks like a straight line, in the first plot (in each group of diagrams), does not remain straight if it is enlarged enough. If one computes their dimension will find, as a result, a decimal number.

In addition, lots of sets of parameters *a, b, ω* can be used in order to draw chaotic attractors for the Van der Pol equation. The observant reader would have noticed that all sets of parameters, chosen above, correspond to the "smudged" regions of diagrams *4.8a, 4.11* and *4.10*, which interfere in periodic intervals with different periods. Therefore chaotic dynamics occur (mainly) in bifurcation points[x].

### *x-time series and exponential divergence*

Next we plot the x-time series corresponding to the Poincarè sections showed above. In order to demonstrate the sensitive dependence in the initial conditions we plot a group of two diagrams (for each set of parameters *a, b* and *ω*), with neighbor initial conditions. Even it in the second diagram the initial conditions are infinitesimally altered (compared with the first ones) the resultant diagrams are obviously different, as they diverge exponentially in a relatively short time interval.

More precisely, in the first trajectory (in both groups of diagrams below) the initial conditions are set as:
$x(0) = 0.5, y(0) = 0$ and in the second one are:
$x'(0) = x(0) + 10^{-5}, y'(0) = y(0) + 10^{-5}$

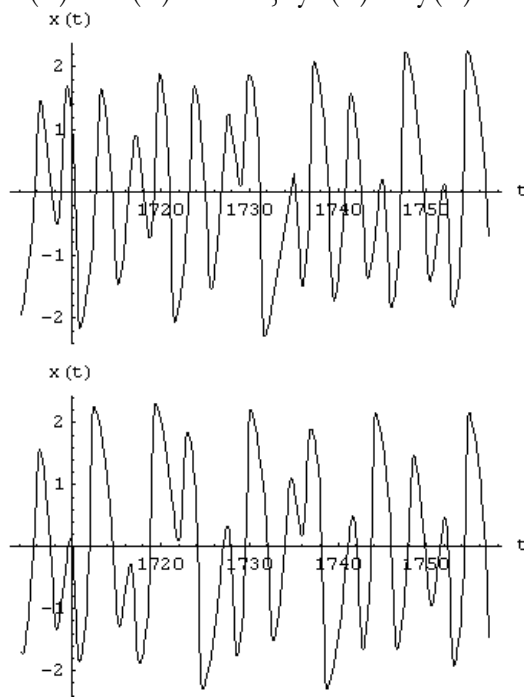

*Fig. 4.17* x-time series for slightly altered initial conditions, with *a=3, b=5, ω=1.788*

---

[x] This notation has also emphasized by Van der Pol and Van der Mark in their historical work B. Van der Pol and J. Van der Mark, Nature 120, 363 (1927).

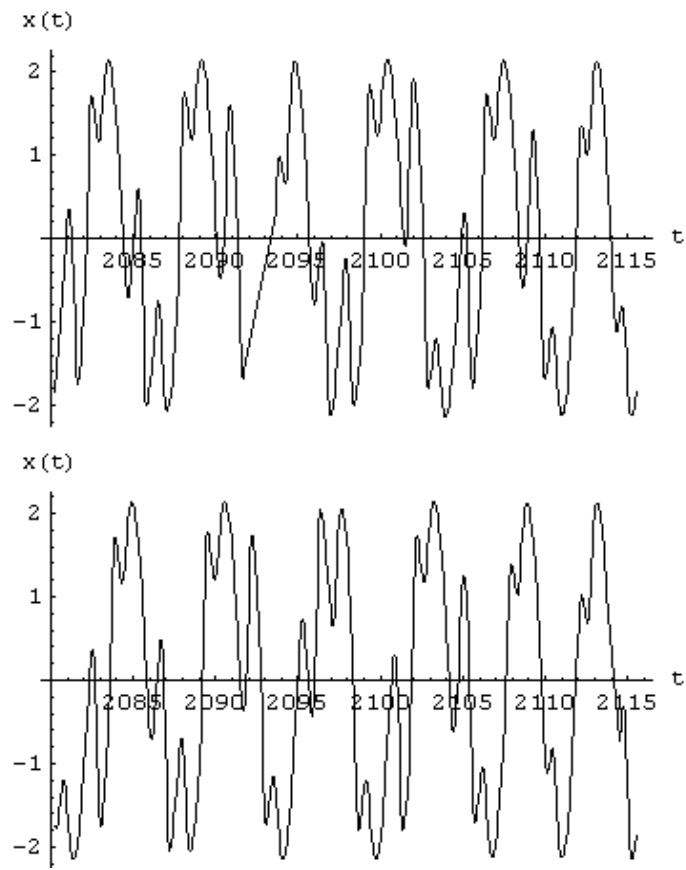

*Fig. 4.18* x-time series for slightly altered initial conditions, with *a*=5, *b*=25, ω=4.455. Exponential divergence is shown.

### *Fourier Spectra*

Last in order to investigate the (chaotic) behavior of the Van der Pol equation we use the *Fourier Spectra.*

Fourier analysis is a powerful tool, which indicates the main (or all) frequency (-ies) from which a periodic time series (i.e. function, signal or trajectory) consists of. A Fourier spectrum is a diagram, whose horizontal axis represents the frequency and the vertical one the amplitude of each of the frequencies. In fact, a Fourier spectrum decomposes a time series in its component frequencies and shows signatures at one (or more) fundamental frequency and at integer multiples (harmonics) of that, which may be infinite. In the case that our signal is periodic, its Fourier spectrum will consist of one fundamental frequency and all the harmonics of it. When the signal is quasi-periodic its spectrum will consist of two (at least) fundamental frequencies, whose ratio is an irrational number, and all of their harmonics. So in these two cases spectrums will consist of discrete signatures. Finally, when the signal is chaotic the spectrum will show signatures at more than one fundamental frequencies and moreover a continuous background will emerge in the diagram.

Now let's see what are the "signatures" of the solutions, studied above, at Fourier Spectra.

First we choose the values for parameters *a, b* and *ω* so as to correspond to a periodic, an almost periodic and a chaotic solution, respectively. Then we plot the Fourier spectrum for each of them (again in a programme constructed in Mathematica and which can be found in appendix), shown in figures 4.19-4.21.

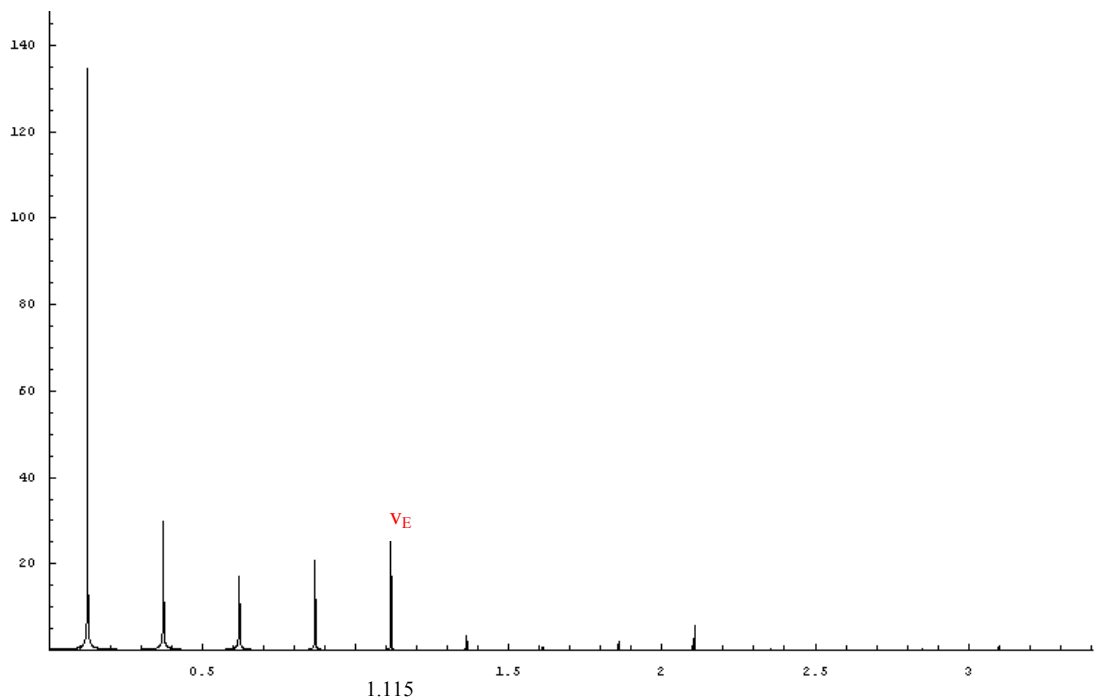

*Fig.4.19* Fourier Spectrum of a solution of Van der Pol system, for values of parameters: *a=5, b=40, ω=7,* corresponding to a periodic solution

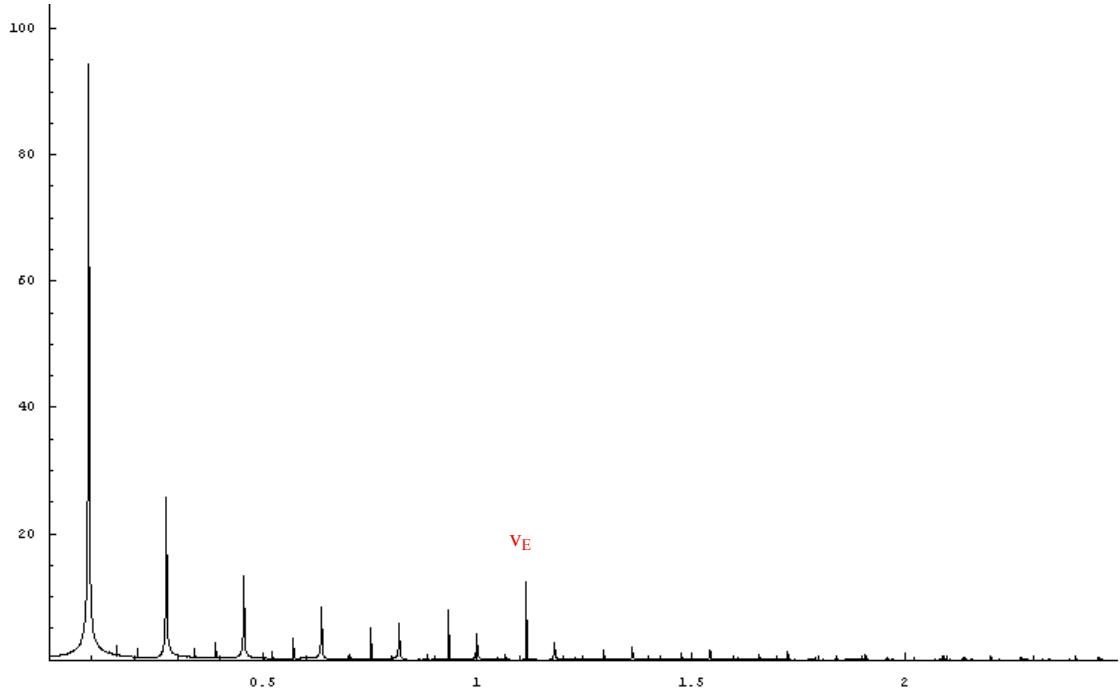

*Fig.4.20* Fourier Spectrum of a solution of Van der Pol system, for values of parameters: *a=5, b=15, ω=7,* corresponding to an almost-periodic solution

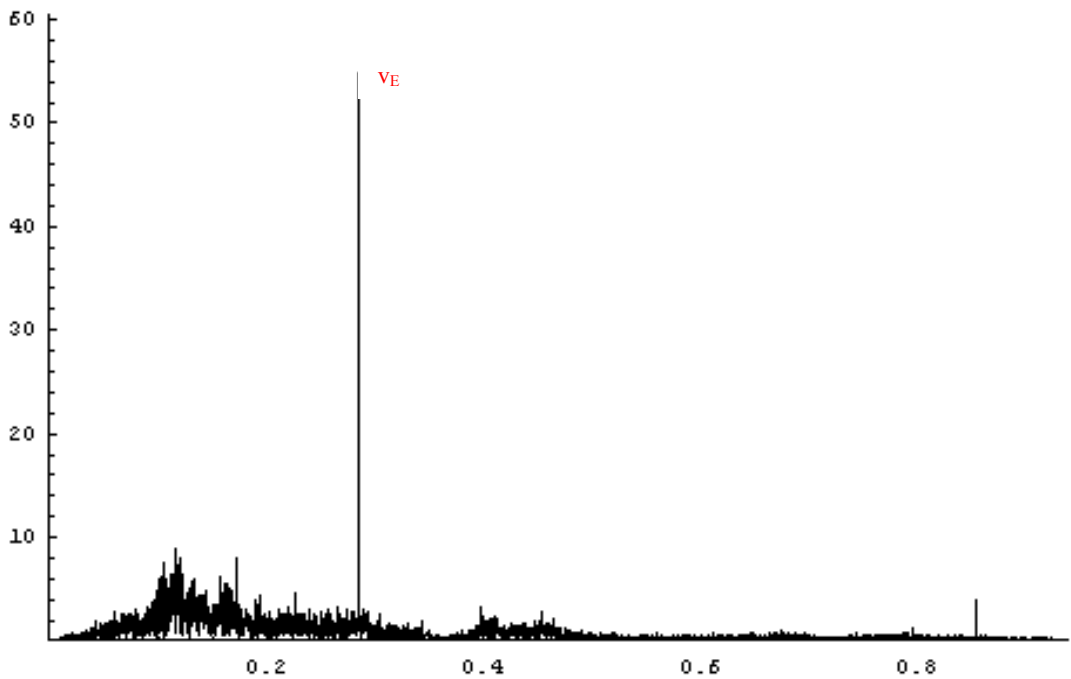

*Fig.4.21* Fourier Spectrum of a solution of Van der Pol system, for values of parameters: *a=3, b=5, ω=1.788,* corresponding to a chaotic solution

In the first diagram we set the parameters to: *a=5, b=40* and *ω=7*. It is obvious that the corresponding Fourier diagram depicts a periodic solution. The external frequency is equal to $v_E = \omega/2\pi = 1.115$ (in the horizontal axis of Fourier diagram the frequency v is presented, instead of the angular frequency *ω*). This frequency shows a signature in the spectrum (marked in the diagram) and its amplitude is 33.7, which is equal to the 24.6% of the amplitude of the frequency $v_{max}$ which has the greatest amplitude. The aforementioned frequency is a submultiple of $v_E$ ($v_E=9\ v_{max}$) or, in other words, a subharmonic of it. Furthermore, we mark that the total number of all frequencies arising in this Fourier spectrum, with amplitude greater than the 0.5% of the amplitude of $v_{max}$, is nine.

For a fixed value of b, there is either one fixed sub-harmonic with period $(2n+1)2\pi/\omega$, either two sub-harmonics co-exist with periods $(2n-1)2\pi/\omega$ and $(2n+1)2\pi/\omega$ [30].

In the second diagram the parameters are set equal to *a=5, b=15* and *ω=7* (the external frequency $v_E$ remains the same). This is the case of an almost periodic solution. The power spectrum shows signatures at both frequencies (of the external force and the free oscillation), whose ratio is equal to an irrational number. The amplitude of $v_E$ is now equal to the 13% of the amplitude of $v_{max}$; it is fairly smaller than was in the previous case. All frequencies, shown in this Fourier spectrum, are harmonics of either the first frequency (the external) or the second one (the frequency of the free oscillation). The total number of all frequencies (with amplitude greater than the 0.5% of the amplitude of $v_{max}$) is forty two.

Last, in the third diagram the parameters are set to *a=3, b=5* and *ω=1.788*. This is the case of a chaotic solution. We notice a high peak in the diagram, which is the signature of the external frequency ($v_E = \omega/2\pi = 0.284$). Contrary to the previous cases, a continuum background is shown. This indicates that a chaotic solution consists of an uncountable infinity of periods. The total number of all frequencies (with amplitude greater than the 0.5% of the amplitude of $v_{max}$) is in this case huge; nine hundred and twenty five.

### *3D Fourier Spectra*

In the Van der Pol system we have seen that solutions pass through periodic and non periodic regions, as a parameter of the system varies. We call them the locked and the drifting case, respectively. In the locked case, the frequencies of the peaks in the spectrum should remain fixed over a finite range of parameter values. In the drifting case, at least some frequencies should vary continuously. In this case the frequency ratios (of internal oscillator peaks to drive frequency) are necessarily passing through irrational values—such motion with two incommensurate frequencies is the quasi-periodic motion.

To demonstrate these ideas we plot the 3-dimensioned Fourier spectra with respect to parameter *b*, (*fig. 4.22a and b*).

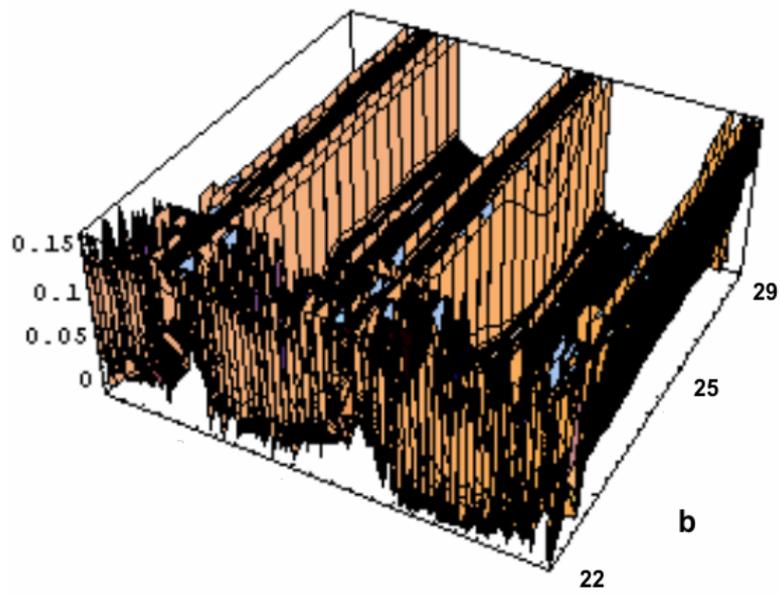

*Fig. 4.21a.* Fourier Spectrum of a solution of Van der Pol system for parameter *a*= 5, ω=7 and 22<*b*<29. The system is passing through locked and drifting regions

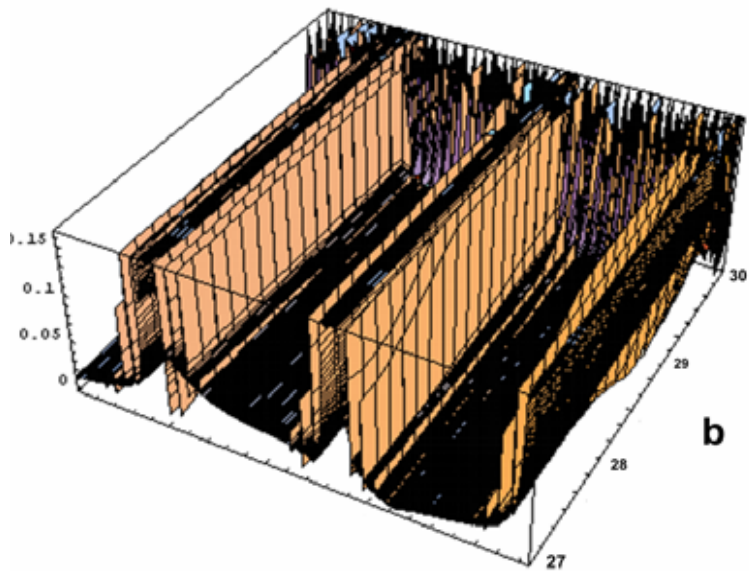

*Fig. 4.21b.* Fourier Spectrum of a solution of Van der Pol system for parameter *a*= 5, ω=7 and 27<*b*<30. The system is passing through locked and drifting regions

*Listening to …Van der Pol!*

The last part of the numerical study of the behavior of the Van der Pol system is an uncommon treatment of the solutions and their Fourier spectra, obtained above.

The basic idea is very simple. We know that a sound wave is an ordinary time-series, with the demand its fundamental frequency to vary from 16 Hz to 20 kHz (which is the audible spectrum). So, it can be analyzed in a Fourier series. A musical tone is a periodic function in time, so if we plot its Fourier spectrum we will notice a signature at a fundamental frequency $f_0$, which characterizes the pitch of the musical tone. A practical way (common to acoustics) to identify and discriminate between musical tones is to plot their Fourier spectra and then compare them.

So, why not follow the inverse procedure? Since we have the Fourier spectra for all kind of solutions (a periodic, an almost periodic and a chaotic) it is easy to combine the frequencies and their harmonics of each spectrum to one "playable" function[xi]. In other words we create a sound for every Fourier spectrum, such that we can distinguish between different solutions (periodic and non periodic) by their sound. The problem here is that the fundamental frequency of our "functions" does not fall into the audible spectrum (f≤1.2 Hz). To achieve that we normalize all the frequencies by a constant $k=10^3$ (further details about this procedure can be found in appendix).

It can easily be understood that the results cannot be presented in-print, but can be found in the accompanying cd.

Each resulting sound (for a periodic, a non periodic and a chaotic solution) differs from the others, just like their Fourier spectra. In the periodic case the result sounds like a musical tone; there is one fundamental frequency and a few harmonics. In the almost periodic, the resulting sound is a bit more complicated than the periodic one; more fundamental frequencies occur. Last, the chaotic case sounds like a noise; thousand of frequencies (that are not harmonics and form the continuum background) sound simultaneously.

Next we present the time series of the corresponding sound of a periodic, a non periodic and a chaotic solution of Van der Pol system.

---

[xi] Actually we do not create a function, but a list of points, computed numerically from the corresponding Fourier spectrum

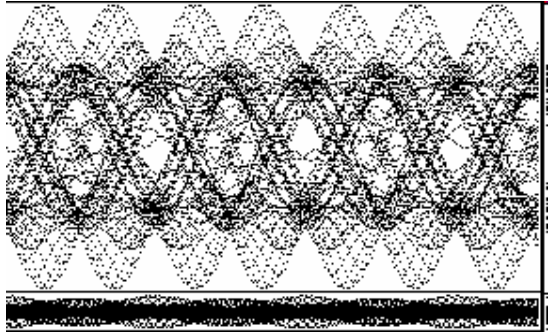 periodic sound
*Fig. 4.22* Sound time-series, corresponding to a periodic solution (*a=5, b=40, ω=7*)

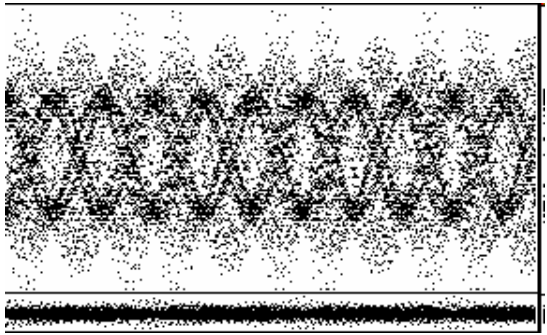 almost periodic sound
*Fig. 4.23* Sound time-series, corresponding to an almost periodic solution (*a=5, b=15, ω=7*)

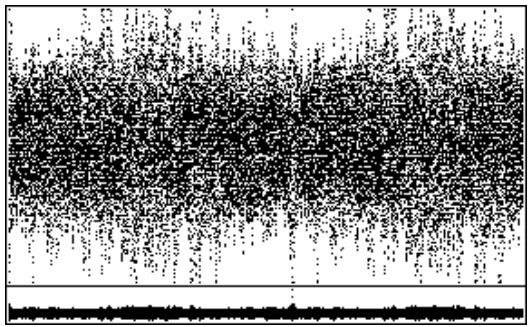 chaotic sound
*Fig. 4.24* Sound time-series, corresponding to a chaotic solution (*a=3, b=5, ω=1.788*)

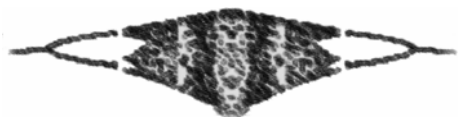

# PART C - APPENDIX

## Chapter 5
### ADDENDUM

### Van der Pol's transformation

In the theory of nonlinear oscillations the method of **van der Pol** is concerned with obtaining approximate solutions for equations of the type
$$\ddot{x} + x = \varepsilon f(x, \dot{x})$$
In particular for the *van der Pol* equation we have
$$\varepsilon f(x, \dot{x}) = (1 - x^2)\dot{x}$$
which arises in studying triode oscillations (Pol26a). *Van der Pol* introduces the transformation $(x, \dot{x}) \to (a, \phi)$ by
$$x = a\sin(t + \phi)$$
$$y = -a\cos(t + \phi)$$
The equation for *a* can be written as
$$\frac{da^2}{dt} = \varepsilon a^2 (1 - \tfrac{1}{4} a^2) + \ldots$$
where the dots stand for higher order harmonics. Omitting the terms represented by the dots, as they have zero average, *van der Pol* obtains an equation which can be integrated to produce an approximation of the amplitude *a*.

Note that the transformation $x = a\sin(t + \phi)$ is an example of *Lagrange's* "Variation des constantes." The equation for the approximation of *a* is the secular equation of *Lagrange* for the amplitude. Altogether *van der Pol's* method is an interesting special example of the perturbation method described by *Lagrange* in (Lag88a).

One might wonder whether *van der Pol* realized that the technique which he employed is an example of classical perturbation techniques. The answer is very probably affirmative. *Van der Pol* graduated in 1916 at the University of Utrecht with main subjects physics, he defended his doctorate thesis in 1920 at the same university. In that period and for many years thereafter the study of mathematics and physics at the Dutch universities involved celestial mechanics which often contained some perturbation theory. A more explicit answer can be found in (Pol20a) on the amplitude of triode vibrations; on page 704 *van der Pol* states that the equation under consideration "is closely related to some problems which arise in the analytical treatment of the perturbations of planets by other planets." This seems to establish the relation of *van der Pol's* analysis for triodes with celestial mechanics. [21]

# Poincarè's Vision of Chaos

We used several methods in this work in order to show strange attractors, occurring in the non autonomous van der Pol equation. More powerful is Poincarè's vision of chaos as the interplay of local instability (unstable periodic orbits) and global mixing (intertwining of their stable and unstable manifolds). In a chaotic system any open ball of initial conditions, no matter how small, will in finite time overlap with any other finite region and in this sense spread over the extent of the entire asymptotically accessible phase space. Once this is grasped, the focus of theory shifts from attempting to predict individual trajectories (which is impossible) to a description of the geometry of the space of possible outcomes, and evaluation of averages over this space.

Successive trajectory intersections with a Poincarè section, a d-dimensional hypersurface or a set of hypersurfaces P embedded in the (d+1)-dimensional phase space M, define the Poincarè return map P(x), a d-dimensional map of form
$$x' = P(x) = f^{\tau(x)}(x), \quad x', x \in P$$
Here the first return function $\tau(x)$ - sometimes referred to as the ceiling function - is the time of flight to the next section for a trajectory starting at x. The choice of the section hypersurface P is altogether arbitrary. The hypersurface can be specified implicitly through a function U(x) that is zero whenever a point x is on the Poincarè section. [32]

# Symbolic dynamics, basic notions

In this section we collect the basic notions and definitions of symbolic dynamics. The reader might prefer to skim through this material on first reading, return to it later as the need arises.

**Shifts**

We associate with every initial point $x_0 \in M$ the **future itinerary**, a sequence of symbols $S_+(x_0) = s_1 s_2 s_3 \cdots$ which indicates the order in which the regions are visited. If the trajectory $x_1, x_2, x_3, \ldots$ of the initial point $x_0$ is generated by

$$x_{n+1} = f(x_n), \quad (11.15)$$

then the itinerary is given by the symbol sequence

$$s_n = s \text{ if } x_n \in M_s. \quad (11.16)$$

Similarly, the **past itinerary** $S_-(x_0) = \cdots s_{-2} s_{-1} s_0$ describes the history of $x_0$, the order in which the regions were visited before arriving to the point $x_0$. To each point $x_0$ in the dynamical space we thus associate a bi-infinite itinerary

$$S(x_0) = (s_k)_{k \in \mathbb{Z}} = S_- . S_+ = \cdots s_{-2} s_{-1} s_0 . s_1 s_2 s_3 \cdots. \quad (11.17)$$

The itinerary will be finite for a scattering trajectory, entering and then escaping $M$ after a finite time, infinite for a trapped trajectory, and infinitely repeating for a periodic trajectory.

The set of all bi-infinite itineraries that can be formed from the letters of the alphabet $A$ is called the **full shift**

$$A^{\mathbb{Z}} = \{(s_k)_{k \in \mathbb{Z}} : s_k \in A \text{ for all } k \in \mathbb{Z}\}. \quad (11.18)$$

The jargon is not thrilling, but this is how professional dynamicists talk to each other. We will stick to plain English to the extent possible. We refer to this set of all conceivable itineraries as the **covering** symbolic dynamics. The name *shift* is descriptive of the way the dynamics acts on these sequences. As is clear from the definition (11.16), a forward iteration

$$x \to x' = f(x)$$

shifts the entire itinerary to the left through the "decimal point". This operation, denoted by the shift operator $\sigma$,

$$\sigma(\cdots s_{-2} s_{-1} s_0 . s_1 s_2 s_3 \cdots) = \cdots s_{-2} s_{-1} s_0 s_1 . s_2 s_3 \cdots, \quad (11.19)$$

demoting the current partition label $s_1$ from the future $S_+$ to the "has been" itinerary $S_-$. The inverse shift $\sigma^{-1}$ shifts the entire itinerary one step to the right. A finite sequence $b = s_k s_{k+1} \cdots s_{k+n_b-1}$ of symbols from $A$ is called a **block** of length $n_b$. A phase space trajectory is **periodic** if it returns to its initial point after a finite time; in the shift space the trajectory is periodic if its itinerary is an infinitely repeating block $p^\infty$. We shall refer to the set of periodic points that belong to a given periodic orbit as a **cycle**

$$p = \overline{s_1 s_2 \ldots s_{n_p}} = \left\{ x_{s_1 s_2 \ldots s_{n_p}}, x_{s_2 \ldots s_{n_p} s_1}, \ldots x_{s_{n_p} s_1 \ldots s_{n_p - 1}} \right\}. \quad (11.20)$$

By its definition, a cycle is invariant under cyclic permutations of the symbols in the repeating block. A bar over a finite block of symbols denotes a periodic itinerary with infinitely repeating basic block; we shall omit the bar whenever it is clear from the context that the trajectory is periodic.

Each **cycle point** is labeled by the first $n_p$ steps of its future itinerary. For example, the 2nd cycle point is labelled by

$$x_{s_2 \ldots s_{n_p} s_1} = x_{\overline{s_2 \ldots s_{n_p} s_1} \overline{s_2 \ldots s_{n_p} s_1}}$$

A **prime** cycle *p* of length *n_p* is a single traversal of the orbit; its label is a block of *n_p* symbols that cannot be written as a repeat of a shorter block.
([32], pp 189-190)

## Synchronization and Entrainment

The history of synchronization goes back to the 17th century when the famous Dutch scientist Christiaan Huygens reported on his observation of synchronization of two pendulum clocks which he had invented shortly before. This invention of Huygens essentially increased the accuracy of time measurement and helped him to tackle the longitude problem. During a sea trial, he observed the phenomenon that he briefly described in his memoirs Horologium Oscillatorium (The Pendulum Clock, or Geometrical Demonstrations Concerning the Motion of Pendula as Applied to Clocks)

'. . . It is quite worths noting that when we suspended two clocks so constructed from two hooks imbedded in the same wooden beam, the motions of each pendulum in opposite swings were so much in agreement that they never receded the least bit from each other and the sound of each was always heard simultaneously. Further, if this agreement was disturbed by some interference, it reestablished itself in a short time. For a long time I was amazed at this unexpected result, but after a careful examination finally found that the cause of this is due to the motion of the beam, even though this is hardly perceptible.'

According to a letter of Huygens to his father, the observation of synchronization was made while Huygens was sick and stayed in bed for a couple of days watching two clocks hanging on a wall. Interestingly, in describing the discovered phenomenon, Huygens wrote about 'sympathy of two clocks' (le phé'nomé'ne de la sympathie, sympathie des horloges). Besides an exact description, he also gave a brilliant qualitative explanation of this effect of mutual synchronization; he correctly understood that the conformity of the rhythms of two clocks had been caused by an imperceptible motion of the beam. In modern terminology this would mean that the clocks were synchronized in antiphase due to coupling through the beam. In the middle of the nineteenth century, in his famous treatise The Theory of Sound, Lord Rayleigh described an interesting phenomenon of synchronization in acoustical systems:

'When two organ-pipes of the same pitch stand side by side, complications ensue which not unfrequently give trouble in practice. In extreme cases the pipes may almost reduce one another to silence. Even when the mutual influence is more moderate, it may still go so far as to cause the pipes to speak in absolute unison, in spite of inevitable small differences.'

Thus, Rayleigh observed not only mutual synchronization when two distinct but similar pipes begin to sound in unison, but also the related effect of oscillation death, when the coupling results in suppression of oscillations of interacting systems. Being, probably, the oldest scientifically studied nonlinear effect, synchronization was understood only in the 1920s when E. V. Appleton and **B. Van der Pol** systematically—theoretically and experimentally—studied synchronization of triode generators. This new stage in the investigation of synchronization was related to the development of electrical and radio physics (now these fields belong to engineering). On 17 February 1920 W. H. Eccles and J. H. Vincent applied for a British Patent confirming their discovery of the synchronization property of a triode generator—a rather simple electrical device based on a vacuum tube that produces a periodically alternating electrical current. The frequency of this current oscillation is determined by the parameters of the elements of the scheme, e.g. of the capacitance. In their experiments, Eccles and Vincent coupled two generators which had slightly different frequencies and demonstrated that the coupling forced the systems to vibrate with a common frequency. A few years later Edward Appleton and Balthasar van der Pol replicated and extended the experiments of Eccles and Vincent and made the first step in the theoretical study of this effect. Considering the simplest case, they showed that the frequency of a generator can be entrained, or synchronized, by a weak external signal of a slightly different frequency. These studies were of great practical importance because triode generators became the basic elements of radio communication systems. The synchronization phenomenon was used to stabilize the frequency of a powerful generator with the help of one which was weak but very precise. To conclude the historical introduction, we cite the Dutch physician Engelbert Kaempfer who, after his voyage to Siam in 1680 wrote:

'The glow-worms . . . represent another shew, which settle on some Trees, like a fiery cloud, with this surprising circumstance, that a whole swarm of these insects, having taken possession of one Tree, and spread themselves over its branches, sometimes hide their Light all at once, and a moment after make it appear again with the utmost regularity and exactness . . ..'

This very early observation reports on synchronization in a large population of oscillating systems. The same physical mechanism that makes the insects to keep in sync is responsible for the emergence of synchronous clapping in a large audience or onset of rhythms in neuronal populations. We end our historical excursus in the 1920s. Since then many interesting synchronization phenomena have been observed and reported in the literature; some of them are mentioned below. More importantly, it gradually became clear that diverse effects which at first sight have nothing in common, obey some universal laws. Modern concepts also cover such objects as rotators and chaotic systems; in the latter case one distinguishes between different forms of synchronization: complete, phase, master-slave, etc.

A great deal of research carried out by mathematicians, engineers, physicists and scientists from other fields, has led to the development of an understanding that, say, the conformity of the sounds of organ pipes or the songs of the snowy tree cricket is not occasional, but can be understood within a unified framework. It is important to emphasize that synchronization is an essentially nonlinear effect. In contrast to many classical physical problems, where consideration of nonlinearity gives a correction to a linear theory, here the account of nonlinearity is crucial: the phenomenon occurs only in the so-called self-sustained systems. [33]

# The Van der Pol equation in Circuit

Our first figure shows an RLC circuit, which contains a voltage source that produces **E(t)** volts, an **R**-ohm resistor, an **L**-henry inductor, and a **C**-farad capacitor.

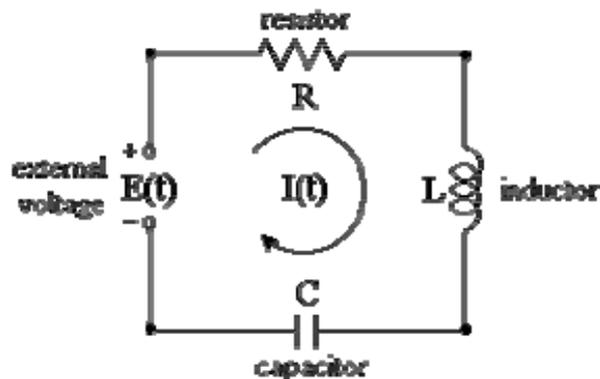

For purposes of this module, we assume the voltage source is a battery, i.e., **E(t)** is a constant **E**. (The circuit also contains a switch -- not shown -- that determines when the measurement of time starts.) When the switch is closed, a current **I(t)** (measured in amperes) begins to flow. **I(t)** is also the rate of change of the charge **Q(t)** on the capacitor, measured in coulombs. According to Kirchhoff's Voltage Law, current in the circuit is modeled by the equation

$$L\frac{dI}{dt} + RI + \frac{1}{C}Q = E.$$

If we differentiate both sides of this equation, we find the second-order linear differential equation for the current function:

$$L\frac{d^2I}{dt^2} + R\frac{dI}{dt} + \frac{1}{C}I = 0.$$

This is the familiar constant-coefficient, homogeneous equation that represents a damped harmonic oscillator.

If we write **V = V(t) = Q(t)/C** to represent the voltage drop at the capacitor, we may also represent the circuit by a first-order system of equations:

$$\frac{d}{dt}\begin{bmatrix} I \\ V \end{bmatrix} = \begin{bmatrix} -R/L & -1/L \\ 1/C & 0 \end{bmatrix}\begin{bmatrix} I \\ V \end{bmatrix} + \begin{bmatrix} E/L \\ 0 \end{bmatrix}$$

Now we turn our attention to a type of circuit -- from an early radio receiver -- studied in the 1920's by <u>Balthazar van der Pol</u>. This circuit is an RLC loop, but with the passive resistor of Ohm's Law replaced by an active element. In the 1920's this element was an array of vacuum tubes; now it is a semiconductor device. Thus, our circuit becomes

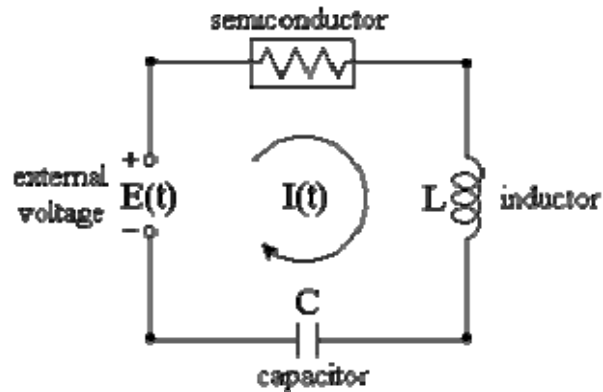

Unlike a passive resistor, which dissipates energy at all current levels, a semiconductor operates as if it were pumping energy into the circuit at low current levels, but absorbing energy at high levels. The interplay between energy injection and energy absorption results in a periodic oscillation in voltages and currents.

We suppose a power supply is attached to the circuit as shown, and the circuit is energized. Then, at time **t = 0**, the external source is switched out, so **E(t) = 0**. We will examine how the voltages and current change from then on. The voltage drop at the semiconductor, instead of being a linear function of **I(t)**, is the nonlinear function **I(I²-a)**, where **a** is a positive parameter. Note that this function is negative for small (positive) values of **I** and positive for larger values of **I**. Since current can flow in either direction in the circuit, all three sign changes of this cubic expression are significant.

For convenience, we will suppose that units are chosen in which **L** and **C** are both **1**. In particular, this means that **Q = V**, the voltage drop at the capacitor, and **I = dV/dt**. Our modified Kirchhoff's Law equation then becomes the (autonomous) van der Pol equation:

$$\frac{dI}{dt} + I(I^2 - a) + V = 0$$

# Modern Applications of the Van der Pol Equation

**We last, briefly present some quite interesting applications of the van der Pol equations, used as a model for much more complicated systems.**

## Neurophysiology

*Modeling the Gastric Mill Central Pattern Generator of the Lobster With a Relaxation-Oscillator Network*
PETER F. ROWAT AND ALLEN I. SELVERSTON, JOURNAL OF NEUROPHYSIOLOGY
Vol. 70, No. 3, September 1993.

The cell model is a generalization and extension of the Van der Pol relaxation oscillator equations. It is described by two differential equations, one for current conservation and one for slow current activation. The model has a fast current that may, by adjusting one parameter, have a region of negative resistance in its current-voltage (I-V) curve. It also has a slow current with a single gain parameter that can be regarded as the combination of slow inward and outward currents.

Cell model
The model for an isolated cell was adapted from the generalization by Lienard (1928) of Van der Pol's relaxation oscillator (1926). It is written as two equations

$$\tau_m \frac{dV}{dt} + F(V) + q = 0$$

$$\tau_s \frac{dq}{dt} = -q + q_\infty(V)$$

where $F(V)$ is given by $F(V) = V - A_f \tanh[(\sigma_f/A_f)V]$
and $q_\infty(V)$ is given by $q_\infty(V) = \sigma_s V$
An alternative definition is also used

$$q_\infty(V) = \begin{cases} \sigma_{in}V & \text{for } V < 0 \\ \sigma_{out}V & \text{for } V > 0 \end{cases}$$

Here V represents the membrane potential, $\tau_m$, represents the membrane time constant, $F(V)$ represents the current-voltage (I-V) curve of an instantaneous, voltage-dependent current, and q represents a slow current with time constant $\tau_s$, and steady-state I-V curve $q_\infty(V)$.

# Continuum media mechanics

## Dynamics of elastic excitable media

Excitable media are usually studied using the model of van der Pol, FitzHugh and Nagumo [van der Pol & van der Mark, 1928; FitzHugh, 1960, 1961; Nagumo et al., 1962]. This model normally includes only di_usive coupling. Originally from physiology and chemistry, excitable media have also captured the attention of physicists and mathematicians working in the area of nonlinear science because of the apparent universality of many features of their complex spatiotemporal properties

The elastic excitable medium model may be written [Cartwright et al., 1997]

$$\frac{\partial^2 \chi}{\partial t^2} = c^2 \frac{\partial^2 \chi}{\partial x^2} - (\chi - \nu t) - \gamma \phi\left(\frac{\partial \chi}{\partial t}\right)$$

where, in the language of frictional sliding, x(x,t) represents the time-dependent local longitudinal deformation of the surface of the upper plate in the static reference frame of the lower plate, $\phi(\partial\chi/\partial t) = (\partial\chi/\partial t)^3/3 - \partial\chi/\partial t$ is the friction function, as the dashed line in Fig. 1(b), measures the magnitude of the friction, c is the longitudinal speed of sound, and v represents the pulling velocity or slip rate.

Here we focus on the properties of the propagating front regime with special emphasis on the selection mechanisms for the front velocity and spatial configuration. We suppose a solution of the type ψ(x,t) = f(z), where z = x/v + t, and v is the front velocity. This together with a further rescaling leads to

$$\frac{d^2 f}{dz^2} + \mu(f^2 - 1)\frac{df}{dz} + f = \nu,$$

which is the van der Pol equation with the nonlinearity rescaled by $\mu = \gamma/\sqrt{1 - c^2/v^2}$. The propagating fronts are then periodic solutions of the van der Pol equation. The parameter *μ* is undefined until the value of the front velocity *v* is chosen. However, we know that the period of the solution is a function T = T(*μ*) of *μ*.

**Relativistic Parametrically Forced van der Pol Oscillator**
Y. Ashkenazy, C. Goren and L. P. Horwitz, chao-dyn/9710010 v2 2 Apr 1998

A manifestly relativistically covariant form of the van der Pol oscillator in 1+1 dimensions is studied. In this paper, a manifestly covariant dynamical system is studied with no Hamiltonian structure, i.e., a covariant form of the van der Pol oscillator. The equation for the nonrelativistic externally forced van der Pol oscillator is

$$\ddot{x} + \alpha(x^2 - 1)\dot{x} + kx = g\cos\omega t,$$

where the right hand side corresponds to an external driving force. Although there is no Hamiltonian which generates this equation, we may nevertheless consider its relativistic generalization in terms of a relative motion problem, for which there is no evolution function K (the relativistic invariant evolution function K is analogous to the nonrelativistic Hamiltonian). The relativistic generalization of this equation is of the form

$$\ddot{x}^\mu + \alpha(\rho^2 - 1)\dot{x}^\mu + x^\mu = 0,$$

or, in terms of components,

$$\ddot{x} + \alpha(x^2 - t^2 - 1)\dot{x} + kx = 0$$
$$\ddot{t} + \alpha(x^2 - t^2 - 1)\dot{t} + kt = 0,$$

where x, t are the relative coordinates of the two body system. The term $\dot{x}$ can be understood as representative of friction due to radiation, as for damping due to dipole radiation of two charged particles, proportional to the relative velocity $\dot{x}^\mu$ in the system. To study the existence of chaotic behavior on this system, we add a driving force‡ to the system (it already contains dissipation intrinsically) in such a way that it does not provide a mechanism in addition to the dissipative terms for the change of "angular momentum" (this quantity is conserved by the second equation with $a=0$).

$$M^{01} = x^0 p^1 - x^1 p^0 = m(t\dot{x} - x\dot{t}).$$

To achieve this, we take a driving force proportional to xμ, so that the system of the equations become

$$\ddot{x} + \alpha(x^2 - t^2 - 1)\dot{x} + kx = fx\cos\omega\tau$$
$$\ddot{t} + \alpha(x^2 - t^2 - 1)\dot{t} + kt = ft\cos\omega\tau.$$

## PROGRAMMING IN MATHEMATICA

We last present the programs that were made in Mathematica and C++, in order to obtain all the above diagrams.

### Autonomous Equations - Limit Cycles

```
ClearAll["Global`*"];
dsol = DSolve[r'[t] == (a/2) r[t] (1 - (1/4) r[t]^2), r[t], t];
r = r[t] /. dsol[[2]];
r1 = r /. {C[1] → r0};
Print["γενική  λύση  r(t)=", r1]
```

$$\gamma\varepsilon\nu\iota\kappa\acute{\eta}\ \lambda\acute{\upsilon}\sigma\eta\ r(t) = \frac{2\,e^{\frac{a\,t}{2}}}{\sqrt{-e^{8\,r0} + e^{a\,t}}}$$

```
ClearAll["Global`*"];
dsol = DSolve[{r'[t] == (a/2) r[t] (1 - (1/4) r[t]^2), r[0] == r0}, r[t], t];
r = r[t] /. dsol[[2]];
Print["r(t)=", r]
```

Solve::ifun : Inverse functions are being used by Solve, so some
   solutions may not be found; use Reduce for complete solution information. More…

Solve::ifun : Inverse functions are being used by Solve, so some
   solutions may not be found; use Reduce for complete solution information. More…

$$r(t) = \frac{2\,e^{\frac{a\,t}{2}}}{\sqrt{-1 + e^{a\,t} + \frac{4}{r0^2}}}$$

```
r2=r/.{a→0.1,r0→1};
Plot[r2,{t,150,225}]
```

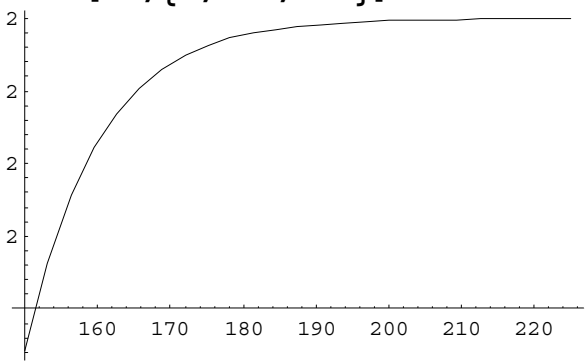

```
For[
 a = 0, a ≤ 7, a = a + 0.1,
 If[
  a == 0.1 || a == 1.4 || a == 6,
  f[a_, x0_, tmax_] :=
   NDSolve[{x'[t] == y[t], y'[t] == -x[t] - a * (x[t]^2 - 1) * y[t], x[0] == x0, y[0] == 0},
    {x, y}, {t, 0, tmax}];
  s1 = f[a, 0.5, 60];
s2 = f[a, 4, 60];
  Plot[{x[t] /. s1, x[t] /. s2}, {t, 0, 60},
    PlotStyle → {{RGBColor[1, 0, 0]}, {RGBColor[0, 0, 1]}}, AxesLabel → {t, "x_1,x_2"}]
   ParametricPlot[{{x[t], y[t]} /. s1, {x[t], y[t]} /. s2}, {t, 0, 60},
    PlotStyle → {{RGBColor[1, 0, 0]}, {RGBColor[0, 0, 1]}}, AxesLabel → {x_1, x_2},
    PlotRange → All];
  Print["a=", a];
 ]
]
```

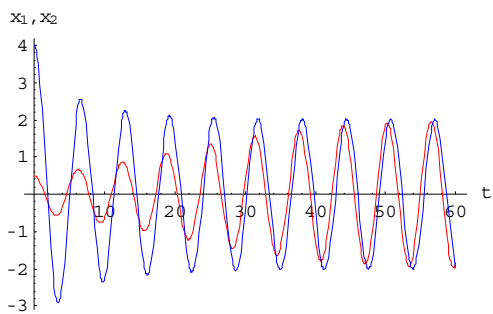

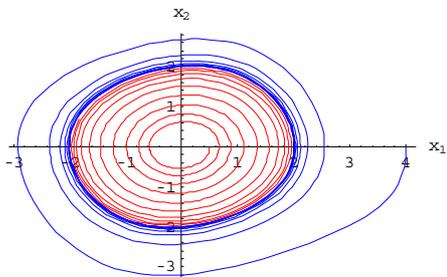

a= 0.1

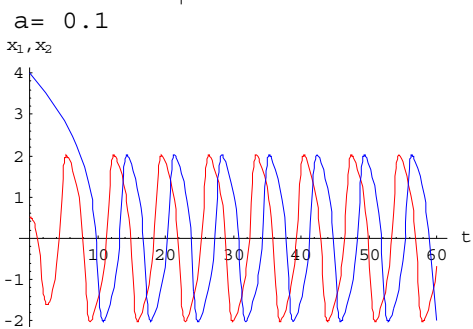

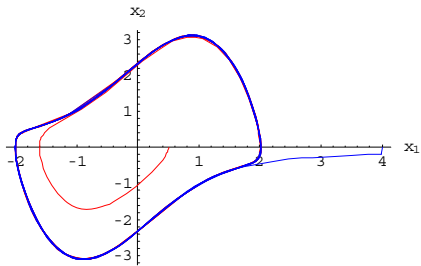

a= 1.4

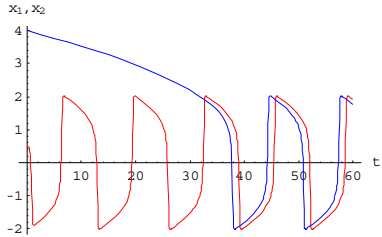

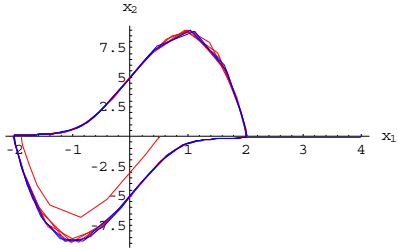

a= 6.

```
Clear["Global`*"]
<< Graphics`PlotField`
f1 = y;
f2 = -x - a (x^2 - 1) y;

a = 0.4;
vfield = PlotVectorField[{f1, f2}, {x, -3, 3}, {y, -3, 3}, DisplayFunction → Identity];
f[a_, x0_, tmax_] := NDSolve[{x'[t] == y[t], y'[t] == -x[t] - a (x[t]^2 - 1) y[t], x[0] == x0, y[0] == 0},
    {x, y}, {t, 0, tmax}, MaxSteps → Infinity, PrecisionGoal → 12, AccuracyGoal → 12];
s1 = f[a, 0.5, 200];
s2 = f[a, 4, 200];
Plot[{x[t] /. s1, x[t] /. s2}, {t, 0, 60},
 PlotStyle → {{Dashing[{0.04, 0.02}], Thickness[0.0065]},
    {Dashing[{0.04, 0.006}], Thickness[0.005]}}, AxesLabel → {t, "x_1,x_2"}]
PhaseSpace = ParametricPlot[{{x[t], y[t]} /. s1, {x[t], y[t]} /. s2}, {t, 0, 60},
    PlotStyle → {Thickness[0.007], {RGBColor[1, 0, 0]}}, AxesLabel → {x_1, x_2}, PlotRange → All];
Show[PhaseSpace, vfield, PlotRange → {{-3, 3}, {-3, 3}}, Axes → Automatic,
  AxesLabel → {"x", "y"}, DisplayFunction → $DisplayFunction];
```

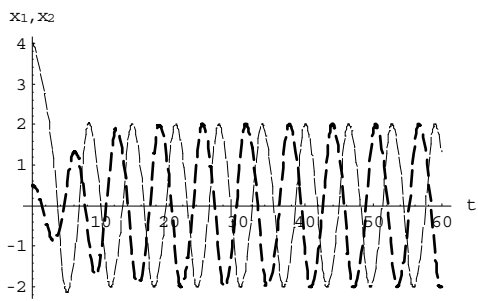

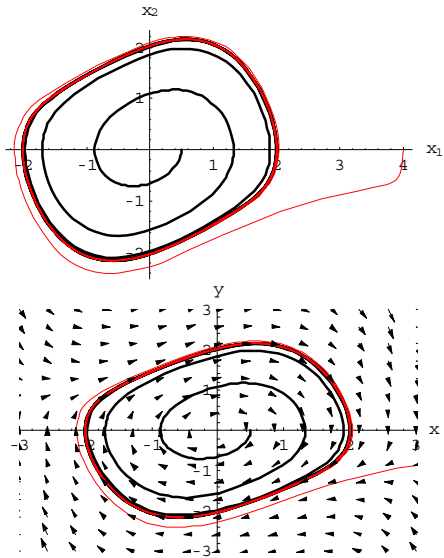

## Relaxation Oscillations

```
Clear["Global`*"]
 <<Graphics`PlotField`
 f1 = a (y - (x^3/3 - x));
 f2 = -x/a;

a = 8;
vfield = PlotVectorField[{f1, f2}, {x, -3.5, 3.5}, {y, -1.5, 1.5}, DisplayFunction → Identity];
f[a_, x0_, tmax_] :=
    NDSolve[{x'[t] == a*(y[t] - (x[t]^3/3 - x[t])), y'[t] == -x[t]/a, x[0] == x0, y[0] == 0},
    {x, y}, {t, 0, tmax}];
s1 = f[a, 0.5, 60];
s2 = f[a, 4, 60];
plot3 = Plot[y = (x^3/3) - x, {x, -4, 4}, PlotStyle → {RGBColor[0, 0, 1]},
    PlotRange → {{-2, 4}, {-4.5, 4.5}}];
Plot[{x[t] /. s1, x[t] /. s2}, {t, 0, 60},
 PlotStyle → {{Dashing[{0.04, 0.02}], Thickness[0.0065]},
    {Dashing[{0.04, 0.006}], Thickness[0.005]}}, AxesLabel → {t, "x_1,x_2"}]
PhaseSpace = ParametricPlot[{{x[t], y[t]} /. s1, {x[t], y[t]} /. s2}, {t, 0, 60},
    PlotStyle → {Thickness[0.007], {RGBColor[1, 0, 0]}}, AxesLabel → {x_1, x_2}, PlotRange → All];
Show[PhaseSpace, vfield, plot3, PlotRange → {{-3, 3}, {-1.3, 1.3}}, Axes → Automatic,
  AxesLabel → {"x", "y"}, DisplayFunction → $DisplayFunction];
```

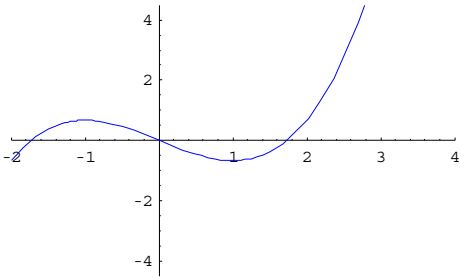
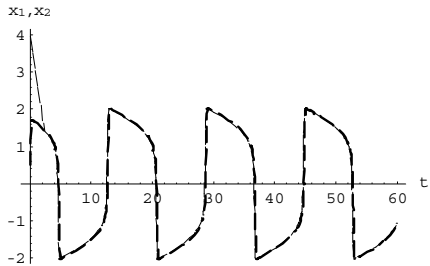
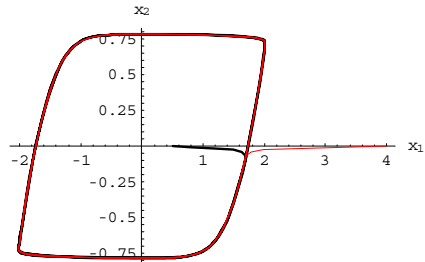
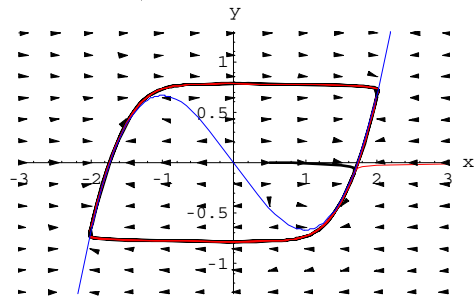

```
Clear[d, ρ, t1];
f = (3 - d) e^(-ρ (t-t1)) - d Cos[t];
f1 = 1;
d = 0.4;
ρ = .1;
t1 = 2.43;
tmin = -10;
tmax = 25;
Plot[{f, f1}, {t, tmin, tmax},
 PlotStyle → {{RGBColor[1, 0, 0]}, {RGBColor[0, 0, 1]},}]
f2 = f /. t → 0
```

# Non Autonomous Equations
# Levinson's Solutions

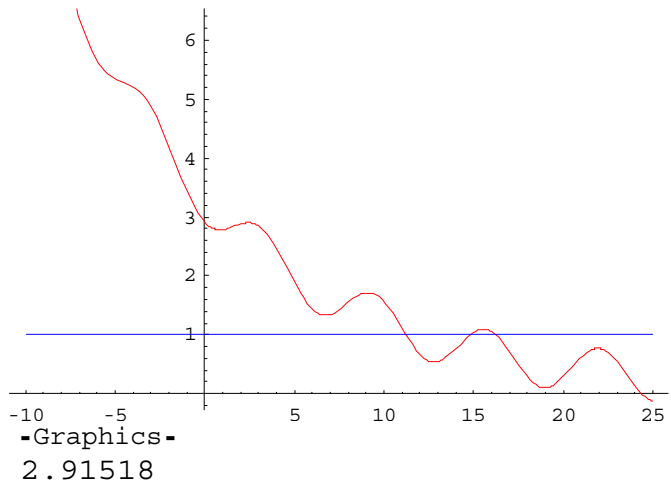

-Graphics-
2.91518

## Non Averaged Solutions

```
Z1[t] = e / w (b x1[t] + (1 - x1[t]^2) x2[t] + p Cos[w t + a]) Cos[w t]
```

$$\frac{e \cos[t w] \; (p \cos[a + t w] + b \, x1[t] + (1 - x1[t]^2) \, x2[t])}{w}$$

```
Z2[t] = -e / w (b x1[t] + (1 - x1[t]^2) x2[t] + p Cos[w t + a]) Sin[w t]
```

$$-\frac{e \sin[t w] \; (p \cos[a + t w] + b \, x1[t] + (1 - x1[t]^2) \, x2[t])}{w}$$

```
Z11[t]=Z1[t]/.{x1[t]→z1[t] Sin[w t]+z2[t] Cos[w
t],x2[t]→w (z1[t] Cos[w t]-z2[t] Sin[w
t]),w→1,e→0.1,p→0.2,a→0,b→0.3}
```

0.1 Cos[t] (0.2 Cos[t] + 0.3 (Sin[t w] z1[t] + Cos[t w] z2[t]) +
   w (Cos[t w] z1[t] − Sin[t w] z2[t]) (1 − (Sin[t w] z1[t] + Cos[t w] z2[t])$^2$))

```
Z22[t]=Z2[t]/.{x1[t]→z1[t] Sin[w t]+z2[t] Cos[w
t],x2[t]→w (z1[t] Cos[w t]-z2[t] Sin[w
t]),w→1,e→0.1,p→0.2,a→0,b→0.3}
```

−0.1 Sin[t] (0.2 Cos[t] + 0.3 (Sin[t w] z1[t] + Cos[t w] z2[t]) +
   w (Cos[t w] z1[t] − Sin[t w] z2[t]) (1 − (Sin[t w] z1[t] + Cos[t w] z2[t])$^2$))

```
w=10;
dsol=NDSolve[{z1'[t]==Z11[t],z2'[t]==Z22[t],z1[0]==1,z2[0]
==1},{z1,z2},{t,-
10,500},MaxSteps→Infinity,AccuracyGoal→12,PrecisionGoal→1
2]

z1t=z1[t]/.dsol〚1〛;
```

```
z2t=z2[t]/.dsol[[1]];

Plot[{z1t,z2t},{t,0,5}];
Plot[{z1t,z2t},{t,0,20}];
Plot[{z1t,z2t},{t,0,300}];
ParametricPlot[{z1t,z2t},{t,0,400},PlotRange→{{-4,4},{-4,4}},PlotPoints→2000,Frame→True,FrameLabel→{"z1","z2"}];
ParametricPlot[{z1t,z2t},{t,0,400},PlotRange→{{1,2},{-1,1}},PlotPoints→2000,Frame→True,FrameLabel→{"z1","z2"},PlotLabel->"ZooM"];
ParametricPlot[{z1t,z2t},{t,0,400},PlotRange→{{1.7,2},{-.4,.4}},PlotPoints→2000,Frame→True,FrameLabel→{"z1","z2"},PlotLabel->"ZoooM"]
```

NDSolve::ndsz : At t == 219.6497245698631`, step
  size is effectively zero; singularity or stiff system suspected. More…

{{z1→InterpolatingFunction[{{-10.,219.65}},<>],z2→InterpolatingFunction[{{-10.,219.65}},<>]}}

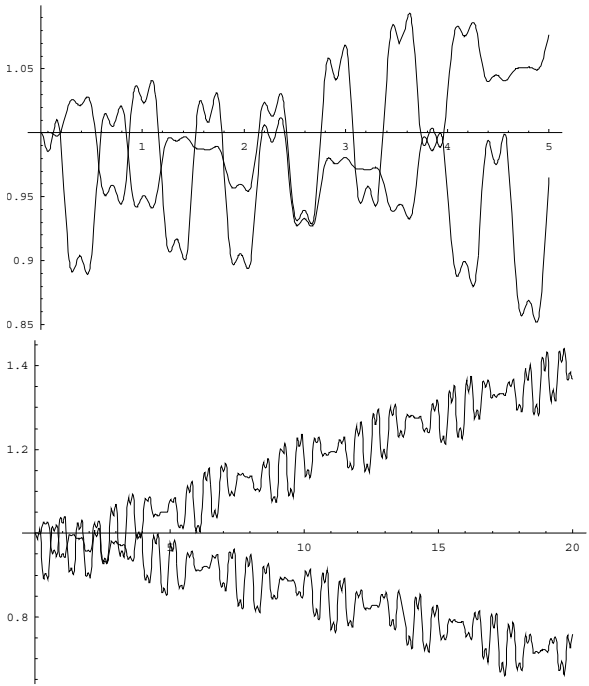

InterpolatingFunction::dmval : Input value {225.919} lies outside the
  range of data in the interpolating function. Extrapolation will be used. More…

InterpolatingFunction::dmval : Input value {222.758} lies outside the
  range of data in the interpolating function. Extrapolation will be used. More…

InterpolatingFunction::dmval : Input value {221.01} lies outside the
  range of data in the interpolating function. Extrapolation will be used. More…

General::stop :
 Further output of InterpolatingFunction::dmval will be suppressed during this calculation. More…

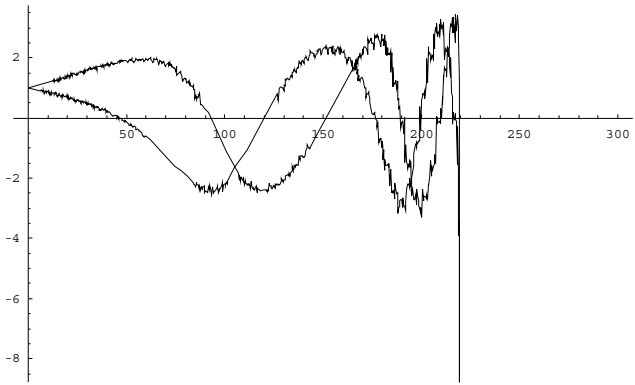

InterpolatingFunction::dmval : Input value {219.755} lies outside the
    range of data in the interpolating function. Extrapolation will be used. More…

InterpolatingFunction::dmval : Input value {219.755} lies outside the
    range of data in the interpolating function. Extrapolation will be used. More…

InterpolatingFunction::dmval : Input value {219.662} lies outside the
    range of data in the interpolating function. Extrapolation will be used. More…

General::stop :
 Further output of InterpolatingFunction::dmval will be suppressed during this calculation. More…

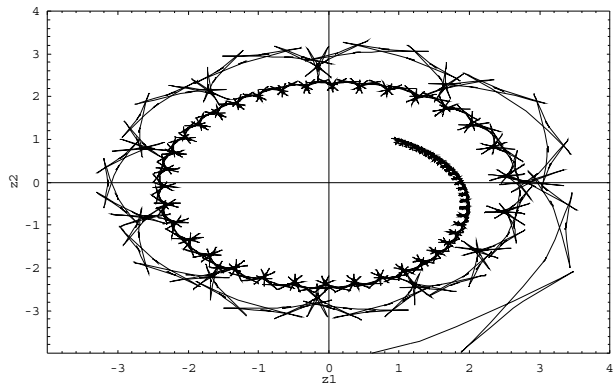

InterpolatingFunction::dmval : Input value {219.755} lies outside the
    range of data in the interpolating function. Extrapolation will be used. More…

InterpolatingFunction::dmval : Input value {219.755} lies outside the
    range of data in the interpolating function. Extrapolation will be used. More…

InterpolatingFunction::dmval : Input value {219.662} lies outside the
    range of data in the interpolating function. Extrapolation will be used. More…

General::stop :
 Further output of InterpolatingFunction::dmval will be suppressed during this calculation. More…

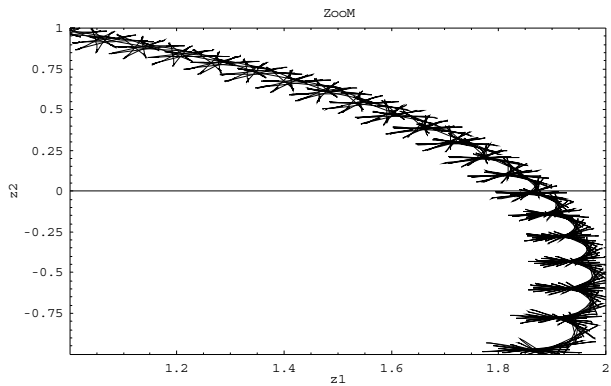

```
InterpolatingFunction::dmval : Input value {219.755} lies outside the
    range of data in the interpolating function. Extrapolation will be used. More…

InterpolatingFunction::dmval : Input value {219.755} lies outside the
    range of data in the interpolating function. Extrapolation will be used. More…

InterpolatingFunction::dmval : Input value {219.662} lies outside the
    range of data in the interpolating function. Extrapolation will be used. More…

General::stop :
 Further output of InterpolatingFunction::dmval will be suppressed during this calculation. More…
```

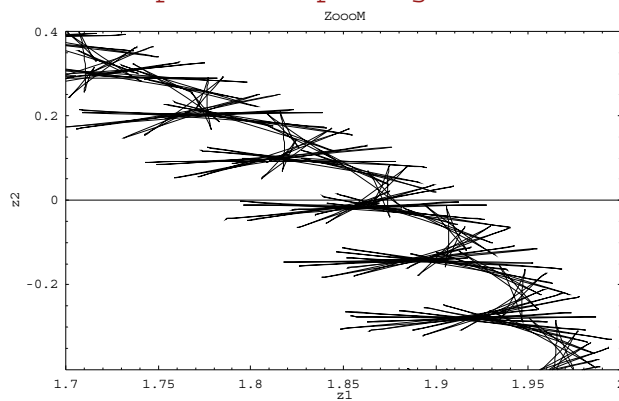

-Graphics-

```
ClearAll["Global`*"];
sol = Solve[r^3 - 4 r - ((4 b) / (a (1 + a σ))) == 0, r];
sol1 = sol[[1]] // Simplify;
r1 = r /. sol1;
r11 = r1 /. {a → 1, b → 0.5};
r11 = r1 /. {a → 1, b → 0.5};
Plot[{Abs[r11], r = 2}, {σ, -3, 1}, PlotStyle → {{RGBColor[0, 0, 0]}, {RGBColor[1, 0, 0]}},
 PlotRange → {{-3, 1}, {0.5, 5}}, AxesOrigin → {-1, 1}, AxesLabel → {"σ", "r"}]
```

```
ClearAll::clloc : Cannot clear local variable σ. More…
```

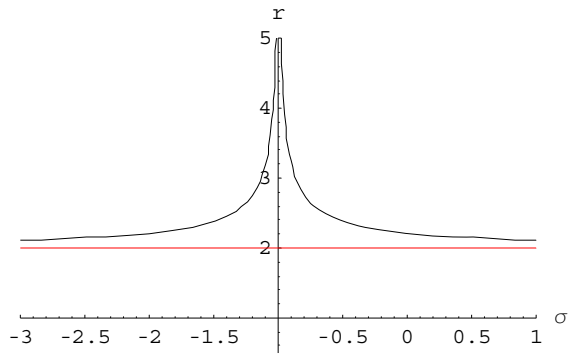
-Graphics-

```
ClearAll["Global`*"]
a=5;
b=15;
ω=7;
T=2π/ω
tmax=5000 T;
tmin=tmax-22 T;
dsol=NDSolve[{x'[t]==y[t],y'[t]==-x[t]-a*(x[t]^2-1)*y[t]+b
Cos[θ[t]],θ'[t]==ω,x[0]==0,y[0]==0,θ[0]==0},{x,y,θ},{t,0,tma
x},MaxSteps→Infinity,AccuracyGoal→12,PrecisionGoal→12];

xt=x[t]/.dsol[[1]];

yt=y[t]/.dsol[[1]];

θt=θ[t]/.dsol[[1]];

ParametricPlot3D[{xt,yt,θt/ω},{t,tmin,tmax},AxesLabel→{x,
y,t},PlotRange→{{-6,6},{-8,8},{tmin,tmax}},
PlotPoints→40000];
```
$$\frac{2\pi}{7}$$

## Phase Portraits & Poincarè Maps

General::spell1 : Possible spelling error: new symbol name "θt" is similar to existing symbol "θ". More…

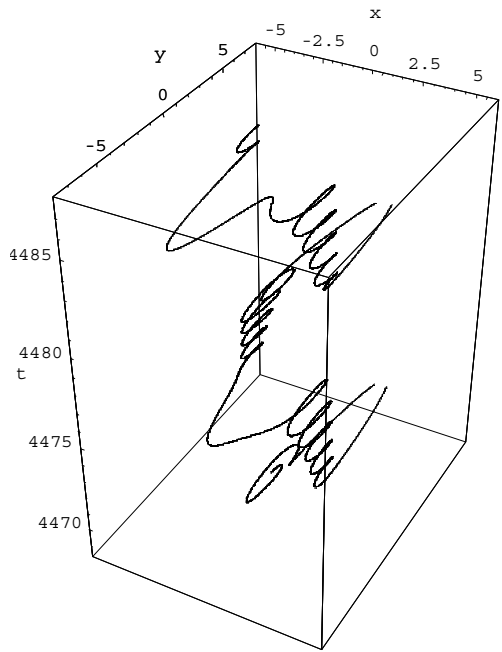

(* Van der Pol $\ddot{x} = -x - a(x^2-1)\dot{x} + b\cos t$

$\dot{x} = y - a\left(\frac{x^3}{3} - x\right)$

$\dot{y} = -x + b\cos t$

$\dot{\theta} = 1$ *)

```
ClearAll["Global`*"]
(*System Parameters*)
tmax = 10000 T; tmin = tmax - 100 T;
a = 5;
b = 50;
ω = 7;
T = 2 π / ω;
(*Initial Conditions *)
x0 = 0;
y0 = 0;
initial = {x[0] == x0, y[0] == y0};
(*System Equation*)
f1 = x'[t] - y[t];
f2 = y'[t] + x[t] + a (x[t]^2 - 1) y[t] - b Cos[ω t];
dsol = NDSolve[{f1 == 0, f2 == 0, x[0] == x0, y[0] == y0}, {x, y}, {t, 0, tmax}, MaxSteps → Infinity,
   AccuracyGoal → 12, PrecisionGoal → 12];
xt = x[t] /. dsol[[1]];
yt = y[t] /. dsol[[1]];
ParametricPlot[{xt, yt}, {t, tmin, tmax}, PlotRange → {{-3, 3}, {-10, 10}}, PlotPoints → 2000,
 Frame → True, FrameLabel → {"x", "y"}, PlotLabel → {"a=", a, "  b=", b}]
Print["a/b=", a / b]
```

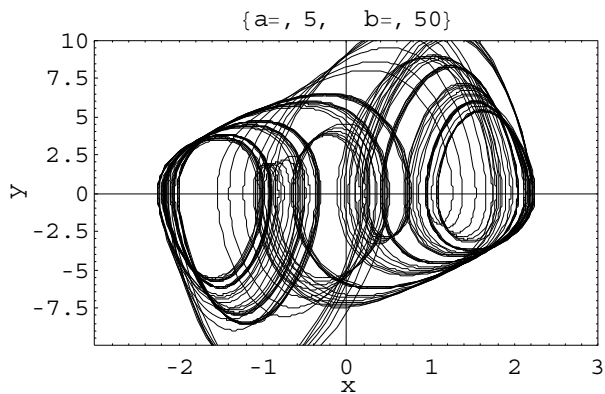

$a/b = \dfrac{1}{10}$

```
ClearAll["Global`*"];
(*System Parameters*)
a = 5;
b = 50;
ω = 7;
T = 2 π / ω;
tmax = 15000 T; tmin = tmax - 100 T;
(*αλλαξτε τον χρονο tmax σε μικροτερο για να τρεξει πιο γρηγορα *)
(*Initial Conditions*)
x0 = 0.5;
y0 = 0;
initial = {x[0] == x0, y[0] == y0};
(*Van der Pol Equation*)
sysdeq = {f1 = x'[t] - y[t], f2 = y'[t] + x[t] + a (x[t]^2 - 1) y[t] - b Cos[ω t]};
(*Solution Trajectory*)
dsol = NDSolve[{f1 == 0, f2 == 0, x[0] == x0, y[0] == y0}, {x, y}, {t, 0, tmax}, MaxSteps → Infinity];
xt = x[t] /. dsol[[1]];
yt = y[t] /. dsol[[1]];
dsoll = NDSolve[{f1 == 0, f2 == 0, x[0] == x0 + 0.01, y[0] == y0 + 0.01}, {x, y}, {t, 0, tmax},
    MaxSteps → Infinity, AccuracyGoal → 12, PrecisionGoal → 12];
xtt = x[t] /. dsoll[[1]];
(* Poincare Section *)
data = {};
For[t = 0, t ≤ tmax, t += T,
 AppendTo[data, {N[xt], N[yt]}]
]
ListPlot[data, PlotRange → {{-3, 3}, {-10, 10}}, Frame → True, Axes → False,
  PlotStyle → {PointSize[0.01]}, PlotLabel → {"a=", a, " b=", b , "ω=", ω}];
ListPlot[data, PlotRange → {{-0.6, 0}, {1.5, 3}}, Frame → True, Axes → False,
  PlotStyle → {PointSize[0.012]}, PlotLabel → {"enlargement" }];
Print["a/b=", a / b];
Plot[xt, {t, tmin + 50 T, tmax}, AxesLabel → {"t", "x(t)"}];
Plot[xtt, {t, tmin + 50 T, tmax}, AxesLabel → {"t", "x(t)"}];
```

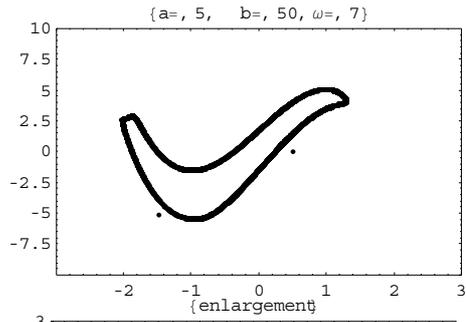

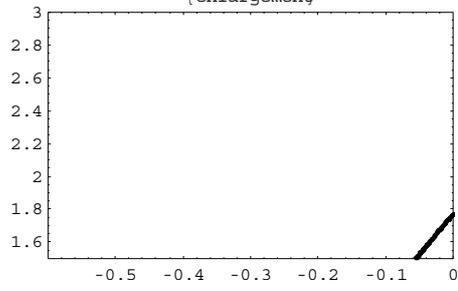

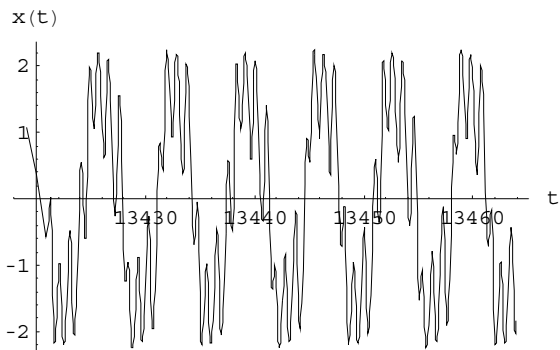

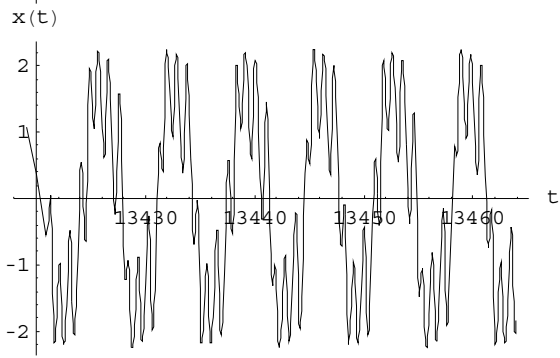

# BIFURCATIONS

```
ClearAll["Global`*"];
(*System Parameters*)
tmax = 10000 T; tmin = tmax - 50 T;
(*Initial Conditions *)
x0 = 0;
y0 = 0;
b = 5;
ω = 7;
T = 2 π / ω;
initial = {x[0] == x0, y[0] == y0};
(*System Equation*)
f1 = x'[t] - y[t];
f2 = y'[t] + x[t] + a (x[t]^2 - 1) y[t] - b Cos[ ω t];
data = {};
 For[
  a=0.01,a≤80,a+=0.1,

dsol=NDSolve[{f1==0,f2==0,x[0]==x0,y[0]==y0},{x,y},{t,0,tmax
},MaxSteps→Infinity,AccuracyGoal→12,PrecisionGoal→12];
  xt=x[t]/.dsol[[1]];
  yt=y[t]/.dsol[[1]];
  For[t=tmin,t≤tmax,t+=T,
    AppendTo[data,{a,N[xt]}]
    ];
  Clear[t]
  ]
ListPlot[data,PlotRange→{{-0.5,81},{-
2.5,2.5}},Frame→True,
Axes→False,PlotStyle→{PointSize[0.006]},PlotLabel→{"
bifurcation diagram ","b=",b,"ω=",ω}];
```

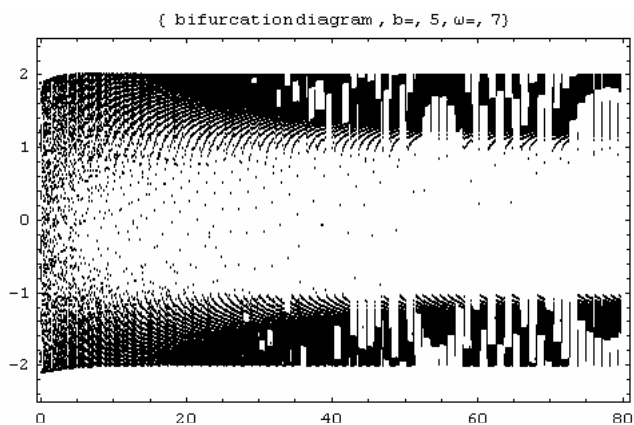

```
ClearAll["Global`*"];
(*System Parameters*)
tmax = 10000 T; tmin = tmax - 50 T;
(*Initial Conditions *)
x0 = 0;
y0 = 0;
a = 5;
ω = 7;
T = 2 π / ω;
initial = {x[0] == x0, y[0] == y0};
(*System Equation*)
f1 = x'[t] - y[t];
f2 = y'[t] + x[t] + a (x[t]^2 - 1) y[t] - b Cos[ ω t];
data = {};
 For[
  b=0.01,b≤80,b+=0.1,
  dsol=NDSolve[{f1==0,f2==0,x[0]==x0,y[0]==y0},{x,y},{t,0,tmax},MaxSteps→Infinity,AccuracyGoal→12,PrecisionGoal→12];
  xt=x[t]/.dsol[[1]];
  yt=y[t]/.dsol[[1]];
  For[t=tmin,t≤tmax,t+=T,
    AppendTo[data,{b,N[xt]}]
    ];
  Clear[t]
  ]
ListPlot[data,PlotRange→{{-0.5,81},{-2.5,2.5}},Frame→True,Axes→False,PlotStyle→{PointSize[0.006]},PlotLabel→{" bifurcation diagram "}];
```

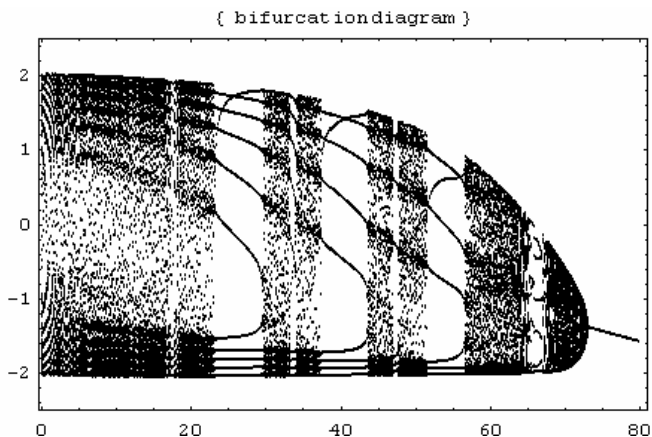

```
ClearAll["Global`*"];
Off[General::spell1];
$HistoryLength=1;
```

```
link=Install["math2.exe"];
 (*System Parameters*)
 tmax = 10000 T;
  tmin = tmax - 50 T;
  (*Initial Conditions *)
  x0 = 0.;
  y0 = 0.;
  a = 5.;
  ω = 7.;
  T = 2 π / ω;
  Nr = 10^6;
  initial = {x[0] == x0, y[0] == y0};
  (*System Equation*)
  f1 = x'[t] - y[t];
  f2 = y'[t] + x[t] + a (x[t]^2 - 1) y[t] - b Cos[ ω t];
  data = {};
Null^12
 For[
  b = 0., b ≤ 80, b += 0.1,
  dsol = SolveVDP[a, b, ω, x0, y0, Nr, N[tmax/Nr]];
  For[t = tmin, t ≤ tmax, t += T,
   AppendTo[data, {b, dsol[[Floor[ t/tmax (Nr - 1) + 1]]]}]
  ];
  Clear[t]
 ]
 Length[data]
 40851
 data2=Table[{0, 0}, {i, 1, Length[data]}];
For[i=1,i≤Length[data],i++,

  data2[[i,1]]=data[[i, 1]];

   data2[[i, 2]]=data[[i, 2]]

  ]
 ListPlot[data2,PlotRange→{{-0.5,81},{-
2.5,2.5}},Frame→True,
Axes→False,PlotStyle→{PointSize[0.003]},PlotLabel→{"
bifurcation diagram "}];
```

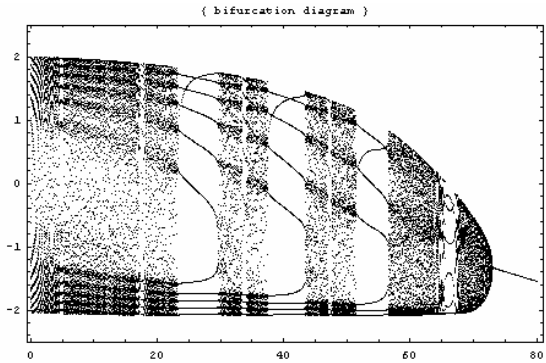

```
data3=Table[Table[data2[[51(i-1)+j,2]], {j, 1, 51}], {i, 1,

Length[data2]/51}];

data4 = {};
For[i = 1, i ≤ 801, i++,
 list = {};
 For[j = 1, j ≤ 51, j++,
  If[Count[list, _?(((data3[[i, j]] < (#+ 10^-3)) && (data3[[i, j]] > (#- 10^-3))) &)] > 0, , AppendTo[list, data3[[i, j]]]];
 ];
 AppendTo[data4, {i/10, Length[list]}];
]
 ListPlot[data4, PlotJoined→True];
```

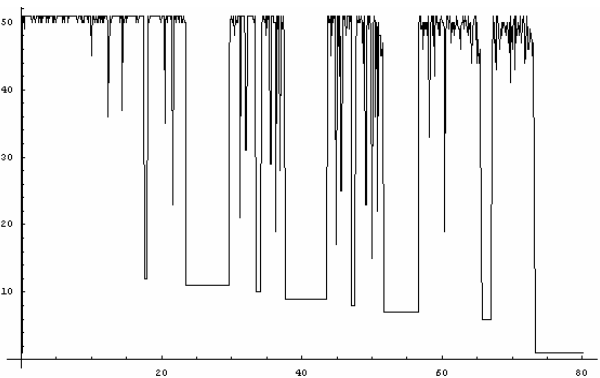

```
Uninstall[link];
 SessionTime[]
 1695.7343750
```

**Bifurcation with respect to ω. Notice the transformation of the system equations!**

```
ClearAll["Global`*"]
tmax = 5000; tmin = tmax - 50;
(*Initial Conditions *)
x0 = 0
y0 = 0;
a = 3;
b = 5;
initial = {x[0] == x0, y[0] == y0};
(*System Equation*)
f1 = ω/(2 π) x'[t] - y[t];
f2 = ω/(2 π) y'[t] + x[t] + a (x[t]^2 - 1) y[t] - b Cos[2 π t];
data = {};
 0
 For[
  ω=1,ω≤7,ω+=0.006,
  tmax=5000;tmin=tmax-50;

dsol=NDSolve[{f1==0,f2==0,x[0]==x0,y[0]==y0},{x,y},{t,0,tmax
},MaxSteps→Infinity];
  xt=x[t]/.dsol[[1]];
  For[t=tmin,t≤tmax,t+=1,
    AppendTo[data,{ω,N[xt]}]
    ];
  Clear[t]
  ]
ListPlot[data,PlotRange→{{-0.5,7.5},{-3,3}},Frame→True,
Axes→False,PlotStyle→{PointSize[0.006]},PlotLabel→{"
bifurcation diagram "}];
```

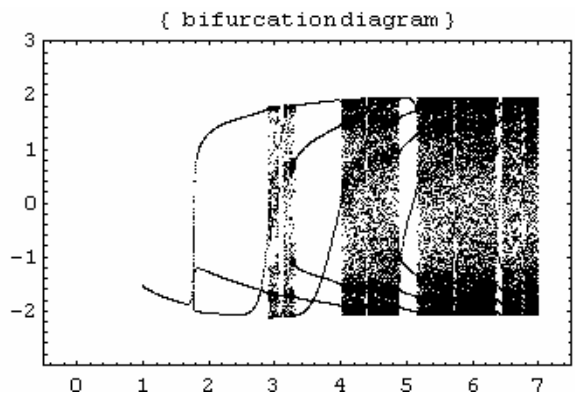

## Period Doubling

(* Van der Pol $\ddot{x} = -x - a(x^2-1)\dot{x} + b\cos t$

$\dot{x} = y - a\left(\frac{x^3}{3} - x\right)$

$\dot{y} = -x + b\cos t$

$\dot{\theta} = 1$ *)

```
ClearAll["Global`*"];
(*System Parameters*)
tmax = 10000 T; tmin = tmax - 100 T;
a = 5;
b = 5;
ω = ω;
T = 2 π / ω;
(*Initial Conditions *)
x0 = 0;
y0 = 0;
initial = {x[0] == x0, y[0] == y0};
(*System Equation*)
f1 = x'[t] - y[t];
f2 = y'[t] + x[t] + a (x[t]^2 - 1) y[t] - b Cos[ω t];
For[
 ω = 2.457, ω ≤ 2.6, ω += 0.003,
 dsol = NDSolve[{f1 == 0, f2 == 0, x[0] == x0, y[0] == y0}, {x, y}, {t, 0, tmax}, MaxSteps → Infinity];
 xt = x[t] /. dsol[[1]];
 yt = y[t] /. dsol[[1]];
 ParametricPlot[{xt, yt}, {t, tmin, tmax}, PlotRange → {{-3, 3}, {-10, 10}},
   PlotPoints → 2000, Frame → True, FrameLabel → {"x", "y"}, PlotLabel → {"a=", a, "  b=", b}]
  Print["a/b=", a / b, "ω=", ω]
]
```

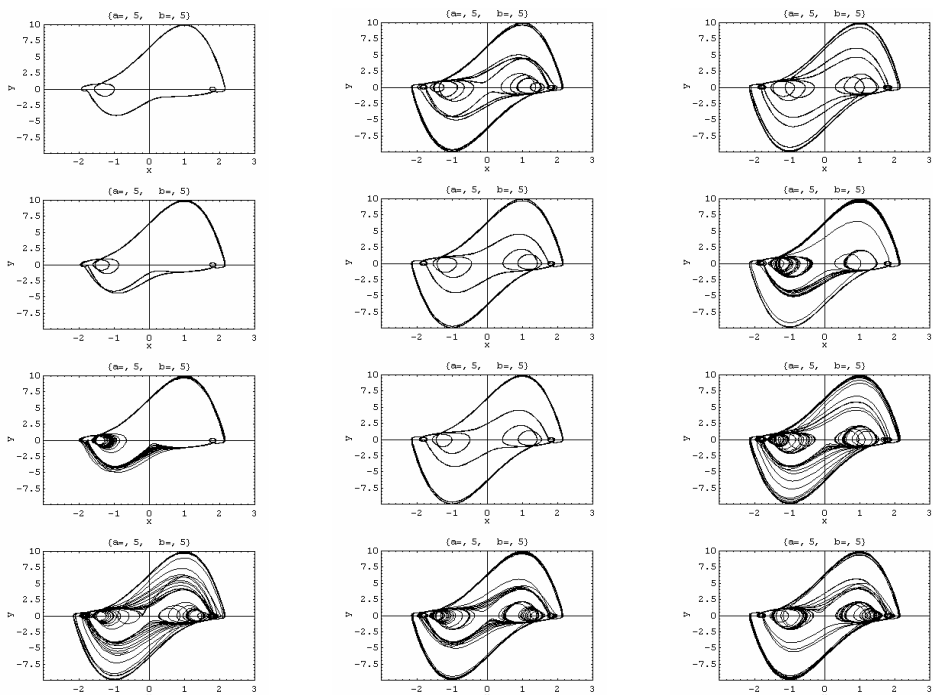

(* Van der Pol $\ddot{x}= -x-a\,(x^2-1)\,\dot{x}+b\cos t$

  $\dot{x}=y-a\left(\frac{x^3}{3}-x\right)$

  $\dot{y}=-x+b\cos t$

  $\dot{\theta}=1$ *)

```
ClearAll["Global`*"];
(*System Parameters*)
tmax = 10000 T; tmin = tmax - 100 T;
a = 3;
b = 5;
ω = 1.788;
T = 2 π / ω;
(*Initial Conditions *)
x0 = 0;
y0 = 0;
initial = {x[0] == x0, y[0] == y0};
(*System Equation*)
f1 = x'[t] - y[t];
f2 = y'[t] + x[t] + a (x[t]^2 - 1) y[t] - b Cos[ω t];
dsol = NDSolve[{f1 == 0, f2 == 0, x[0] == x0, y[0] == y0}, {x, y}, {t, 0, tmax}, MaxSteps → Infinity,
   AccuracyGoal → 12, PrecisionGoal → 12];
xt = x[t] /. dsol[[1]];
yt = y[t] /. dsol[[1]];
ParametricPlot[{xt, yt}, {t, tmin, tmax}, PlotRange → {{-3, 3}, {-10, 10}}, PlotPoints → 2000,
 Frame → True, FrameLabel → {"x", "y"}, PlotLabel → {"a=", a, "  b=", b}]
Print["a/b=", a / b]
```

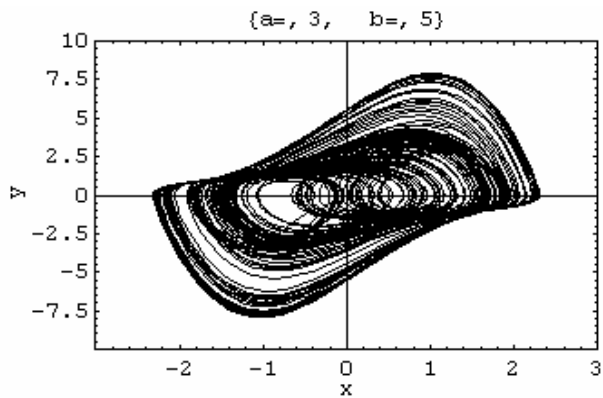

**-Graphics-**

a/b= $\frac{3}{5}$

```mathematica
ClearAll["Global`*"];
(*System Parameters*)
a = 5;
b = 25;
ω = 4.455;
T = 2 π / ω;
tmax = 15000 T; tmin = tmax - 100 T;
(*αλλαξτε τον χρονο tmax σε μικροτερο για να τρεξει πιο γρηγορα *)
(*Initial Conditions*)
x0 = 0.5;
y0 = 0;
initial = {x[0] == x0, y[0] == y0};
(*Van der Pol Equation*)
sysdeq = {f1 = x'[t] - y[t], f2 = y'[t] + x[t] + a (x[t]^2 - 1) y[t] - b Cos[ω t]};
(*Solution Trajectory*)
dsol = NDSolve[{f1 == 0, f2 == 0, x[0] == x0, y[0] == y0}, {x, y}, {t, 0, tmax}, MaxSteps → Infinity,
   AccuracyGoal → 12, PrecisionGoal → 12];
xt = x[t] /. dsol[[1]];
yt = y[t] /. dsol[[1]];
dsoll = NDSolve[{f1 == 0, f2 == 0, x[0] == x0 + 0.01, y[0] == y0 + 0.01}, {x, y}, {t, 0, tmax},
   MaxSteps → Infinity, AccuracyGoal → 12, PrecisionGoal → 12];
xtt = x[t] /. dsoll[[1]];
(* Poincare Section *)
data = {};
For[t = 0, t ≤ tmax, t += T,
 AppendTo[data, {N[xt], N[yt]}]
]
ListPlot[data, PlotRange → {{-2.5, 1.5}, {-5.3, 4.3}}, Frame → True, Axes → False,
  PlotStyle → {PointSize[0.0066]}, PlotLabel → {"a=", a, " b=", b , "ω=", ω}];
ListPlot[data, PlotRange → {{-2, -1.4}, {-3, -0.5}}, Frame → True, Axes → False,
  PlotStyle → {PointSize[0.0022]}, PlotLabel → {"enlargement" }];
ListPlot[data, PlotRange → {{-1.71, -1.56}, {-2.87, -2.15}}, Frame → True, Axes → False,
  PlotStyle → {PointSize[0.0025]}, PlotLabel → {"enlargement"}];
Print["a/b=", a / b];
Plot[xt, {t, tmin + 75 T, tmax}, AxesLabel → {"t", "x(t)"}];
Plot[xtt, {t, tmin + 75 T, tmax}, AxesLabel → {"t", "x(t)"}];
```

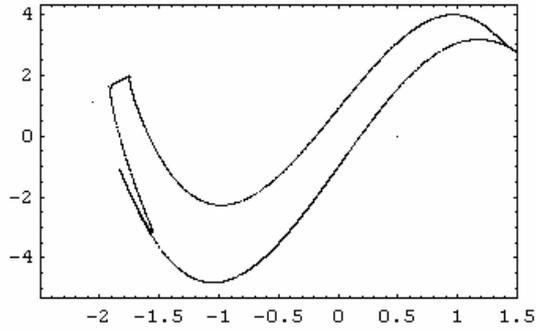
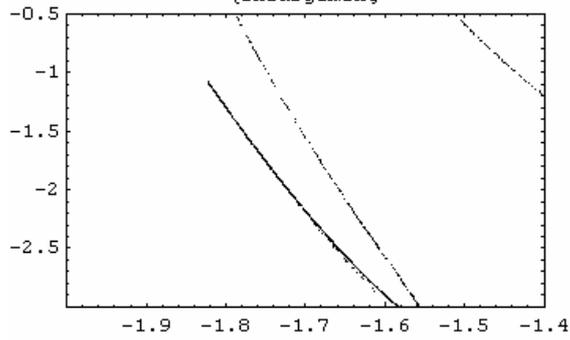
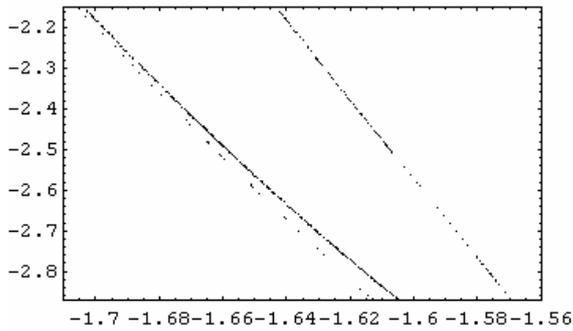
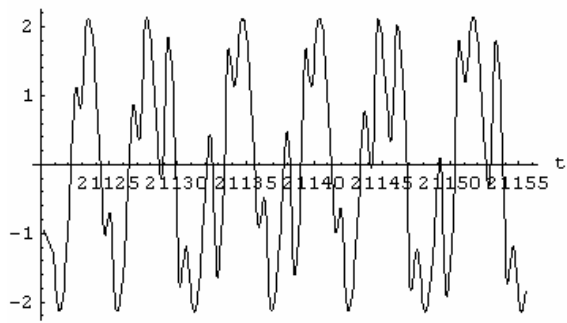
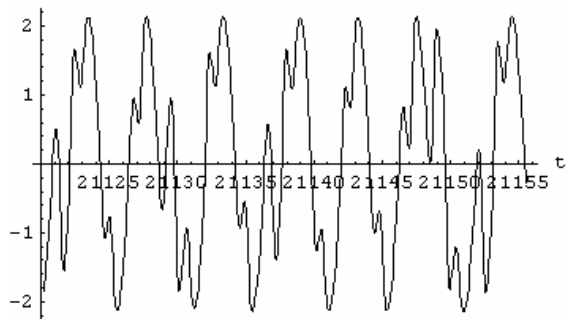

## Fourier Plots

```mathematica
ClearAll["Global`*"];
 (*System Parameters*)
 tmax = 10000 T; tmin = tmax - 1000 T;
  (*Initial Conditions *)
  x0 = 0;
  y0 = 0;
  a = 5;
  ω = 7;
  T = 2 π / ω;
  initial = {x[0] == x0, y[0] == y0};
 (*System Equation*)
  f1 = x'[t] - y[t];
  f2 = y'[t] + x[t] + a (x[t]^2 - 1) y[t] - b Cos[ω t];
  data = {};
 Null^10
 b=15;
dsol=NDSolve[{f1==0, f2==0,initial}, {x, y}, {t, 0, tmax},MaxSteps→∞];

minp = 0.5;
pointsperperiod = 20;
data = {};
For[t1 = tmin, t1 ≤ tmax, t1 += T / pointsperperiod,
  AppendTo[data, N[x[t1] /. dsol]]
 ];
list1 = Flatten[Abs[Fourier[data]]];
list2 = {};
For[i = 1, i ≤ Length[list1] / 2, i++, AppendTo[list2, {N[i/(tmax - tmin)], Abs[list1[[i]]]}]]
list3 := {};
templist = Ordering[list1[[Range[1, Floor[Length[list1] / 2]]]], All, Greater];
tempmax = Max[list1];
For[i = 1, i < Floor[Length[list1] / 2], i++,
  (tk = templist[[i]]; If[(list1[[tk]] ≥ list1[[tk - 1]]) && (list1[[tk]] ≥ list1[[tk + 1]]) && (list1[[tk]] ≥ (minp * tempmax)/100),
    AppendTo[list3, tk];)];
ListPlot[list2, PlotJoined→True, PlotRange→{{0, 1.1 Max[list3]/(tmax - tmin)}, {0, 1.1 tempmax}}];
Print["Οι κύριες συχνότητες (ένταση > ", minp, "% της μέγιστης) και οι εντάσεις τους για b = ", b, " είναι"]
For[i = 1, i ≤ Length[list3], i++, Print[N[list3[[i]]/(tmax - tmin)], ":", N[100 * list1[[list3[[i]]]]/tempmax, 1], "%"]]
```

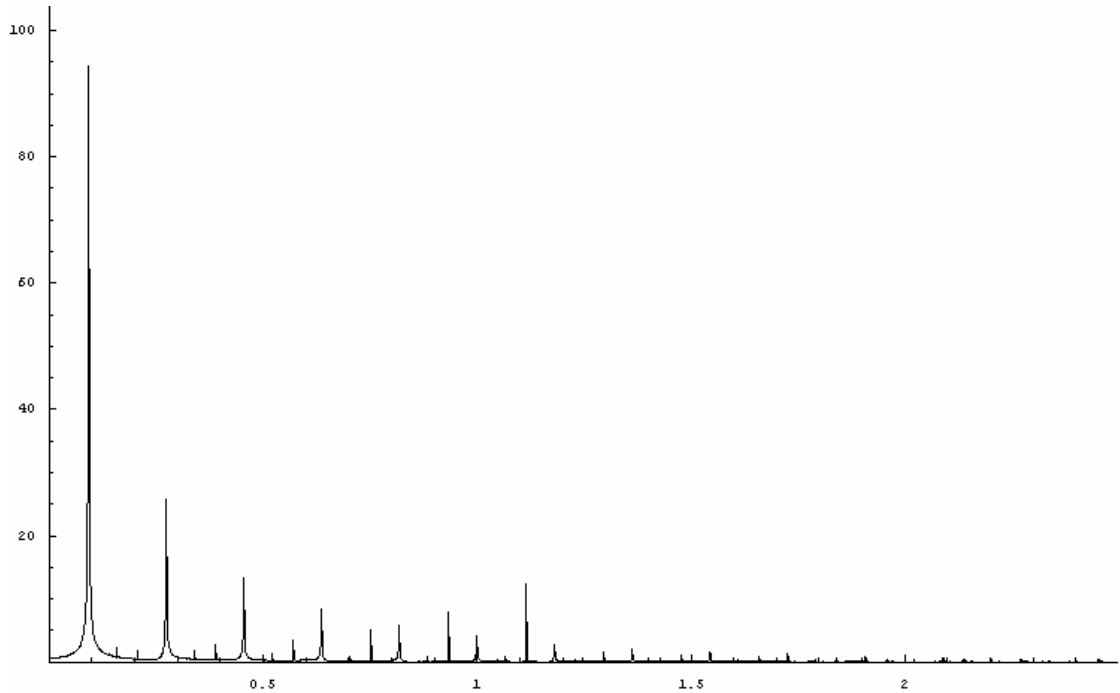

Οι κύριες συχνότητες (ένταση > 0.5 % της μέγιστης) και οι εντάσεις τους για b = 15 είναι
 0.0913549 : 100. %
 0.272951 : 27.185 %
 0.454547 : 14.2245 %
 1.1152 : 13.041 %
 0.636142 : 8.9755 %
 0.933603 : 8.4125 %
 0.817738 : 6.15773 %
 0.752007 : 5.3783 %
 0.999334 : 4.42031 %
 0.570411 : 3.73649 %
 1.18093 : 3.02537 %
 0.388816 : 2.81111 %
 0.1582 : 2.54176 %
 1.36253 : 2.23885 %
 0.20722 : 2.04607 %
 0.339796 : 1.96433 %
 1.29679 : 1.73367 %
 1.54412 : 1.7287 %
 0.521392 : 1.43474 %
 0.141489 : 1.40349 %
 1.72572 : 1.34984 %
 0.0434493 : 1.18262 %
 1.47839 : 1.09139 %
 0.702987 : 1.03415 %
 0.0412211 : 1.02993 %
 1.90731 : 1.02654 %
 0.026738 : 0.933817 %
 1.65999 : 0.848018 %
 0.884583 : 0.844179 %
 1.06618 : 0.8091 %
 0.323085 : 0.788577 %
 0.0222817 : 0.782823 %
 1.24777 : 0.77056 %
 2.08891 : 0.752294 %
 0.00891268 : 0.73586 %
 0.176025 : 0.698008 %
 1.84158 : 0.678442 %
 1.42937 : 0.643756 %
 0.191623 : 0.575389 %
 2.13904 : 0.547822 %
 2.2705 : 0.536447 %
 2.02318 : 0.520179 %

## 3D Fourier Plots

```mathematica
ClearAll["Global`*"];
 (*System Parameters*)
 tmax = 10000 T; tmin = tmax - 50 T;
  (*Initial Conditions *)
  x0 = 0;
  y0 = 0;
  a = 5;
  ω = 7;
  T = 2 π / ω;
  initial = {x[0] == x0, y[0] == y0};
  (*System Equation*)
  f1 = x'[t] - y[t];
  f2 = y'[t] + x[t] + a (x[t]^2 - 1) y[t] - b Cos[ω t];
  data = {};
Null^10

bmin = 22;
bmax = 29;
bsteps = 20;
dsols = {};
For[i = 0, i < bsteps, i++,
  {b = bmin + (bmax - bmin)/(bsteps - 1) i, AppendTo[dsols, NDSolve[{f1 == 0, f2 == 0, initial}, {x, y}, {t, 0, tmax}, MaxSteps → ∞]]}]
```

General :: spell1 : Possible spelling error : new symbol name "bmin" is similar to existing symbol "tmin". More…

General :: spell1 : Possible spelling error : new symbol name "bmax" is similar to existing symbol "tmax". More…

```mathematica
list1={};
For[i=1,i≤bsteps, i++, AppendTo[list1,

Abs[Fourier[Table[x[t]/.dsols[[i]], {t, 0, tmax}]]]]]

 list2={};

For[i=1, i≤bsteps, i++, {list3={}, For[j=1, j<Length[list1[[1]]]/2,

j++, AppendTo[list3, list1[[i, j, 1]]]], AppendTo[list2, list3]}]

  ListPlot3D[list2]
```

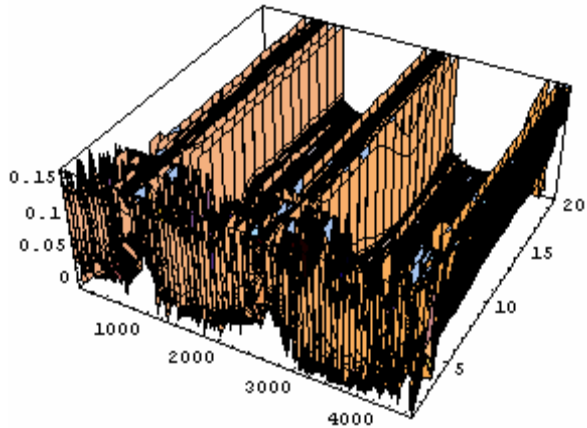

## Play Fourier Lists

```
ClearAll["Global`*"];
Off[General::spell1];
$HistoryLength=1;
link=Install["math2.exe"];
 (*System Parameters*)
 tmax = 10000 T;
 tmin = tmax - 1000 T;
 (*Initial Conditions *)
 x0 = 0.;
 y0 = 0.;
 a = 5.;
 ω = 7.;
 b = 15.;
 T = 2 π / ω;
 Nr = 2 10^6;
 initial = {x[0] == x0, y[0] == y0};
 (*System Equation*)
 x'[t] = y[t];
 y'[t] = -x[t] - a (x[t]^2 - 1) y[t] - b Cos[ω t];
 datai=SolveVDP[a, b, ω, x0, y0,Nr, N[tmax/Nr]];
minp=1;
pointsperperiod=10;
 minp = 4;
 pointsperperiod = 10;
 data = {};
 Module[{t}, For[t = tmin, t ≤ tmax, t += T/pointsperperiod,
    AppendTo[data, datai[[Floor[t/tmax Nr]]]]] ];
```

```
Module[{i, templist, tk, list2},
 list1 = Flatten[Abs[Fourier[data]]];
 list2 = {};
 For[i = 1, i ≤ Length[list1]/2, i++, AppendTo[list2, {N[(i - 1)/(tmax - tmin)], Abs[list1[[i]]]}]];
 list3 = {};
 templist = Ordering[list1[[Range[1, Floor[Length[list1]/2]]]], All, Greater];
 tempmax = Max[list1];
 For[i = 1, i < Floor[Length[list1]/2], i++,
  (tk = IntegerPart[templist[[i]]]; If[(list1[[tk]] ≥ list1[[tk - 1]]) && (list1[[tk]] ≥ list1[[tk + 1]]) && (list1[[tk]] ≥ (minp * tempmax)/100),
    AppendTo[list3, tk];];)];
 ListPlot[list2, PlotJoined → True, PlotRange → {{0, 1.1 Max[list3]/(tmax - tmin)}, {0, 1.1 tempmax}}];
 Print["Οι κύριες συχνότητες (ένταση > ", minp, "% της μέγιστης) και οι εντάσεις τους για b = ", b, " είναι"];
 For[i = 1, i ≤ Length[list3], i++, Print[N[list3[[i]]/(tmax - tmin)], ":", N[100 * list1[[list3[[i]]]]/tempmax, 1], "%"]];
]
```

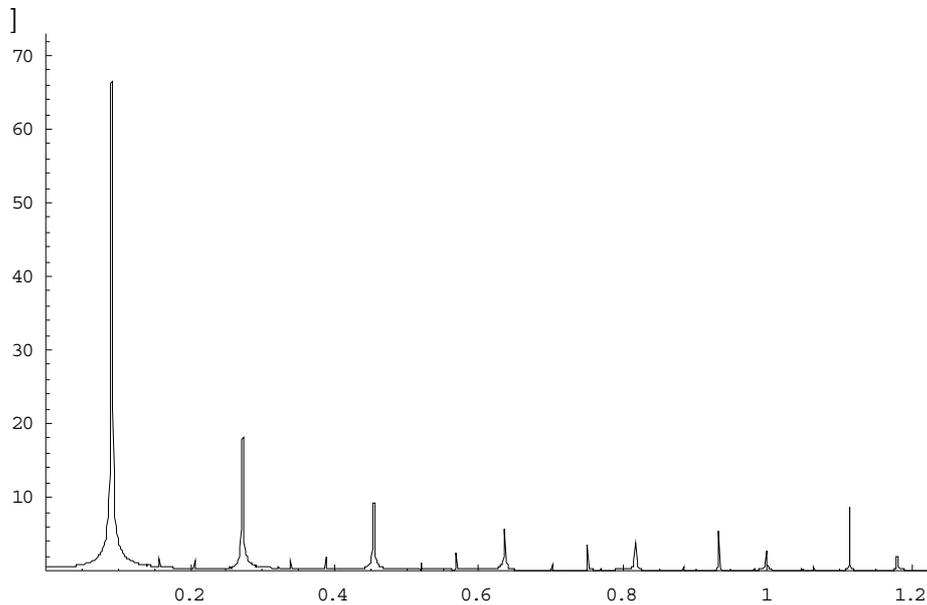

```
Οι κύριες συχνότητες (ένταση > 4 % της μέγιστης) και οι εντάσεις
τους για b = 15. είναι
0.092469 : 100. %
0.274065 : 27.4009 %
0.455661 : 13.886 %
1.1152   : 12.9809 %
0.636142 : 8.58138 %
0.933603 : 8.36459 %
0.817738 : 5.83309 %
0.752007 : 5.35049 %
0.999334 : 4.14991 %
```

**Uninstall[link];**

```
kanon = 10000;
hxos[t_] := Sum[list1[[list3[[i]]]]/tempmax Sin[kanon list3[[i]]/(tmax - tmin) t], {i, 1, Length[list3]}]
hxos2 = Compile[{t}, hxos[t]];
N[Evaluate[hxos2[1]]]
1.11794
Export["file1.wav", Play[hxos2[t], {t, 0, 4}]];
```

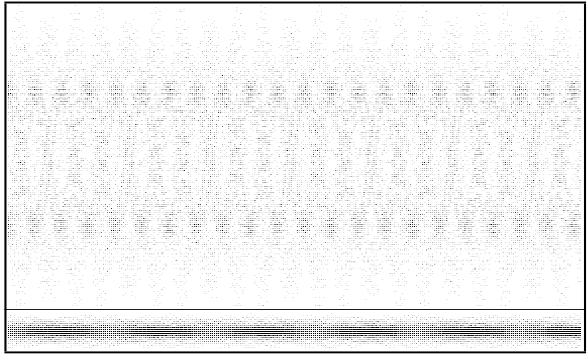

**N[$\frac{2\pi}{7}$ 11]**
9.87358
**N[1/0.1024]**
9.76563

### math2.exe

// math2.cpp : Defines the entry point for the console application.
//

```
#include "stdafx.h"
#include "mathlink.h"
#include <math.h>
#include <malloc.h>

int N, bst;
float ttot, ix0, ixt0, a, b, b2, w;
double Dt;

inline double g(double t, double x, double y)
{
            return (-x - a*(pow(x, 2) - 1)*y + b*cos(w*t));
}

void calculateRK(double x[], double y[])
{
            double t=0, k1, k2, k3, k4, j1, j2, j3, j4;
            int n, nx;
            for (n=0; n<N-1; n++)
            {
               nx=2*n;
               k1=y[n];
               j1=g(t, x[nx+1], y[n]);

               k2=y[n]+(Dt*j1/2);
```

```
                j2=g(t+(Dt/2), x[nx+1]+(k1*Dt/2), y[n]+(j1*Dt/2));
                k3=y[n]+(Dt*j2/2);
                j3=g(t+(Dt/2), x[nx+1]+(k2*Dt/2), y[n]+(j2*Dt/2));
                k4=y[n]+(Dt*j3);
                j4=g(t+Dt, x[nx+1]+(k3*Dt), y[n]+(j3*Dt));
                x[nx+1+2]=x[nx+1]+(Dt/6)*(k1 + 2*k2 + 2*k3 + k4);
                y[n+1]=y[n]+(Dt/6)*(j1 + 2*j2 + 2*j3 + j4);
                t=n*Dt;
                x[nx+0]=t;
            }
            x[nx+2]=(N-1)*Dt;
    }
    void solvevdp(double pa, double pb, double pw, double px0, double pxt0, int pN, double pDt)
    {
                printf(".");
                a = pa;
                b = pb;
                w = pw;
                ix0 = px0;
                ixt0 = pxt0;
                N = pN;
                Dt = pDt;
                double *x, *xt;
                long dims[2];//dme
                dims[0]=pN;//dme
                dims[1]=2;//dme
                //x = (double*)malloc(2*N*sizeof(double));
                x = new double[2*pN];
                xt = new double[pN];
                x[0, 1]=px0;
                xt[0]=pxt0;
                calculateRK(x, xt);
                MLPutRealArray(stdlink, x, (longp_st)&dims, NULL, 2);//dme
    //          MLPutRealList(stdlink, (doublep_nt)x, N);
    //          MLPutRealList(stdlink, (doublep_nt)x, N/2);
                delete []x;
                delete []xt;
    }
    int main(int argc, _TCHAR* argv[])
    {
    //          solvevdp(5, 0, 7, 0.1, 0.1, 1000000, 0.001);
                return MLMain(argc, (charpp_ct)argv);
    }
```

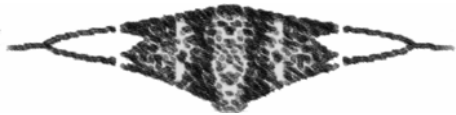

# References - Bibliography

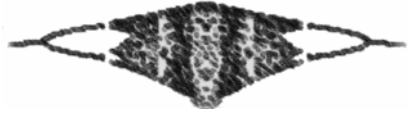